\documentclass[journal]{vgtc}

\usepackage{color}
\usepackage{ifthen}
\usepackage{float}
\usepackage{alltt}
\usepackage{amsmath}
\usepackage{amssymb}
\usepackage{amsthm}
\usepackage{newlfont} 
\usepackage{floatflt}
\usepackage{wrapfig}
\usepackage{fixltx2e}
\usepackage[caption=false]{subfig} 
\usepackage{multirow}
\usepackage{booktabs}
\usepackage{algorithmic}
\usepackage[export]{adjustbox}
\usepackage{mathtools, cuted}
\usepackage{CJKutf8} 

\usepackage{microtype}                 
\PassOptionsToPackage{warn}{textcomp}  
\usepackage{textcomp}                  
\usepackage{mathptmx}                  
\usepackage{times}                     
\usepackage{cite}                      

\usepackage{hyperref}
\usepackage{cleveref}
\hypersetup{draft}
\usepackage[ruled]{algorithm2e} 

\SetAlFnt{\small}
\SetAlCapFnt{\small}
\SetAlCapNameFnt{\small}
\SetAlCapHSkip{0pt}
\IncMargin{-\parindent}
\newcommand{\Caption}[2]{\caption[#1]{{\em #1} #2}}

\newcommand{\new}[1]{{{#1}}}

\definecolor{AudioColor}{rgb}{0.56,0.34,0.62}

\definecolor{VideoColor}{rgb}{0.44,0.66,0.38}

\definecolor{TodoColor}{rgb}{0.9,0.4,0} 

\definecolor{DoneColor}{rgb}{0.1,0.6,1.0} 

\definecolor{figred}{rgb}{1,0,0}
\definecolor{figgreen}{rgb}{0,0.6,0}
\definecolor{figblue}{rgb}{0,0,1}
\definecolor{figpink}{rgb}{1,0.63,0.63}

\newcommand{\pseudocode}{Algorithm}
\floatstyle{plain}
\newfloat{algorithm}{tbhp}{lop}
\floatname{algorithm}{\pseudocode}

\newcommand{\filename}[1]{\url{#1}}
\newcommand{\foldername}[1]{\url{#1}}

\let\oldparagraph\paragraph
\iffalse

\renewcommand{\paragraph}[1]{\oldparagraph{\textbf{#1}.}} 
\else

\renewcommand{\paragraph}[1]{\oldparagraph{{#1}.}}
\fi

\ifdefined\email
\else
\newcommand{\email}[1]{\url{#1}}
\fi

\newcommand{\contentFreq}{f}
\newcommand{\contentFreqVec}{\mathbf{f}}
\newcommand{\contentAmp}{A}
\newcommand{\band}{B}
\newcommand{\displayBand}{\band_{d}}
\newcommand{\retinalBand}{\band_{r}}
\newcommand{\retinalBandSup}{\sup\retinalBand}
\newcommand{\clampedBand}{\bar{\band}}
\newcommand{\eccentricityMask}{E}
\newcommand{\eccentricity}{\hat{P}_{ec}}

\newcommand{\specturmIntegralClampedOverall}{\Phi}
\newcommand{\specturmIntegralClamped}{\hat{\Phi}}
\newcommand{\csf}{s}

\newcommand{\poppingIntensity}{\hat{P}_{op}}
\newcommand{\adaptive}{\hat{P}}
\newcommand{\threeDimAdaptive}{\hat{P}}
\newcommand{\weight}{W}
\newcommand{\dataSize}{D}
\newcommand{\generalLuminance}{L}
\newcommand{\luminance}{L}
\newcommand{\weberWeight}{\omega}

\newcommand{\image}{I}
\newcommand{\imageApprox}{\hat{\image}}
\newcommand{\imageSpaceVec}{\mathbf{x}}
\newcommand{\imageSpaceX}{x}
\newcommand{\imageSpaceY}{y}

\newcommand{\gazeVec}{\mathbf{g}}
\newcommand{\gazeX}{g_x}
\newcommand{\gazeY}{g_y}

\newcommand{\bandStepSize}{b}

\newcommand{\pointContrast}{c}
\newcommand{\contrastSensitivityFilter}{a}

\newcommand{\threeDimUnit}{U}
\newcommand{\LoDLevel}{L}
\newcommand{\twoThreeDimMap}{M}
\newcommand{\timeStamp}{t}

\newcommand{\updatesize}{S}
\newcommand{\weightEccPop}{\lambda}

\newcommand{\cameraVec}{\mathbf{c}}
\newcommand{\network}{\mathcal{N}}

\newcommand{\uniformCondition}{{\bf UNI}\xspace}
\newcommand{\eccCondition}{{\bf ECC}\xspace}
\newcommand{\oursCondition}{{\bf OURS}\xspace}

\graphicspath{
{figs/handdrawn/}
{figs/raster/}
}

\onlineid{1151}
\vgtccategory{Research}

\begin{document}

\title{Instant Reality: Gaze-Contingent Perceptual Optimization for 3D \new{Virtual Reality} Streaming}

\author{Shaoyu~Chen,
	Budmonde~Duinkharjav,
	Xin~Sun,
	Li-Yi~Wei,
	Stefano~Petrangeli,\\
	Jose~Echevarria,
	Claudio~Silva,~\textit{Fellow,~IEEE,}
	Qi~Sun
}
\authorfooter{
	\item S. Chen, B. Duinkharjav, C. Silva, Q. Sun are with New York University\\
	E-mail: \{sc6439, budmonde, csilva, qisun\}@nyu.edu.\\
	\item X. Sun, L. Wei, S. Petrangeli, J. Echevarria are with Adobe Research. \\
	E-mail:\{xinsun, lwei, petrange, echevarr\}@adobe.com.
}

\abstract{
Media streaming, with an edge-cloud setting, has been adopted for a variety of applications such as entertainment, visualization, and design.
Unlike video/audio streaming where the content is usually consumed passively, virtual reality applications require 3D assets stored on the edge to facilitate frequent edge-side interactions such as object manipulation and viewpoint movement.
Compared to audio and video streaming, 3D asset streaming often requires larger data sizes and yet lower latency to ensure sufficient rendering quality, resolution, and latency for perceptual comfort.
Thus, streaming 3D assets faces remarkably additional than streaming audios/videos, and existing solutions often suffer from long loading time or limited quality.

To address this challenge, we propose a perceptually-optimized progressive 3D streaming method for spatial quality and temporal consistency in immersive interactions.
On the cloud-side, our main idea is to estimate perceptual importance in 2D image space based on user gaze behaviors, including where they are looking and how their eyes move.
The estimated importance is then mapped to 3D object space for scheduling the streaming priorities for edge-side rendering.
Since this computational pipeline could be heavy, we also develop a simple neural network to accelerate the cloud-side scheduling process.
We evaluate our method via subjective studies and objective analysis under varying network conditions (from 3G to 5G) and edge devices (HMD and traditional displays), and demonstrate better visual quality and temporal consistency than alternative solutions.
}

\teaser{
\centering
\begin{minipage}{0.23\linewidth}
\subfloat[][Edge view]{\includegraphics[width=0.96\linewidth]{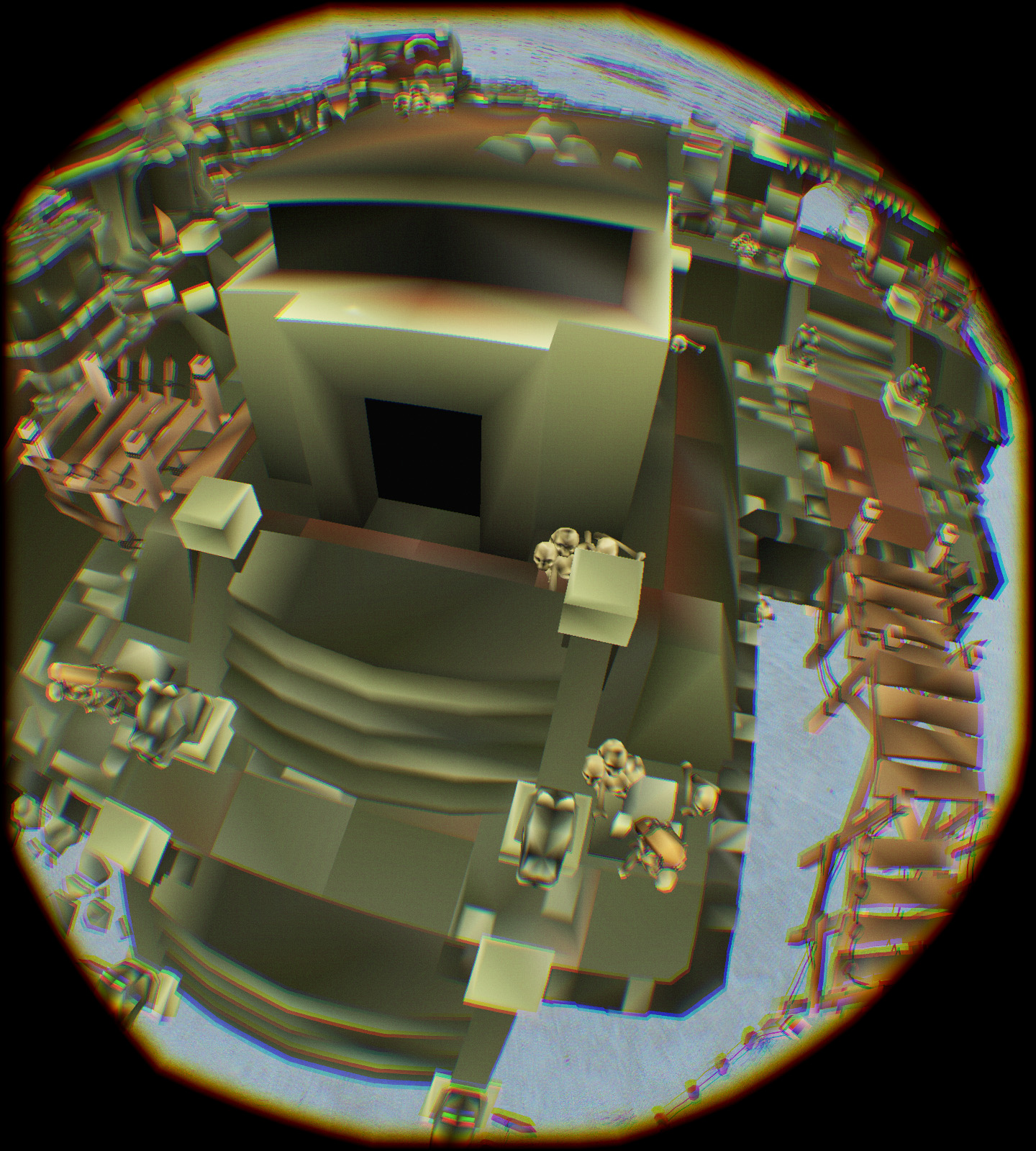}\label{fig:teaser:current}}\\
\subfloat[][Cloud data]{\includegraphics[width=0.96\linewidth]{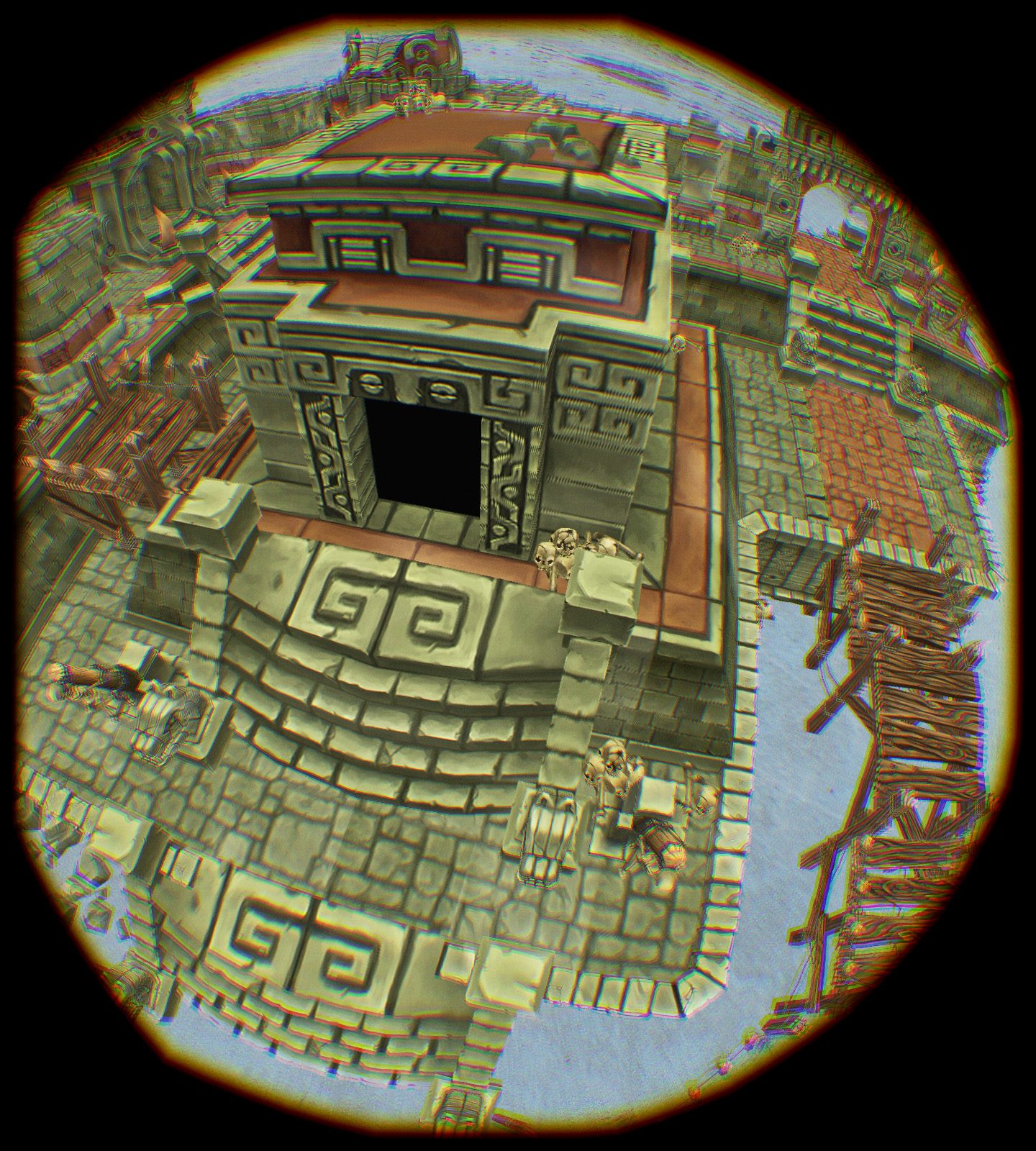}\label{fig:teaser:target}}
\end{minipage}
\begin{minipage}{0.23\linewidth}
\subfloat[][Uniform streaming]{\includegraphics[width=0.96\linewidth]{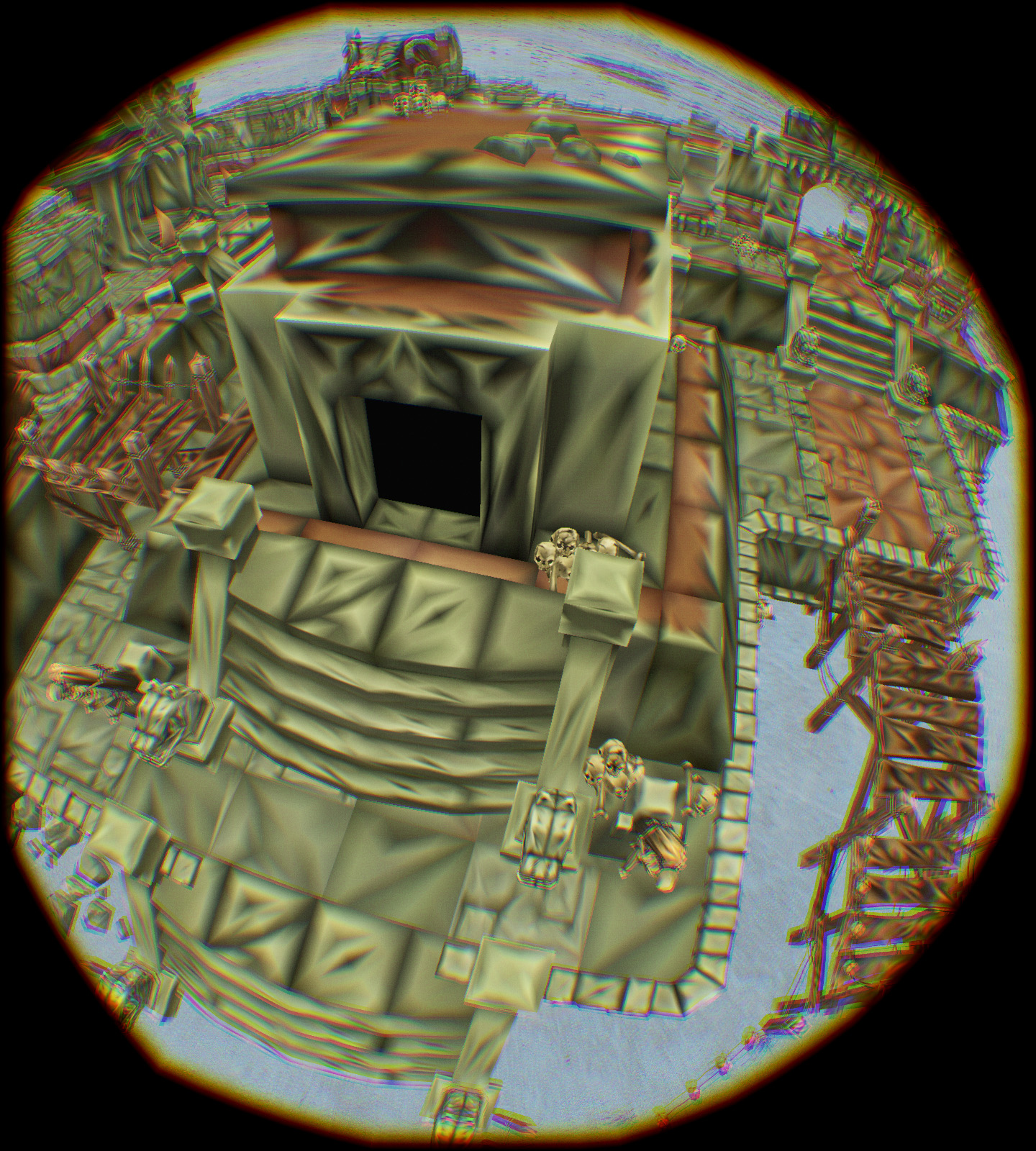}\label{fig:teaser:uniform_hmd}}\\
\subfloat[][Quality for \subref{fig:teaser:uniform_hmd}]{\includegraphics[width=0.96\linewidth]{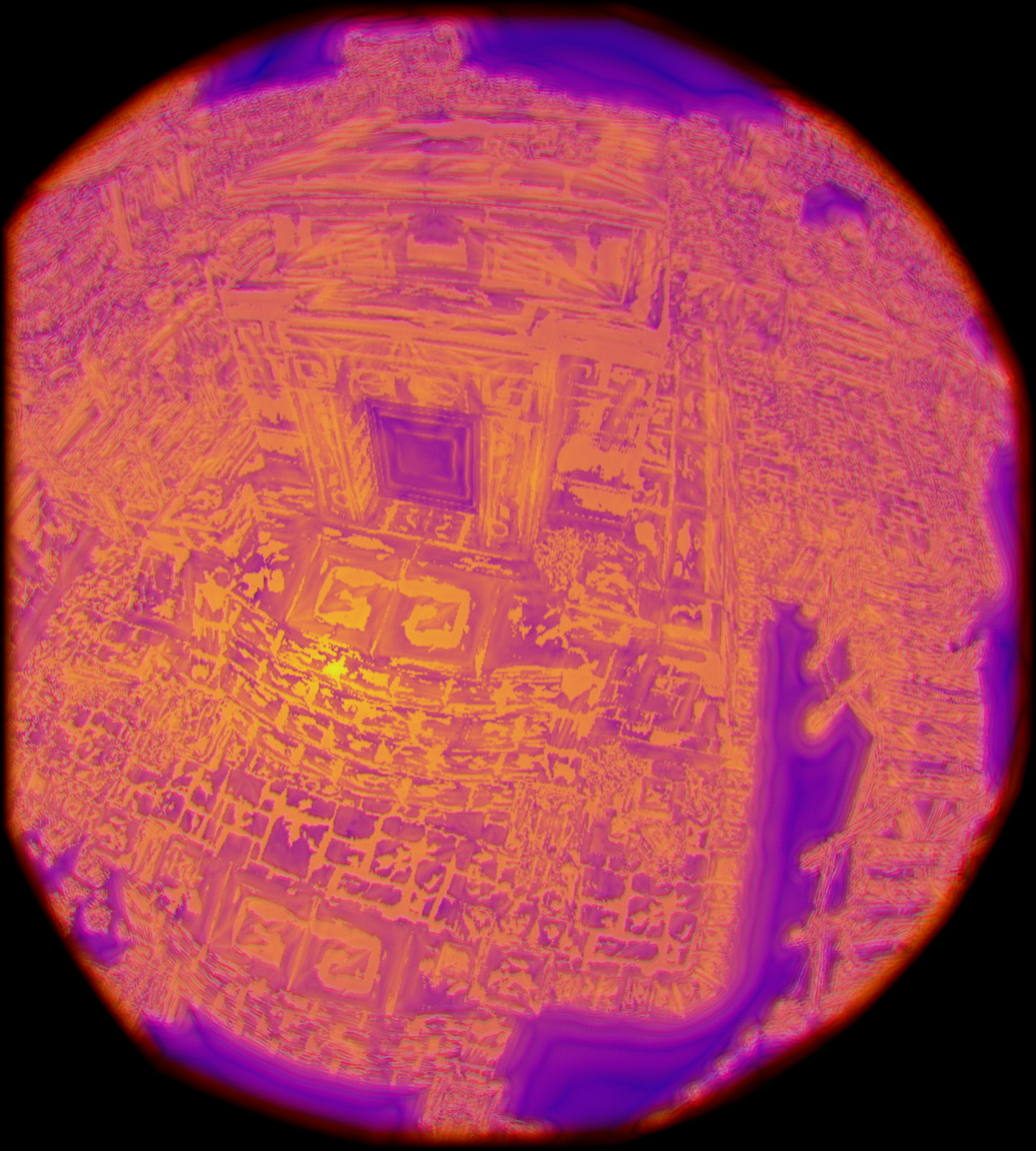}\label{fig:teaser:uniform_perception}}
\end{minipage}
\begin{minipage}{0.23\linewidth}
\subfloat[][Fixation-based streaming]{\includegraphics[width=0.96\linewidth]{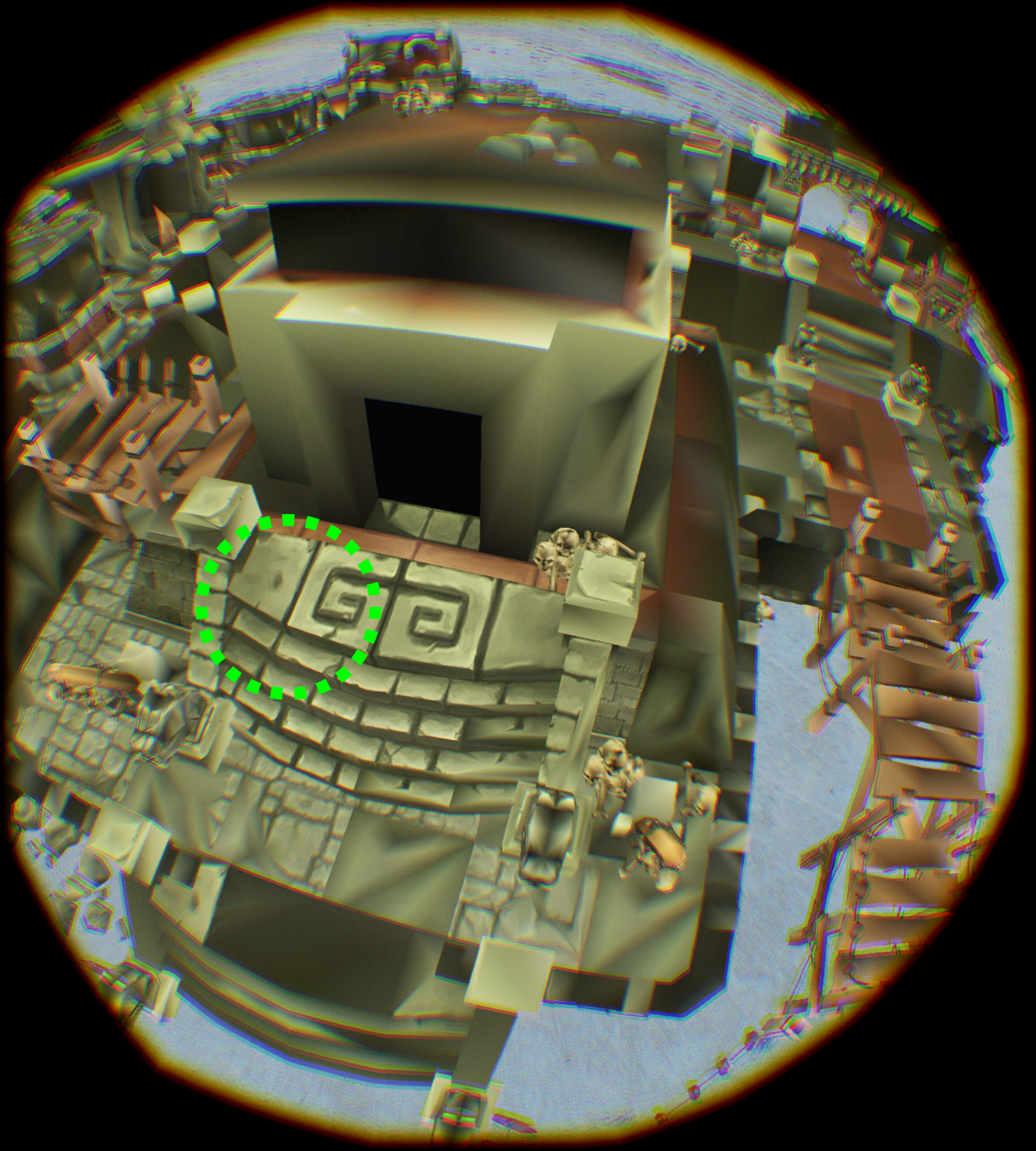}\label{fig:teaser:fix_hmd}}\\
\subfloat[][Quality for \subref{fig:teaser:fix_hmd}]{\includegraphics[width=0.96\linewidth]{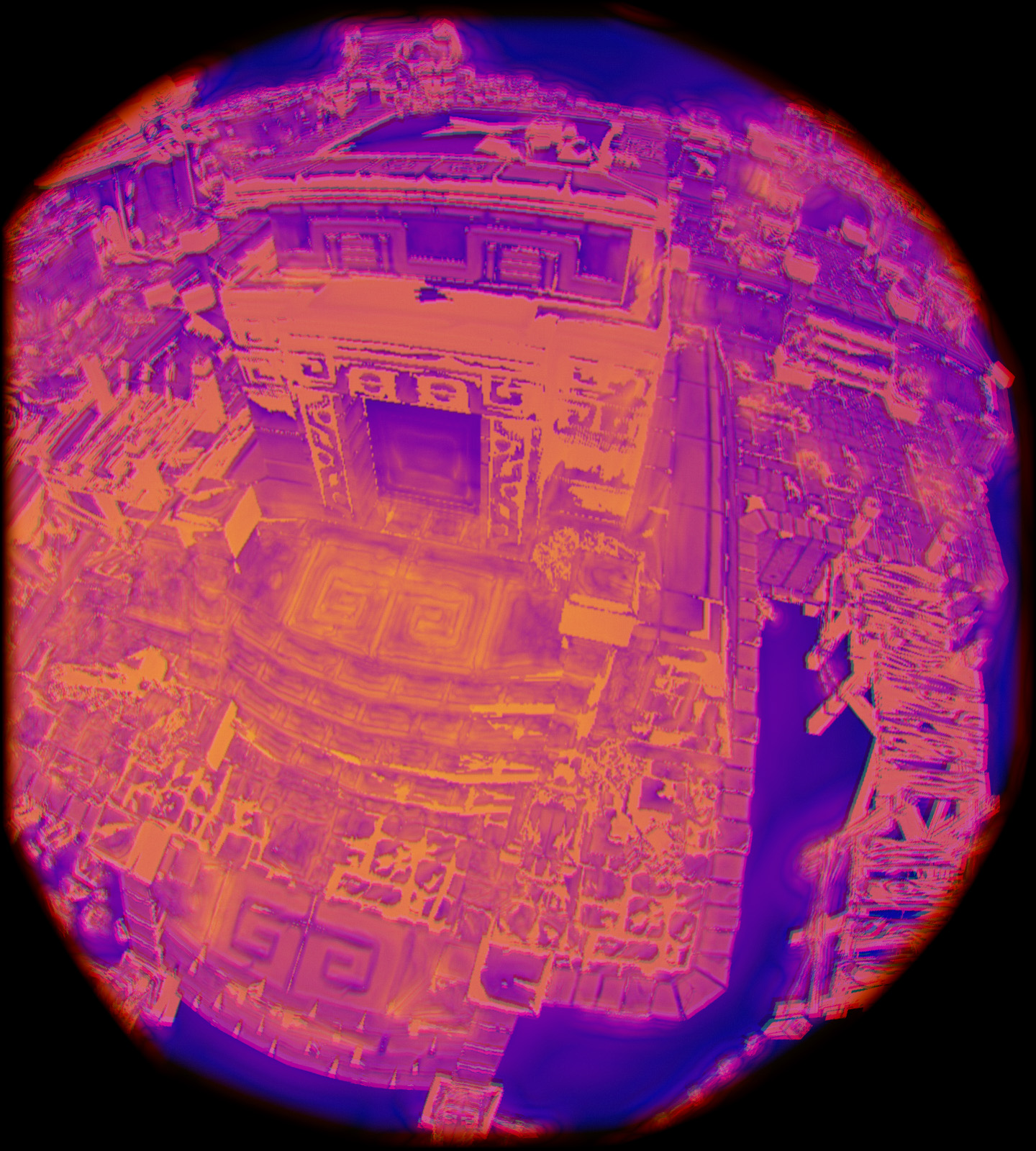}\label{fig:teaser:fix_perception}}
\end{minipage}
\begin{minipage}{0.23\linewidth}
\subfloat[][Saccade-based streaming]{\includegraphics[width=0.96\linewidth]{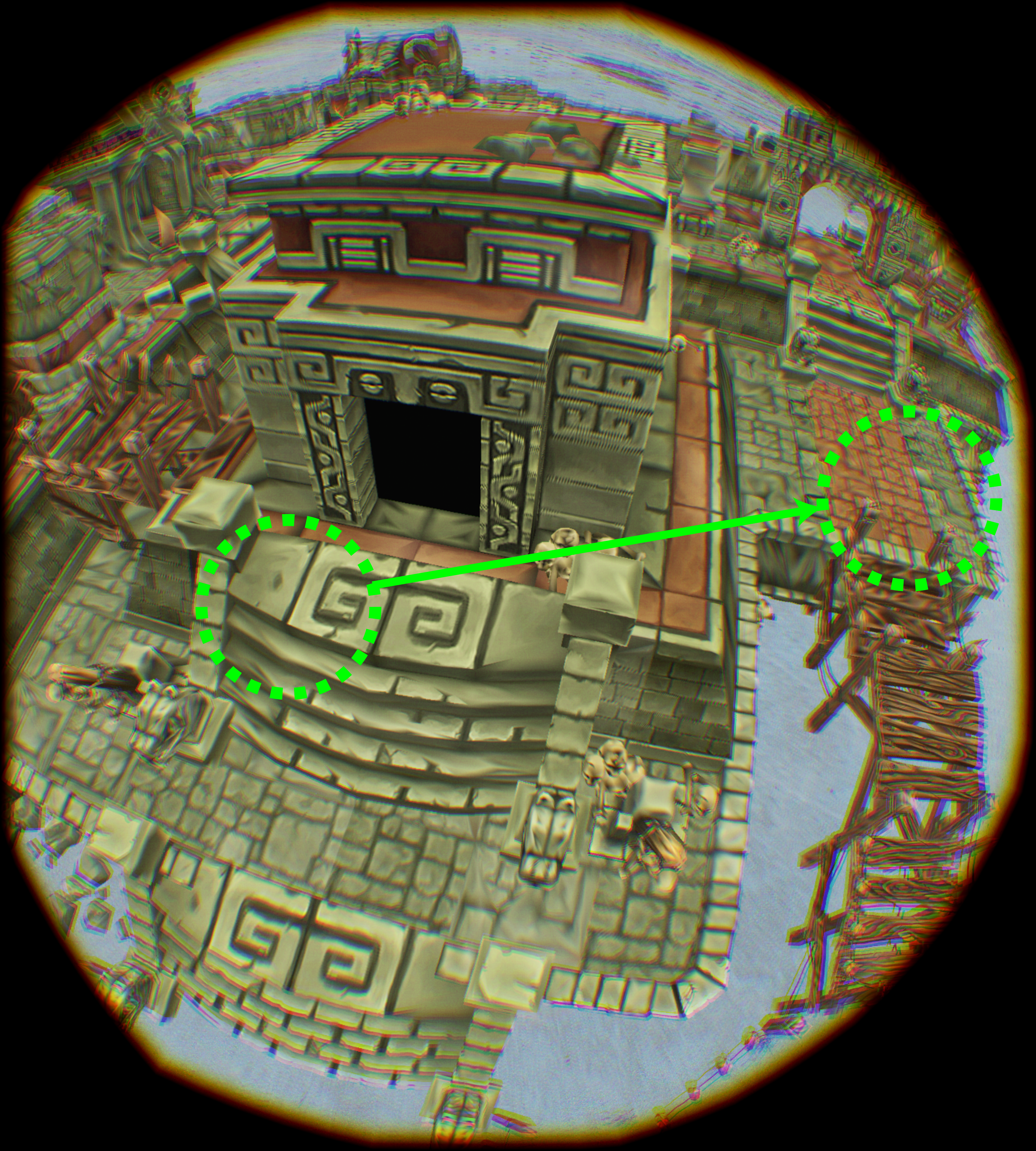}\label{fig:teaser:saccade_hmd}}\\
\subfloat[][Quality for \subref{fig:teaser:saccade_hmd}]{\includegraphics[width=0.96\linewidth]{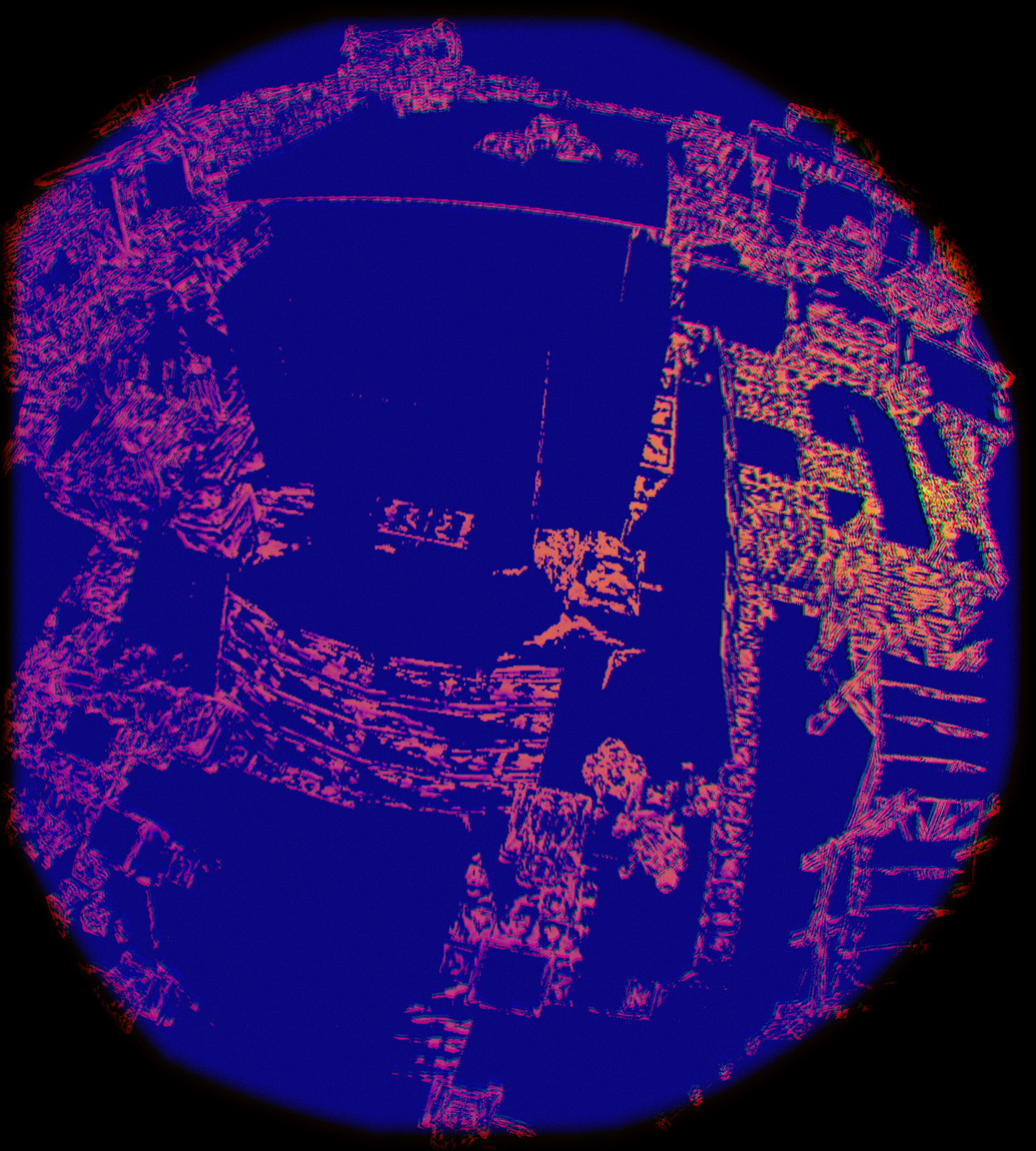}\label{fig:teaser:saccade_perception}}
\end{minipage}
\Caption{Our gaze-contingent immersive 3D assets streaming interface.}
{
Starting from the partially-streamed 3D assets on the edge-side rendered at a given time \subref{fig:teaser:current}, our method streams additional updates from the cloud to the edge for perceptually closer rendering to the full assets stored on the cloud \subref{fig:teaser:target}. 
Standard uniform streaming \subref{fig:teaser:uniform_hmd} evenly updates all visible assets in the scene, causing suboptimal perceptual quality in \subref{fig:teaser:uniform_perception}, which visualizes both the temporal (popping with respect to \subref{fig:teaser:current}) and the spatial (quality with respect to \subref{fig:teaser:target}) perceptual errors.
Brighter colors indicate worse artifacts.
Our method, in contrast, optimizes the subset of the assets to be streamed to the edge for better spatio-temporal quality under the same network bandwidth. 
Our perceptual model considers both eccentricity-based acuity during fixation \subref{fig:teaser:fix_hmd} and temporal masking during eye movement \subref{fig:teaser:saccade_hmd}. 
If the user fixes the gaze \subref{fig:teaser:fix_hmd}, our model prioritizes regions near the gaze point (green circle) while reducing potential popping artifacts.
If a saccade is detected \subref{fig:teaser:saccade_hmd}, our model can safely ignore popping to stream more aggressive updates for further quality improvement \subref{fig:teaser:saccade_perception}.
}
\label{fig:teaser}
}

\maketitle

\section{Introduction}

Thanks to the rapid growth of Internet technologies such as 5G, streaming services have unprecedentedly revolutionized how we access and consume high-quality multimedia, from listening to audios, watching videos, to scientific discovery (e.g., NASA's Eyes\footnote{\new{https://eyes.nasa.gov/}}). The edge-cloud streaming setting significantly reduces edge-side computation and redundant offline storage, allowing for portable and wearable consumer devices.

When it comes down to VR/AR, storing all the assets offline is impractical for dynamic and location-based scenarios. Thus, streaming assets from the cloud and perform the rendering on the edge-side \new{become} an effective solution. 
However, VR streaming poses unique challenges due to the required high resolution, high frame rate, low latency, and stereo rendering. 
Frame-based 2D streaming falls unrealistic given the exceptionally high demands of data volume and responsiveness.
Still, current approaches fail in streaming large high-quality 3D assets, introducing unacceptably long loading times or undesirable visual and interactive artifacts (please refer to our video for examples).
Comparisons of the recent  $250\%$ FLOPS gains in the last two consecutive GPU generations \footnote{https://www.digitaltrends.com/computing/nvidia-rtx-3090-vs-rtx-2080-ti-most-powerful-gaming-gpus-duke-it-out/} versus a smaller growth of only $26\%$ in global internet bandwidth during the same time \footnote{https://blog.telegeography.com/466-tbps-the-global-internet-continues-to-expand} also suggest the main roadblock for immersive 3D experiences lies in the delayed access to 3D assets, instead of computational resource. 

Perceptual mechanisms, such as foveation, have been harnessed in recent efforts to accelerate interactive rendering \new{\cite{Koulieris:2019:NED,Murphy:2001:GCL}}.
However, the acceleration only occurs after all assets are stored at the user-end, thus improving the computation from, but not the transmission of, scene assets (especially complex geometries).
Gaze-contingent effects have been used to optimize the limited network bandwidth for 2D frame-based streaming \cite{Kaplanyan:2019:DNR,Romero:2018:FSV} but not 3D assets.
Existing solutions for 3D content streaming often have globally uniform granularity, such as game levels or camera distance. 
Finer-grained controls can be achieved based on visibility \cite{Hladky:2019:COS} or viewpoint \cite{Schutz:2020:PRR} to optimize level-of-details (LoD).
However, without comprehensive optimization for human perceived quality, the adaptive streaming may cause strong visual artifacts such as temporal popping or low quality with limited network bandwidth. 

To this end, we propose a 3D LoD-based streaming framework that automatically schedules the data transmission priority towards optimal perceptual quality.
Our method is based on modeling human spatio-temporal perception and adapts to varying network conditions.
Our key idea is to estimate perceptual sensitivity based on spatio–temporal user gazes, including foveation, saccade, and popping, so that we can more aggressively optimize static image quality and dynamic frame change, such as image regions outside the visual fovea or quickly glanced over during saccade.
The perceptual importance is then transferred to transmission priorities from the cloud to the edge.
To further enable real-time responses without extra latency caused by the complex frequency domain computations, we also implemented a simple neural-network-based accelerator for fast importance computation.
Our method is general enough to support various data formats, such as meshes, volumes, height fields, and point clouds.
\new{It targets high field-of-view and eye-tracked displays that require interactive streaming, such as cloud-based gaming.}
The framework is also compatible with transmitting geometric transformations in dynamic scenes, as our proof-of-concept shows in the supplementary video.

We conducted both subjective studies (via VR and traditional displays) and objective analysis based on the recorded data, which indicated higher quality and lower artifacts of our method than alternative solutions under identical bandwidth.
We also tested our method under varying latencies by simulating real-world network transmission rates (from 3G to 5G).

In summary, the main contributions of this paper are:
\begin{itemize}
\item a perceptually optimized high-quality and low-latency 3D immersive streaming framework, supporting various 3D computer graphics data formats;

\item a gaze-contingent perceptual model based on foveation, saccade, and popping to depict spatio-temporal visual behaviors during procedural streaming, including static quality acuity and dynamic change suppression;

\item a neural-network-based accelerator for real-time computation of the complex perceptual mechanisms and large data volumes;

\item a series of subjective studies and objective analysis validating our method under varying network conditions.
\end{itemize}

\new{Details on the implementation can be found in the InstantReality Github repository\footnote{\new{https://github.com/chenshaoyu1995/InstantReality}}.}

\section{Related Work}
\label{sec:prior}

\subsection{Perception-Aware Graphics}
With the rapid deployment of eye-tracking and personal electroencephalography (EEG) devices, perceptually-aware computer graphics approaches have drawn extensive attention from various applications.
Gaze-contingent rendering\new{\cite{Weier:2017:PAR}, as a representative example, involves understanding the visual optics\cite{Guenter:2012:F3G,Patney:2016:TFR}, anatomical retina\cite{Sun:2017:PGF}, peripheral perceptual encoding\cite{Brown:2021:EDM} and higher-order neural process\cite{Walton:2021:BBR}}.
Perceptual models can also been deployed in depth perception \cite{Krajancich:2020:ODP,Konrad:2019:GCO}, saliency \cite{Sitzmann:2019:SVH}, color perception \cite{Cohen:2020:LCA}, visual masking \cite{Ferwerda:1997:MVM} , visual attention \cite{Bruckert:2021:WLM}, navigation in VR \cite{Sun:2018:TVR,Hu:2019:RSS}, light\new{/luminance perception \cite{Luidolt:2020:GDS, Wang:2020:FIR, Tursun:2019:LCA}}, and spatio-temporal media \cite{Rashidi:2020:OVS,Serrano:2017:MEC,Schwarz:2009:PVP}. 
However, previous perceptually-based graphics systems focus on optimizing local content than transmission.
In cloud-based 3D scenarios, solutions on perceived low-latency and temporal consistency have been under-explored.
These factors are critical for a smooth and realistic experience, comparing with the less demanding local computation on the user end.

\subsection{Low-Latency Streaming in Cloud-Edge Systems}
\new{Low-latency streaming has been widely practiced in cloud-based 2D content services, such as film/video platforms and live broadcasting, based on factors such as network conditions\cite{Shuai:2015:Icc, Gutterman:2020:MMSys}, foveated vision\cite{Romero:2018:FSV,Kaplanyan:2019:DNR}, viewport and visible area\cite{Shu:2019:FFR,Zhang:2021:DVP}.
With the current frameworks, the rendering occurs on the cloud side, requiring the network to stream each individual frame. The frequent transmission may harm the experience for highly interactive and latency-sensitive VR/AR scenarios. For instance, e-sports, high quality 3D model design, or physics-based simulation.}

\subsection{3D Data Transmission}
How to prioritize 3D primitives is critical to 3D streaming,
which has been studied with little consideration of human perception.
To incorporate the unique structures of 3D meshes, Hladky et al. \cite{Hladky:2019:COS} prioritize the triangle components in the streaming pipeline based on their individual visibility.
In a real cloud-based system, object-wise LoDs shall be determined with respect to the virtual camera views and the available resources (bandwidth) and constraints (reaction time) \cite{Petrangeli:2019:DAS}. It is also possible to perform progressive rendering without building hierarchical structures \cite{Schutz:2020:PRR}.
However, these systems rely on particular data types such as meshes or point clouds. 
\new{A recent advancement to enable a decoupled and efficient rendering-transmission is through irradiance volume data \cite{Stengel:2021:DDS}.}
In highly interactive scenarios, however, the virtual content may comprise various components, such as particles for physics-based simulation, meshes for characters, and point clouds from sensors.
Moreover, under near-eye high field-of-view and high frame rate immersive displays, the systems may introduce the well-known popping artifacts \cite{Luebke:2002:LD3}.
Our system, by analytically modeling human perception, preserves both visual quality and temporal stability.

\section{Method}
\label{sec:method}

Given a limited network bandwidth and a set of scene data, we optimize and determine the streaming priority with regard to individual user's spatio-temporal perception.
The priority is updated according to users' static and dynamic gaze behaviors. 
Our goal is to maximize perceived quality (compared with a fully local rendering) and ensure dynamic smoothness (minimize popping artifacts).

\begin{figure*}
	\centering
	\includegraphics[height=4.5cm]{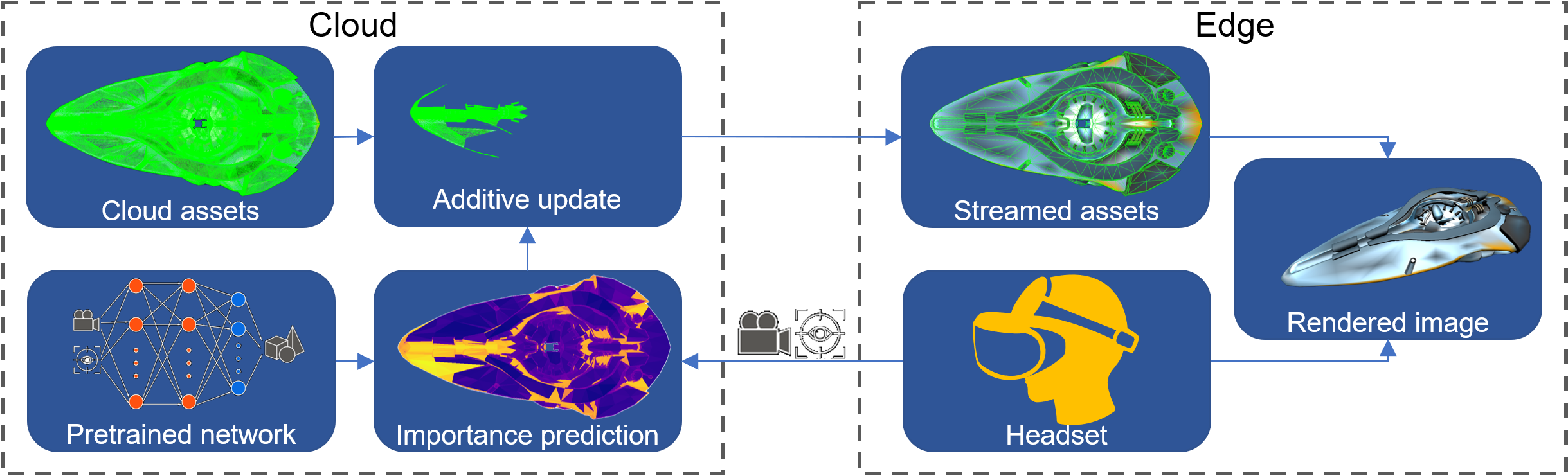}
	\Caption{\new{Overview of the system.}}
	{%
		\new{The parameters of the camera and gaze are sent to cloud, and they are used as the input of the pretrained neural network to predict the perceptual importance of primitives. The image is rendered on edge with the streamed assets.}
	}
	\label{fig:system}
\end{figure*}

We first describe the factors in our spatio-temporal perception model (\Cref{sec:vision}), followed by our method in 2D image space (\Cref{sec:method:2d}) and 3D object space (\Cref{sec:method:3d}).
\Cref{eqn:adaptive} summarizes our key idea.
Finally, to enable the cloud's real-time responses to users' dynamic head/gaze motions, we accelerate the system via a deep neural network (\Cref{sec:method:neural}).
At each moment, the cloud (for computation and storage) receives the gaze and existing content from the edge (for rendering and interaction).
Then, guided by our gaze-contingent and perception/content-aware model, the cloud automatically determines the streaming priority of each asset component based on their added perceptual quality and data size (with regard to a given network bandwidth). \new{\Cref{fig:system} shows an overview of our system.}

\subsection{Modeling Spatio-temporal Vision}
\label{sec:vision}

\paragraph{Spatial visual acuity}
\new{The human visual acuity is foveated, with a deterioration along age \cite{Elliott:1995:VAC} and visual field eccentricity \cite{Patney:2016:TFR,Mantiuk:2021:FVD}.}
The importance of a pixel $\eccentricityMask(\imageSpaceVec)$ is determined by its distance to the retinal eccentricity $r = \sqrt{x^2+y^2}$.
As the retinal eccentricity $r$ increases, the importance function $\eccentricityMask(\imageSpaceVec)$ decreases.
Please refer to \Cref{sec:sup:foveation} for the details of $\eccentricityMask(\imageSpaceVec)$.

\begin{figure}[tb]
	\centering
	\subfloat[4 cycle / im]{
		\includegraphics[width=0.23\linewidth]{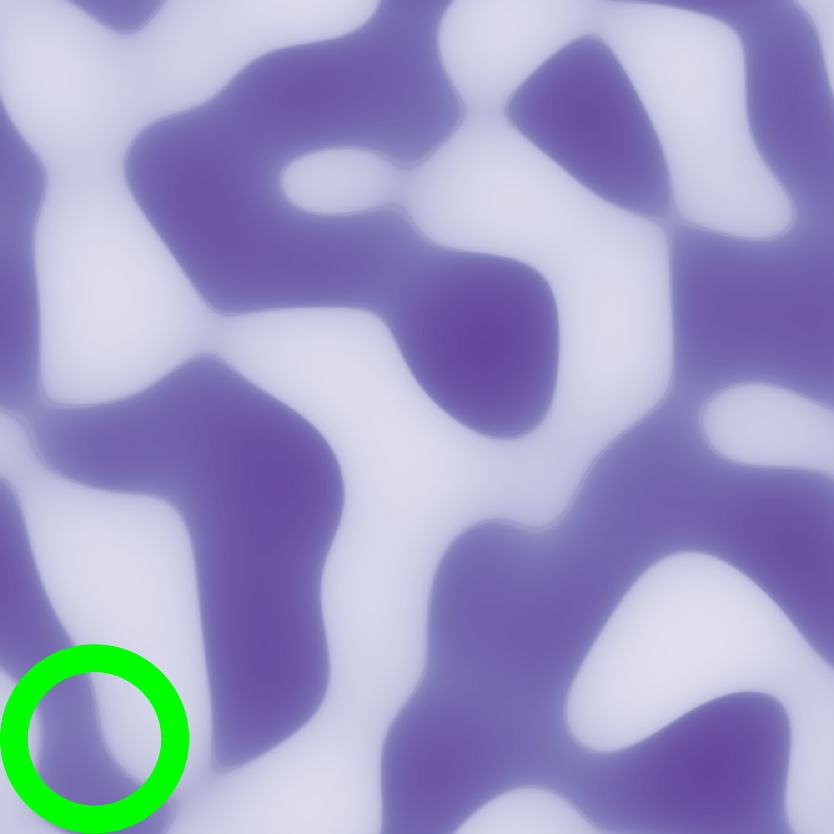}\label{fig:bandpass_filtered_images:2}
	}
	\subfloat[16 cycle / im]{
		\includegraphics[width=0.23\linewidth]{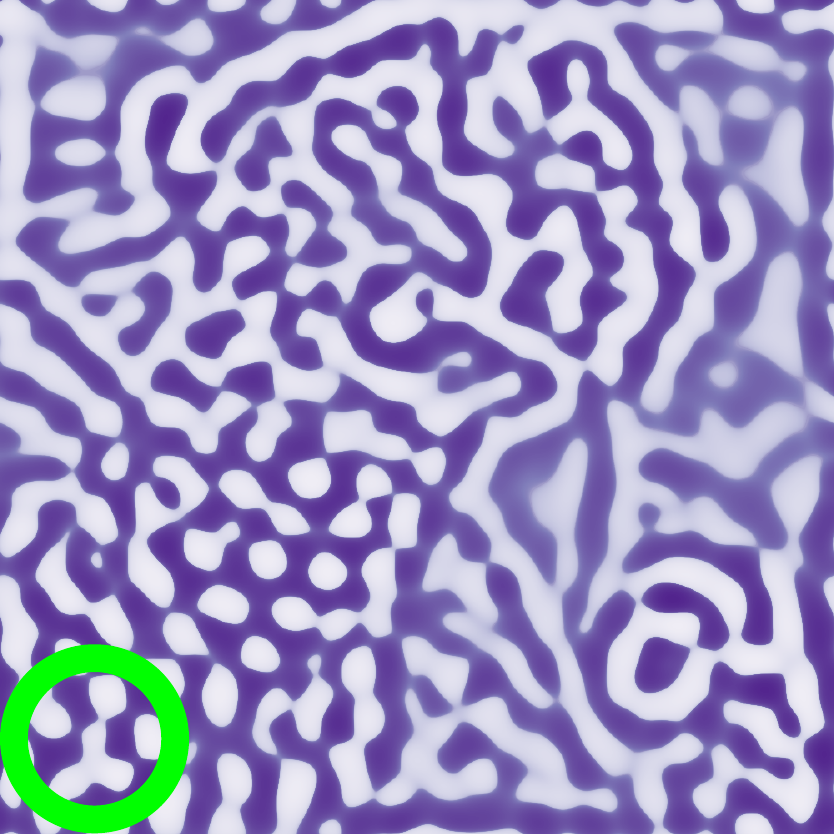}\label{fig:bandpass_filtered_images:4}
	}
	\subfloat[64 cycle / im]{
		\includegraphics[width=0.23\linewidth]{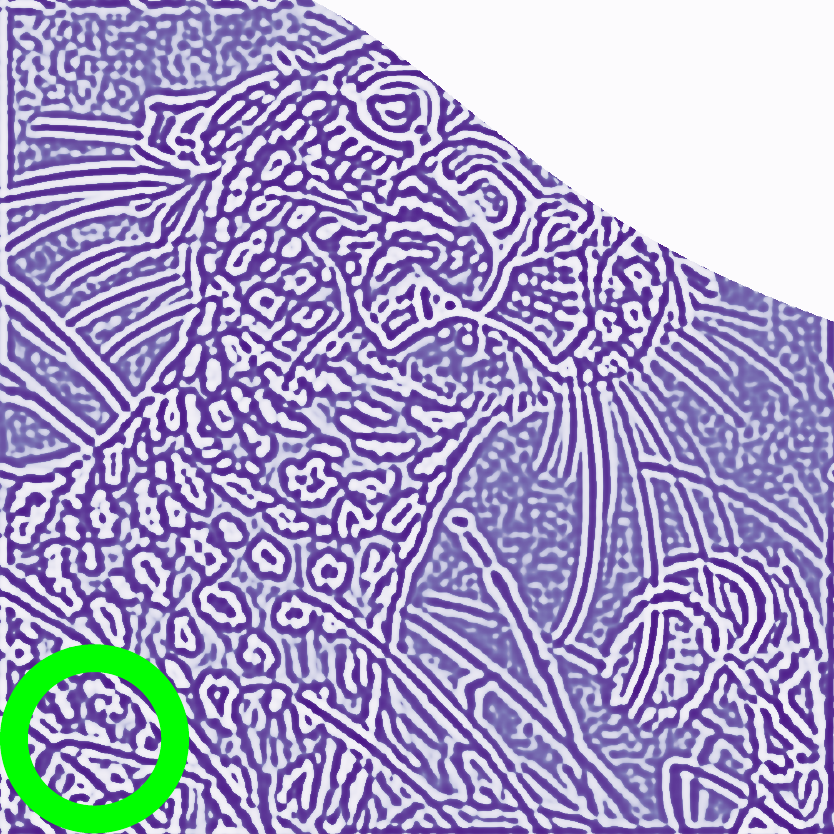}\label{fig:bandpass_filtered_images:6}
	}
	\subfloat[256 cycle / im]{
		\includegraphics[width=0.23\linewidth]{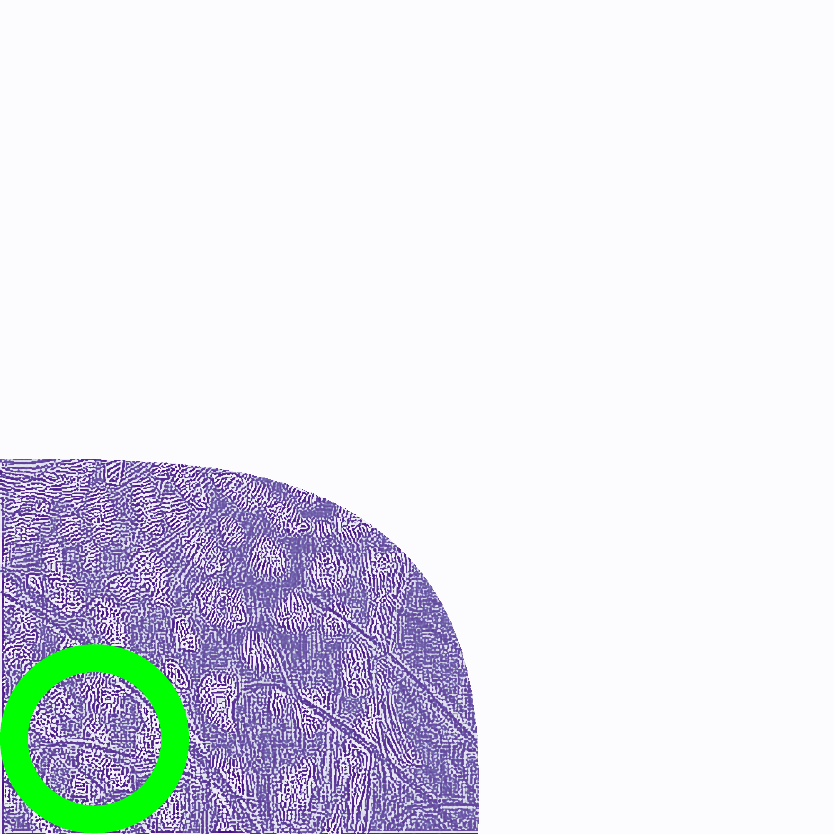}\label{fig:bandpass_filtered_images:8}
	}
	\Caption{Decomposition visualization of bandpass filtered contrast multiplied by contrast sensitivity function of corresponding frequency band ($\pointContrast(\imageSpaceVec,\contentFreqVec_i,\image)$ from \protect\Cref{eq:poppingIntensity} and \protect\Cref{eq:point_contrast} in \Cref{sec:sup:bandpass}).}
    {%
    The green circle indicates gaze position. When $\contentFreqVec_i$ is low (\subref{fig:bandpass_filtered_images:2} and \subref{fig:bandpass_filtered_images:4}), 
    the bandpass filtered contrast is computed over the entire visual field based on visual content. 
    With high $\contentFreqVec_i$ (\subref{fig:bandpass_filtered_images:6} and \subref{fig:bandpass_filtered_images:8}), the periphery sensitivity was clamped by $\retinalBand$, thus the empty content.
	}
	\label{fig:bandpass_filtered_images}
\end{figure}

\paragraph{Static stimuli}
As shown in \Cref{fig:bandpass_filtered_images}, we model the perception of a static image by separating it into a number of frequency bands.
The perception for each frequency band~$\contentFreqVec_i$ is the product of its bandpass-filtered contrast and its contrast sensitivity function.
The foveated retinal band $\retinalBand(\gazeVec,\imageSpaceVec) = \eccentricityMask(\gazeVec-\imageSpaceVec)$ will also affect the perception, results in clamped peripheral part for high frequency content.
Please refer to \Cref{sec:sup:bandpass} for details.

\paragraph{Dynamic stimuli}
In runtime, the level of content LoD constantly changes and the quality is improved gradually,
because more detailed data is streamed continually with the available network bandwidth.
However, abrupt changes within a small time period may cause popping artifacts. Similar to \cite{Schwarz:2009:PVP}, we define the perceived {\it change} as a temporally adapted Weber's contrast considering short-term memory.

\paragraph{Change blindness}
The gaze fixations and the dynamic transitioning motions among them also significantly alter the perceptual quality. 
Besides motion parallax \cite{Serrano:2019:MPR} and smooth pursuit movements to keep tracking on objects, people perform very fast eye movements (a.k.a. saccades) to change among fixations. Due to the fast movement, the visual contrast sensitivity ($\pointContrast$) is suppressed as studied in \cite{Knoell:2011:SPP}. We ran a pilot study and validated the suppressed perception of the popping artifacts along with the sensitivity drop.
Therefore, we adaptively model visual sensitivities w.r.t dynamic gaze behaviors including fixations and saccades, as detailed next in \Cref{sec:method:2d}.
\new{Albert et al. \cite{Albert:2017:LRF} suggested for foveated rendering, an overall system latency of 50-70ms is acceptable when users are specifically tasked with finding artifacts, and the requirements shall be further relaxed for regular tasks.}
Although reading-introduced saccades can be as short as 20ms, both short- and long- peri-saccades are combined with post-saccade suppression that stretches the applicable duration to mitigate popping artifacts. 
Previous literature has shown that saccadic suppression may typically last for at least 100ms even with a 50ms saccade \cite{Ibbotson:2011:VPS}.
Thus, the allowable saccadic suppression durations contain modern mobile network latencies ($\sim100$ms for 3G) \cite{Serrano:2009:LBM}.

\subsection{Model in Screen Space}
\label{sec:method:2d}

\begin{figure*}[thb]
\centering
	\subfloat[current]{
		\includegraphics[width=0.17\linewidth]{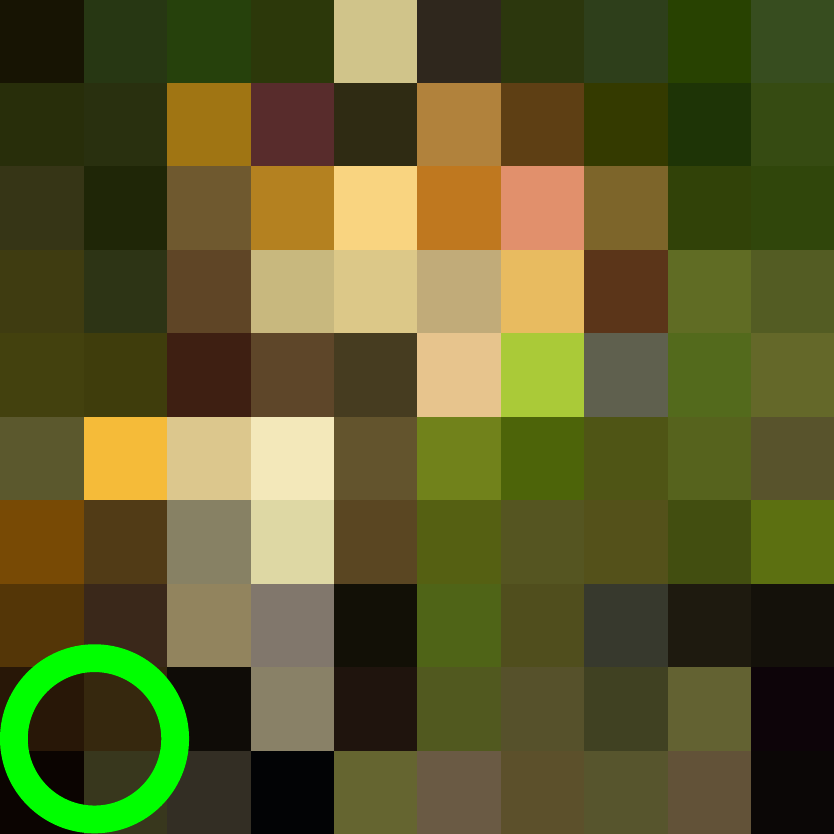}\label{fig:results:2d:current}
	}
	\hspace{0.3em}
	\subfloat[$\eccentricity$]{
		\includegraphics[width=0.17\linewidth]{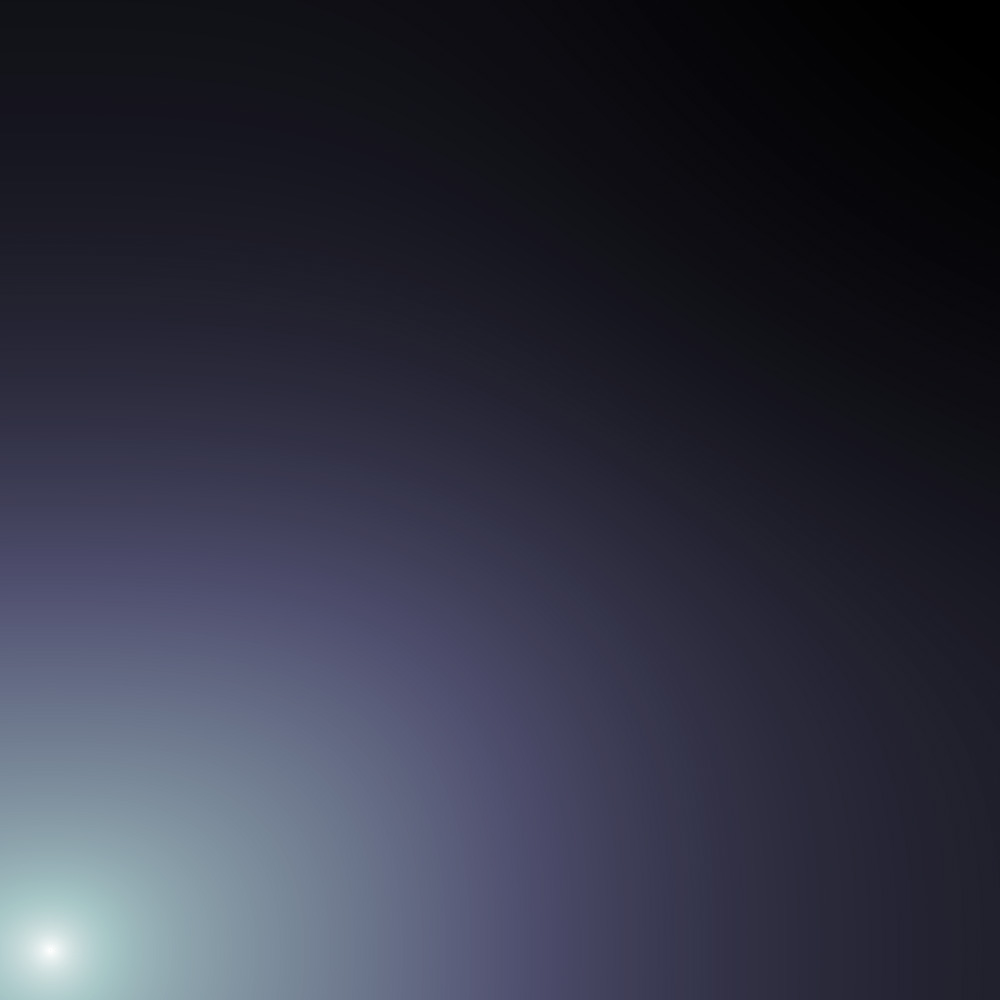}\label{fig:results:2d:eccHeat}
	}
	\subfloat[$\poppingIntensity$]{
		\includegraphics[width=0.17\linewidth]{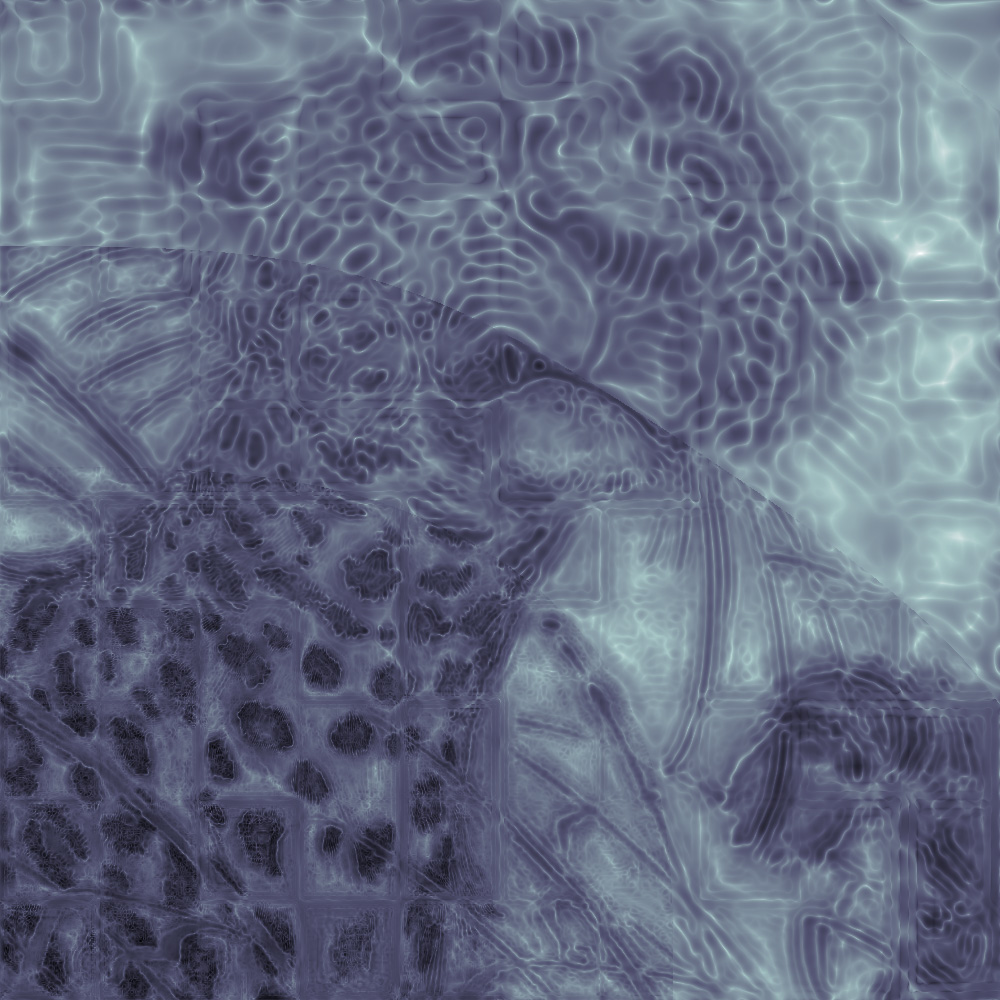}\label{fig:results:2d:popHeat}
	}
	\hspace{0.3em}
	\subfloat[fixation perception]{
		\includegraphics[width=0.17\linewidth]{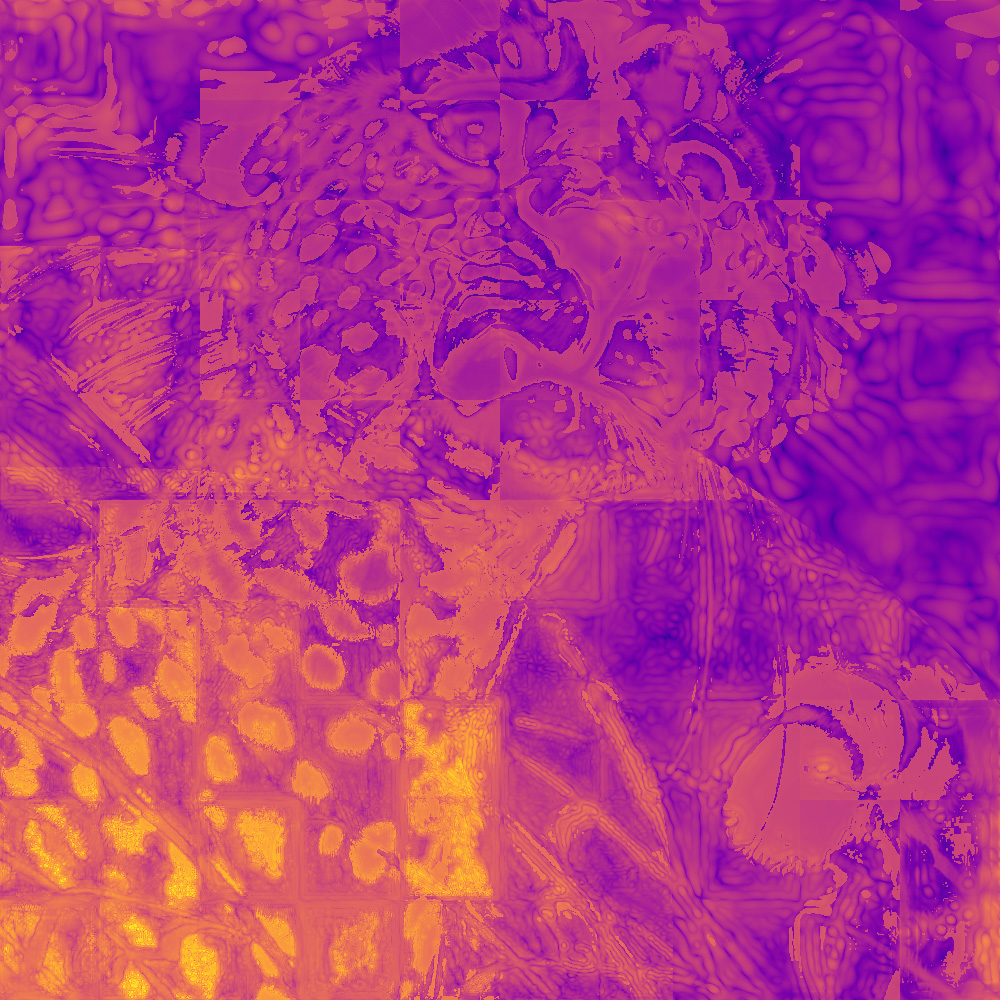}\label{fig:results:2d:fixation}
	}
	\subfloat[saccade perception]{
		\includegraphics[width=0.17\linewidth]{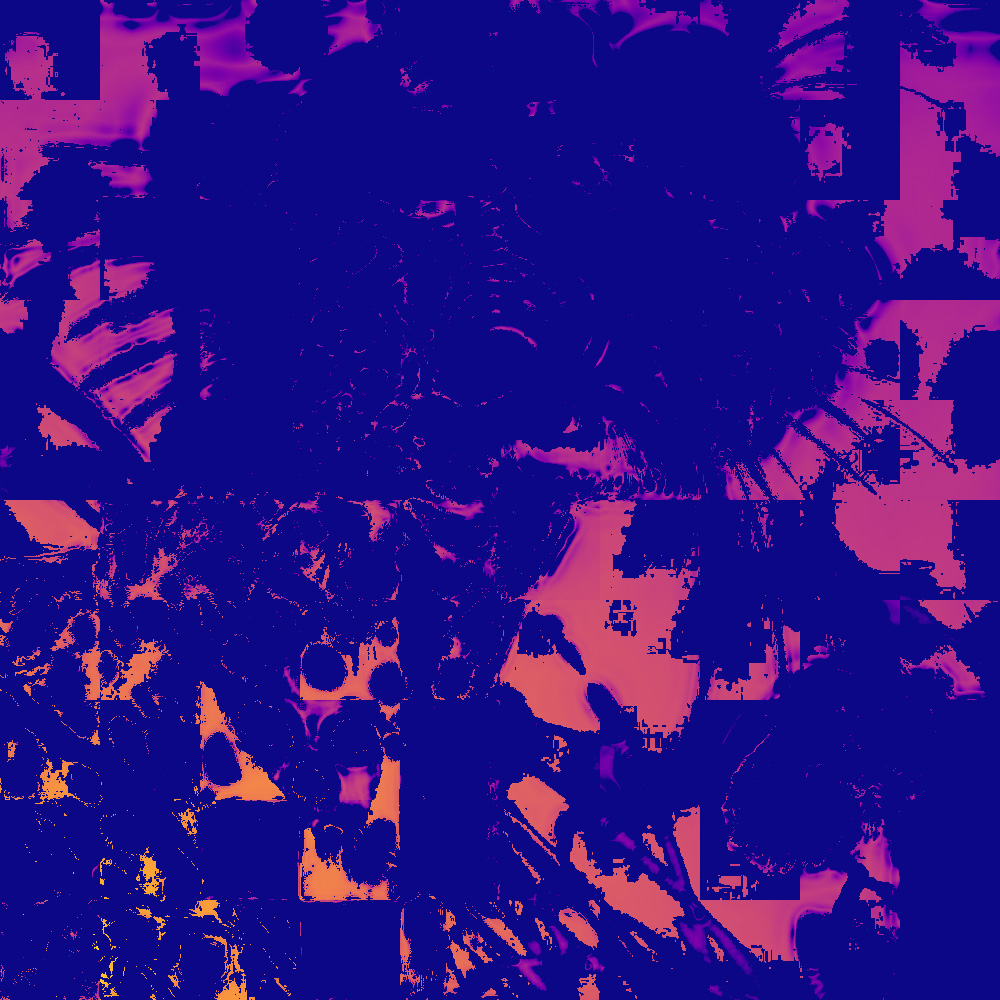}\label{fig:results:2d:saccade}
	}

	\subfloat[target]{
		\includegraphics[width=0.17\linewidth]{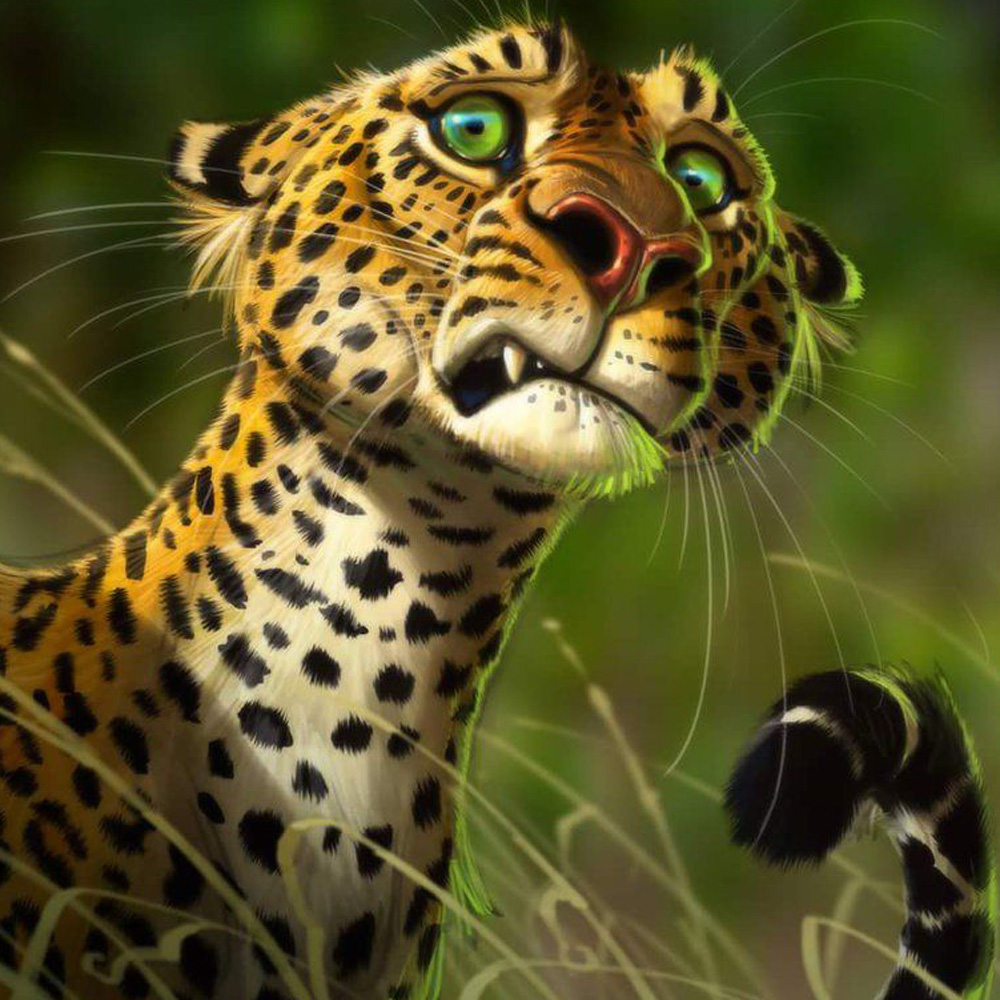}\label{fig:results:2d:target}
	}
	\hspace{0.3em}
	\subfloat[$\eccentricity$-only update]{
		\includegraphics[width=0.17\linewidth]{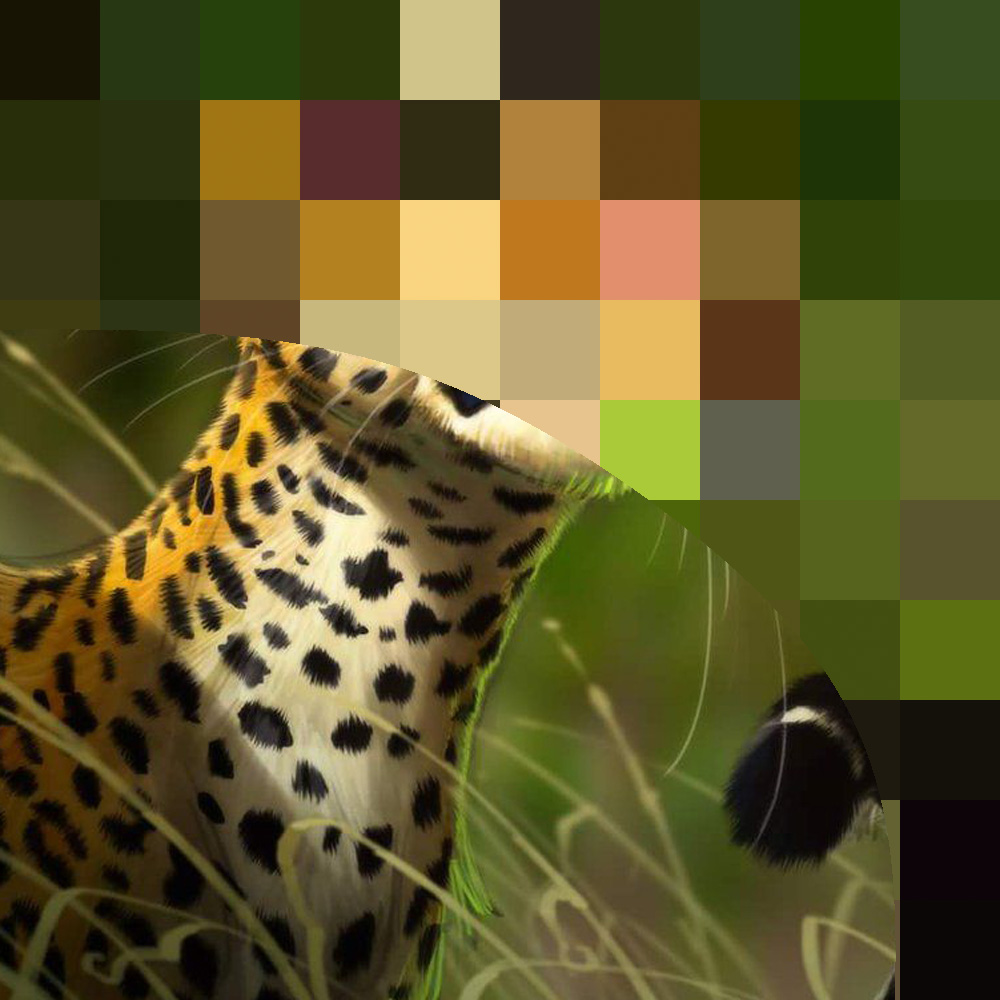}
	}
	\subfloat[$\poppingIntensity$-only update]{
		\includegraphics[width=0.17\linewidth]{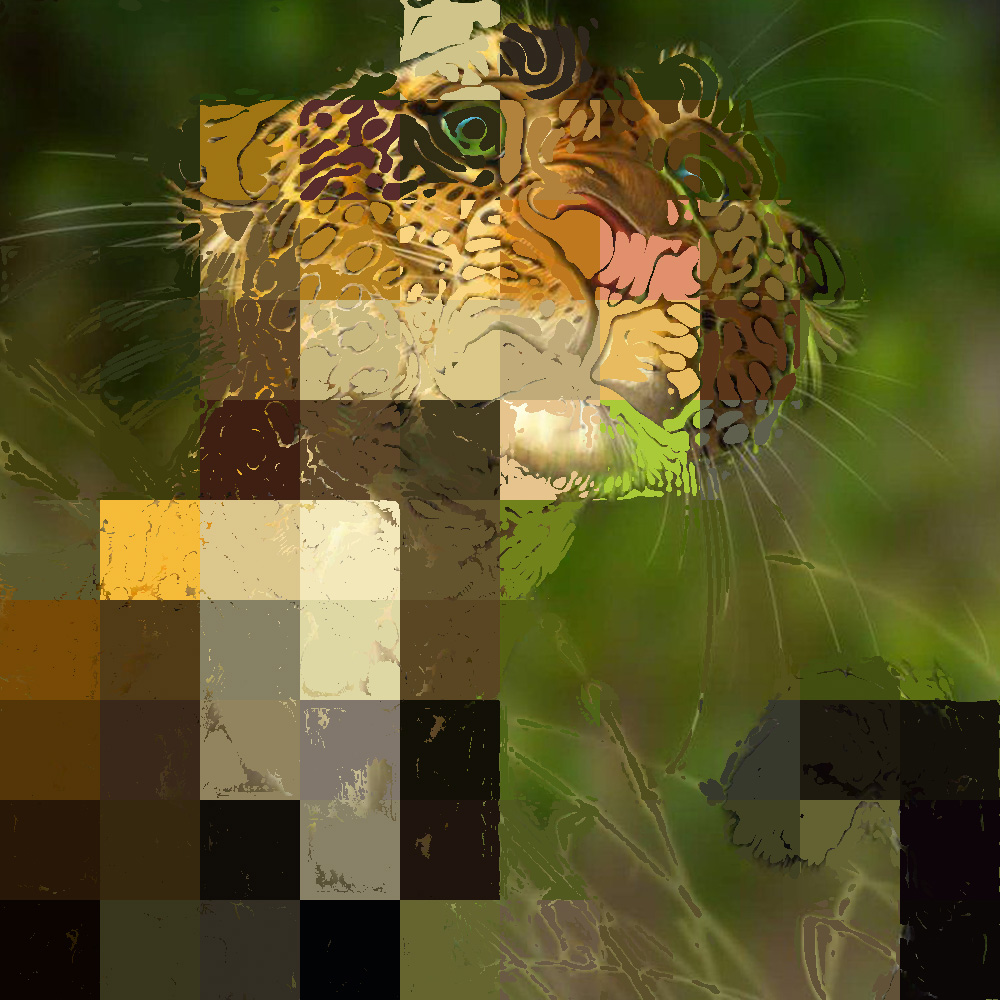}
	}
	\hspace{0.3em}
	\subfloat[fixation update]{
		\includegraphics[width=0.17\linewidth]{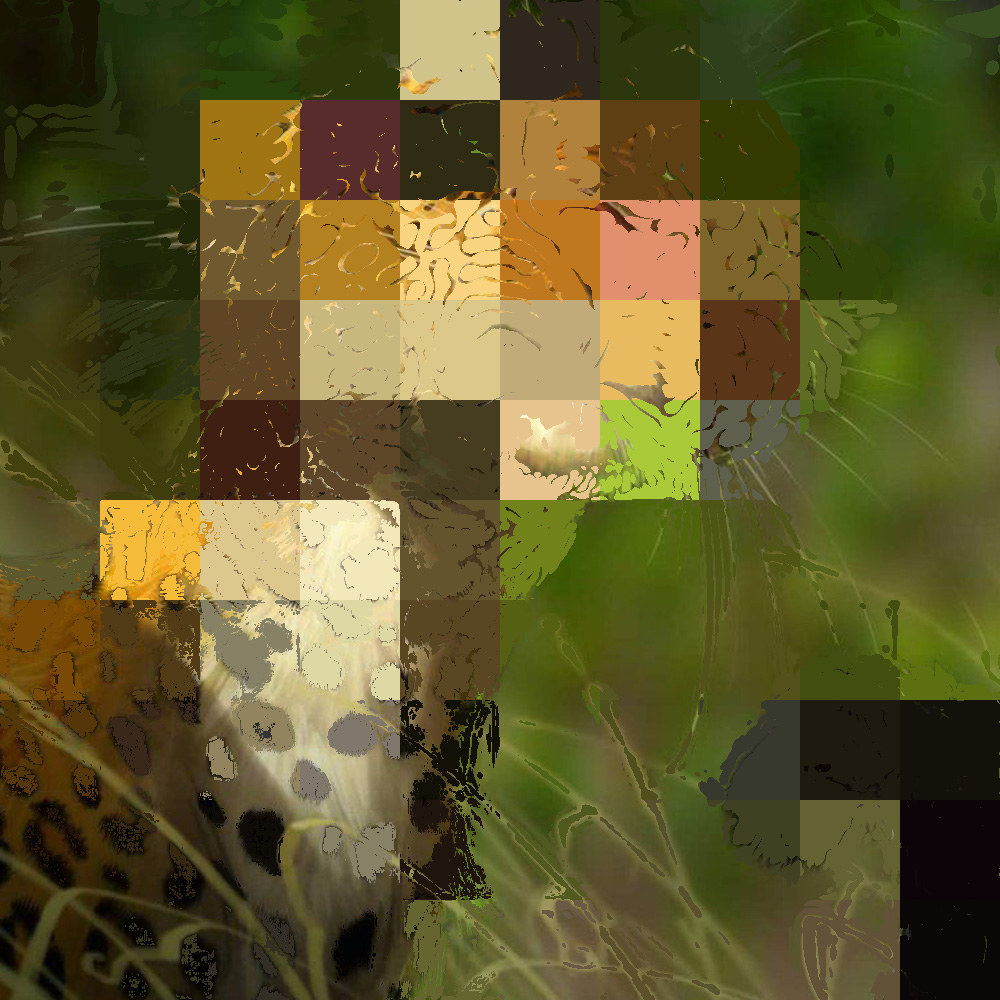}
	}
	\subfloat[saccade update]{
		\includegraphics[width=0.17\linewidth]{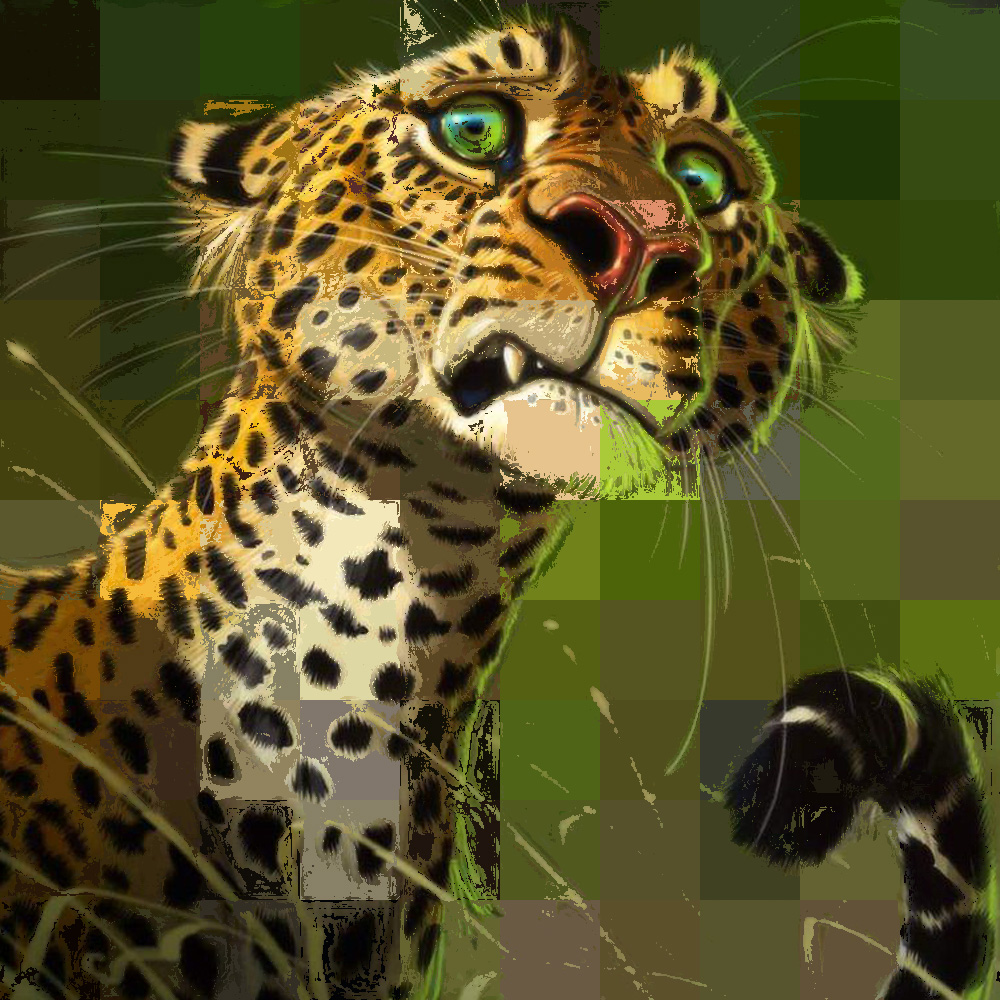}
	}

	\Caption{Visualization of our perceptual model in 2D screen space.}
    {%
Here each pixel is treated as a unit with its own LOD.
	\subref{fig:results:2d:current} is a semi-transmitted low LoD image on the edge and the green circle indicates the gaze.
	\subref{fig:results:2d:target} is the full LoD target on the cloud.
	\subref{fig:results:2d:eccHeat} and \subref{fig:results:2d:popHeat} visualize the importance of the eccentricity in \Cref{eqn:static} and the temporal consistency in \Cref{eq:poppingIntensity}.
	Our model exploits both eccentricity and temporal consistency in \Cref{eqn:adaptive}.
	\subref{fig:results:2d:fixation} and \subref{fig:results:2d:saccade} visualize the corresponding perceptual quality measures considering both static visual quality and dynamic consistency.
    The second row shows the corresponding screen space update.
    }
    \label{fig:results:2d}
\end{figure*}

\paragraph{Spatial acuity and quality}
Due to foveated vision, we prioritize quality and details in the fovea over the periphery.
Thus, the importance of a given pixel $\imageSpaceVec$ under gaze position $\gazeVec$ is computed as:
\begin{align}
\label{eqn:static}
\eccentricity(\gazeVec,\imageSpaceVec) = \eccentricityMask(\gazeVec-\imageSpaceVec),
\end{align}
where $\eccentricityMask$ is the importance function defined in \Cref{eqn:eccentricity_mask} in \Cref{sec:sup:foveation}.
\paragraph{Temporal consistency}
\label{sec:popping}
A major problem from traditional LoD-based procedural rendering is the visual popping effect \cite{Luebke:2002:LD3}.
That is, when the LoD level of an area receives an update, the abrupt visual changes may easily be noticed and distractive to the experience. 
The human visual system perceives LoD-introduced popping artifacts in spatial frequency and retinal velocity \cite{Schwarz:2009:PVP}. 

By extending the perceived {\it change} as a temporally adapted Weber's contrast considering short-term memory to individual frequency band, we obtain the approximated popping (i.e., perceived temporal intensities) between two varied frames ($\image$ and $\image^{\prime}$ in the screen space):
\begin{align}
\begin{split}
\poppingIntensity(\gazeVec,\image,\image^{\prime},\imageSpaceVec) = 
\sum_{i=0}^{\bandStepSize - 1}
\csf\left(\contentFreqVec_i,\luminance\right)\times
\frac{\lvert\pointContrast(\imageSpaceVec,\contentFreqVec_i,\image)-\pointContrast(\imageSpaceVec,\contentFreqVec_i,\image^{\prime})\rvert}
{\lvert\pointContrast(\imageSpaceVec,\contentFreqVec_i,\image)\rvert+\weberWeight}
,
\label{eq:poppingIntensity}
\end{split}
\end{align}
where $\gazeVec$ is the tracked gaze positions in $\image$ and $\image^{\prime}$, respectively. $\weberWeight$ is the controlling parameter that balances for low-intensity stimuli where Weber's law may fail. 
$\csf$ is the contrast sensitivity function given the frequency band $\contentFreqVec_i$ and luminance $\generalLuminance$,
as detailed in \Cref{eq:static_sensitivity} under \Cref{sec:sup:static_vision_model}.
\new{The transition from $\image$ to $\image^{\prime}$ can be from the changes of scene or camera. The corresponding coefficients are omitted here for the brevity of presentation.}
Although the spatial contrast sensitivity function is not reflecting the sensitivity of the human visual system to temporal changes, we have modeled the sensitivity temporal changes by introducing Weber's law on two temporally adjacent frames $\image$ and $\image^{\prime}$.
$\pointContrast$ is the bandpass filtered point contrast given the frequency band $\contentFreqVec_i$,
as detailed in \Cref{sec:sup:bandpass}.
It assumes slow gaze/head motion with fast (90FPS) frame update. Thus, $\gazeVec$ in the two frames are approximately identical. The next paragraph discusses cases when the gaze moves fast (saccade).

\paragraph{Adapting to dynamic gaze behaviors}
As studied by \cite{Knoell:2011:SPP}, our visual sensitivity gets significantly suppressed during saccades. Due to the change blindness, we only perceive weak popping artifacts in this period.
Thus, when saccades are detected, we can instead stream the most noticeable popping elements to reduce the popping intensity after it lands.
This motivates our gaze-behavior-adaptive per-pixel sensitivity by combining both spatial acuity and temporal consistency models:
\begin{align}
\begin{split}
&
\adaptive(\gazeVec,\image,\image^{\prime},\imageSpaceVec) = 
\begin{cases}
\eccentricity(\gazeVec,\imageSpaceVec) - \weightEccPop\poppingIntensity(\gazeVec,\image,\image^{\prime},\imageSpaceVec) & \text{during fixation}\\
\int_{\gazeVec^{\prime} \in \image^{\prime}}\poppingIntensity(\gazeVec^{\prime},\image,\image^{\prime},\imageSpaceVec) \mathrm{d}\gazeVec^{\prime} & \text{during saccade}
\end{cases}
\end{split}
,
\label{eqn:adaptive}
\end{align}
where $\weightEccPop$ is the balance between maximizing foveated perceptual quality and minimizing popping artifacts during fixation. 
A saccade is considered to happen if the gaze speed is greater than 180 deg/sec.
Because of the open challenge in real-time saccade landing prediction \cite{Arabadzhiyska:2017:SLP}, we integrate over the entire visual field while computing the popping for saccade instead of assuming the landing position. This ensures global robustness to any user's attentional changes.
\Cref{fig:results:2d} visualizes individual importance and the resulting image-space update.

We can adapt \Cref{eqn:adaptive} for progressive LoD update from level $i+j$ to $i$ as follows:
\begin{align}
\adaptive(\gazeVec,\image_{i+j},\image_{i},\imageSpaceVec) = \sum_{l=i}^{i+j-1}\adaptive(\gazeVec,\image_{l+1},\image_l,\imageSpaceVec),
\label{eqn:adaptive_impl}
\end{align}
where $\image_i$ represents the image at the i-th LoD.

\subsection{Mapping from 2D Screen to 3D Streaming}
\label{sec:method:3d}
In this section, we present how to extend the screen-space model from \Cref{sec:method:2d} to 3D assets for a real-world cloud-based 3D streaming system.
The 3D assets can have various representations, including triangle meshes, volumes, terrains, or large crowded objects, as shown in \Cref{fig:3dAll}.

\paragraph{Map to 3D assets}
So far, our perceptual model depicting static quality and dynamic artifacts in \Cref{sec:method:2d} applies to individual pixels. The 3D assets, however, comprise non-uniformly distributed content such as depths and connectivities.
We apply a deferred shading algorithm to convert various types of 3D primitives to 2D perception evaluations.

We divide the 3D content based on the coarsest level of LoD. 
We denote an individual computational unit as $\threeDimUnit_i$,
where $i$ is its index among all units.
It can be a coarsest triangle in a 3D mesh,
a largest super-voxel in a volume,
a texel in the coarsest mipmap level of a height/displacement texture,
or a separate object in a swarm scene.
We denote the LoD level of $\threeDimUnit_i$ at time frame $\timeStamp$ as $\LoDLevel_{\threeDimUnit_i, \timeStamp}$.

At time frame $\timeStamp - 1$ when the LoD levels of all units are determined and transmitted to edge already,
we render a framebuffer at cloud side for the whole scene without anti-aliasing to retrieve the unit indices of every pixel,
i.e., a mapping $\twoThreeDimMap_{\timeStamp - 1}: \left\{\imageSpaceVec \right\} \longmapsto \left\{ \threeDimUnit_i \right\}$
from the set of pixels $\left\{\imageSpaceVec \right\}$ to the set of units $\left\{ \threeDimUnit_i \right\}$.

\begin{figure}[htb]
\centering
	\subfloat[2D heatmap]{
	\includegraphics[width=0.23\linewidth]{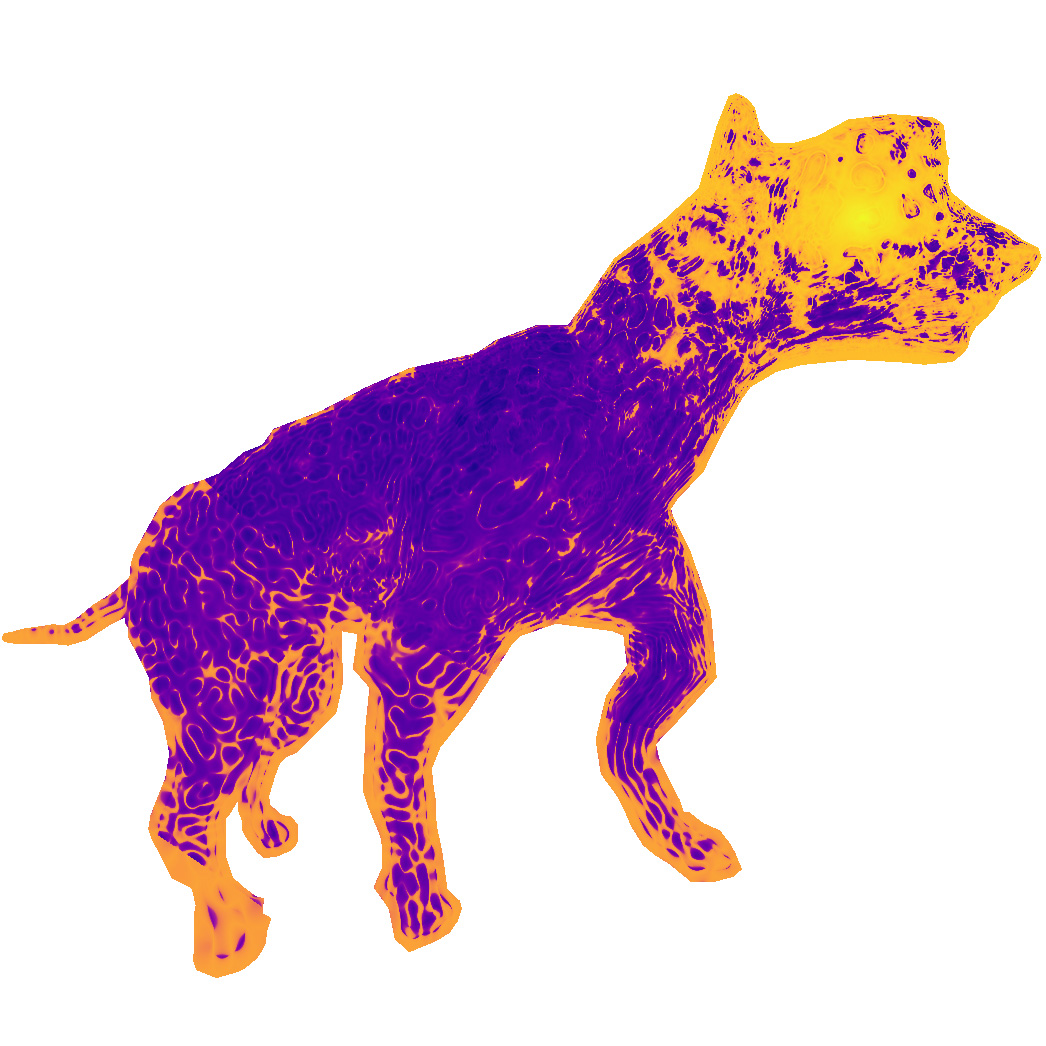}
	\label{fig:mapping:2d}
	}
	\subfloat[3D heatmap]{
		\includegraphics[width=0.23\linewidth]{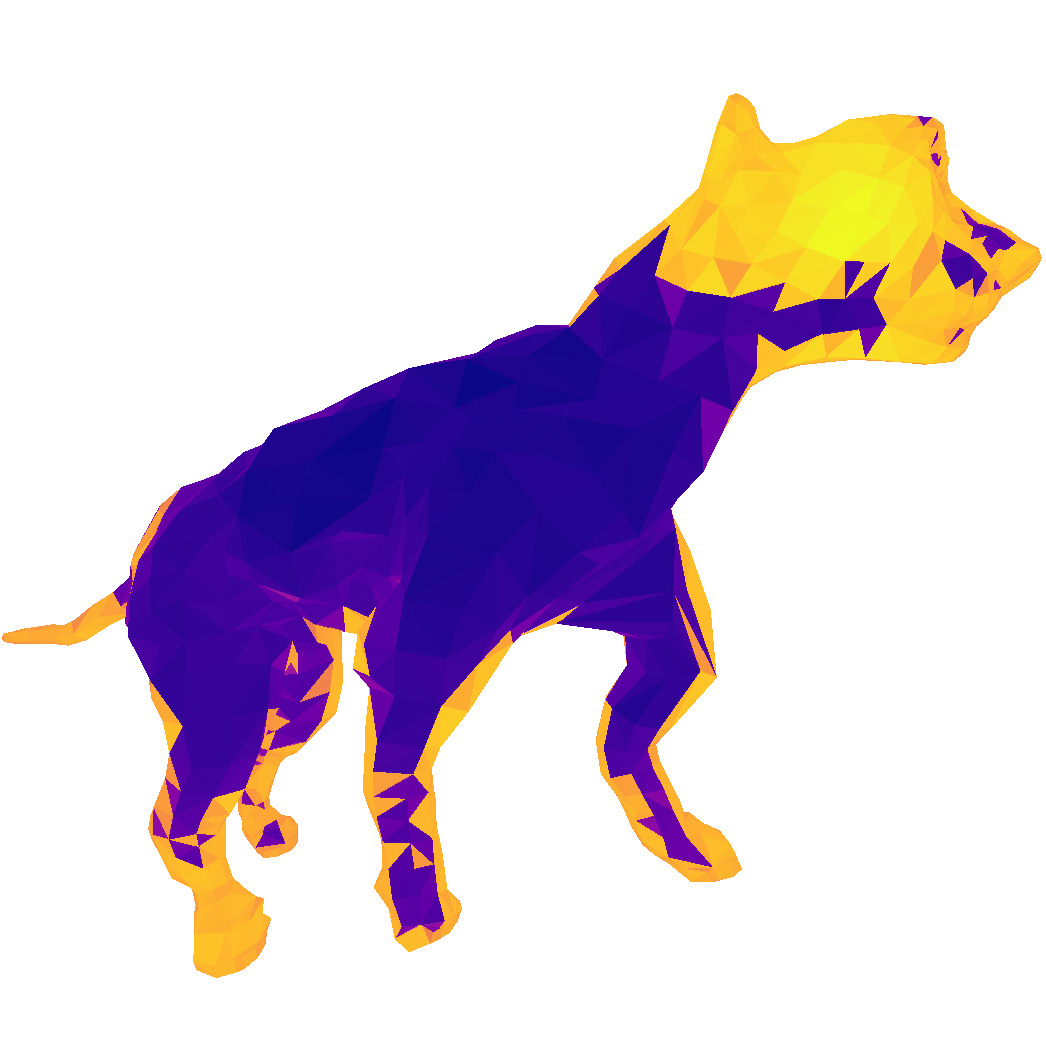}
		\label{fig:mapping:3d}
	}
	\subfloat[before]{
		\includegraphics[width=0.23\linewidth]{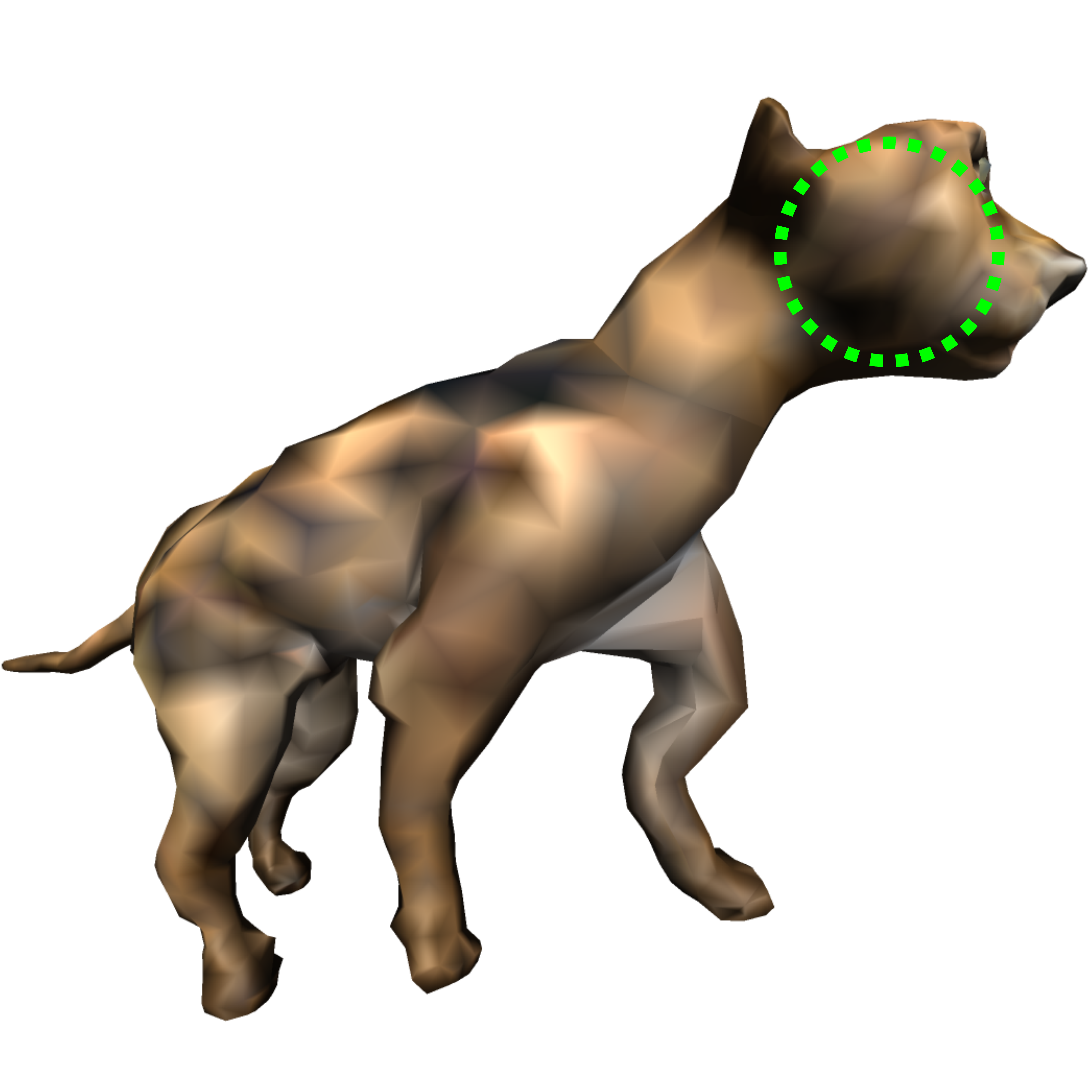}
		\label{fig:mapping:before}
	}
	\subfloat[update]{
		\includegraphics[width=0.23\linewidth]{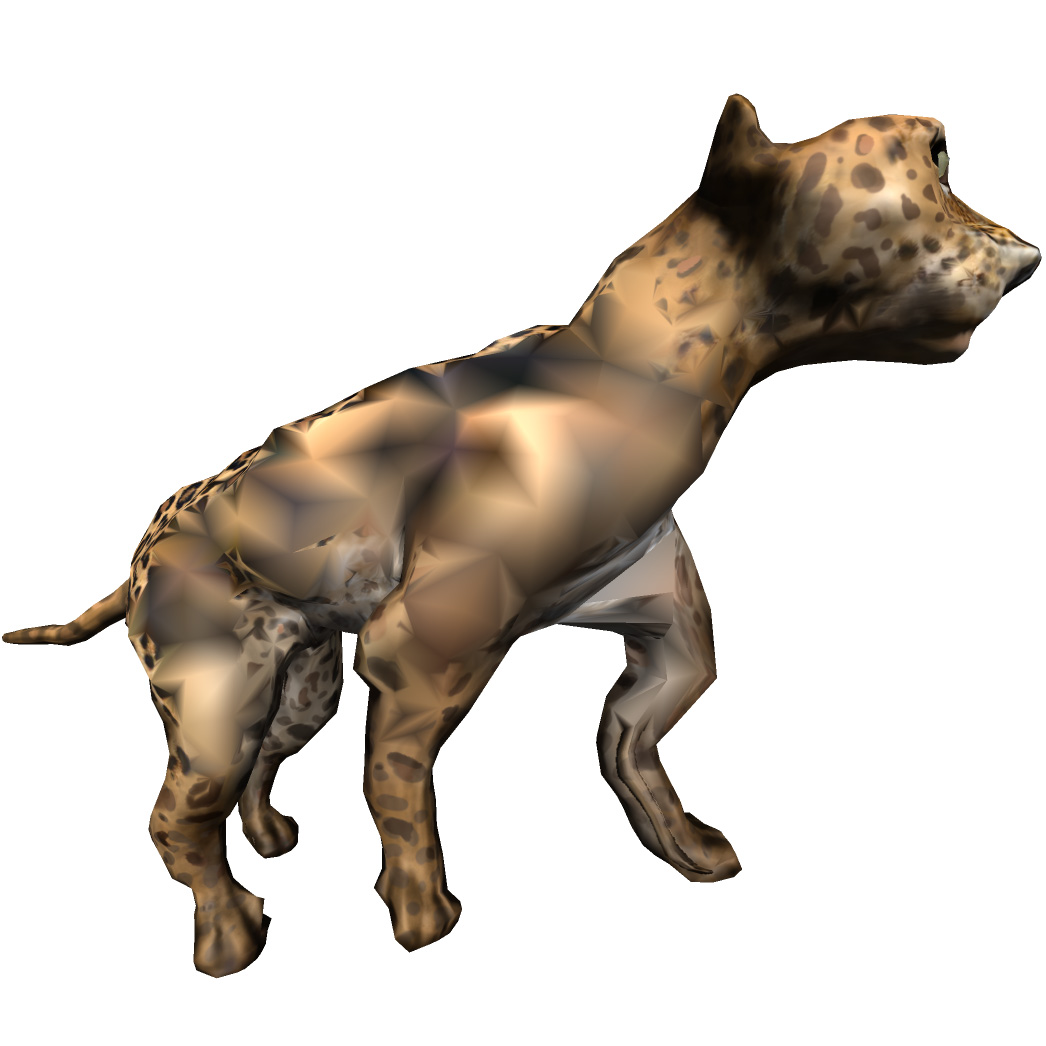}
		\label{fig:mapping:after}
	}
	\Caption{Visualization of mapping importance from 2D pixels to 3D triangles.}
    {%
    \subref{fig:mapping:2d} shows the gaze-aware importance $\adaptive$ in 2D. \subref{fig:mapping:3d} shows mapped importance in 3D.
 	\subref{fig:mapping:before} and \subref{fig:mapping:after} show the object before and after streaming, guided by \subref{fig:mapping:3d}. The green circle indicates the user's gaze fixation position.
    }
    \label{fig:mapping}
\end{figure}

If the LoD level of $\threeDimUnit_i$ is updated to $\LoDLevel_{\threeDimUnit_i, \timeStamp}$ at time frame $\timeStamp$,
we approximate its sensitivity by accumulating all pixels $\twoThreeDimMap_{\timeStamp - 1}\left(\imageSpaceVec\right) = \threeDimUnit_i$ of the unit at time frame $\timeStamp - 1$,
\begin{align}
	& \threeDimAdaptive_{\threeDimUnit_i, \timeStamp}\left(\LoDLevel_{\threeDimUnit_i, \timeStamp}, \gazeVec_{\timeStamp-1}, \twoThreeDimMap_{\timeStamp - 1}\right) \nonumber\\
	& \approx \sum_{\imageSpaceVec \in \twoThreeDimMap^{-1}_{\timeStamp - 1}\left(\threeDimUnit_i\right)} \adaptive\left(\gazeVec_{\timeStamp-1},\image_{\timeStamp-1},\imageApprox_{\timeStamp}\left|\LoDLevel_{\threeDimUnit_i, \timeStamp}\right.,\imageSpaceVec \right),
	\label{eq:unit_sensitivity}
\end{align} 
where $\gazeVec_{\timeStamp - 1}$, $\twoThreeDimMap_{\timeStamp - 1}$, and $\image_{\timeStamp-1}$
are the gaze position, unit mapping, and render image at time frame $\timeStamp - 1$. \Cref{fig:mapping} visualizes the mapping.
The mapping $\twoThreeDimMap$ implicitly represents the camera of each time frame.
It also varies according to the LoD of all units.
The approximation in~\Cref{eq:unit_sensitivity} simplified the evaluation 
by assuming $\twoThreeDimMap_{\timeStamp-1} = \twoThreeDimMap_{\timeStamp}$.
$\imageApprox_{\timeStamp}\left|\LoDLevel_{\threeDimUnit_i, \timeStamp}\right.$ is the render image at time frame $\timeStamp$
if the LoD level of $\threeDimUnit_i$ is $\LoDLevel_{\threeDimUnit_i, \timeStamp}$.
We denote it with a hat because it is an approximation
by assuming the LoD level of other units is not changed between time frames of $\timeStamp - 1$ and $\timeStamp$.
$\threeDimAdaptive_{\threeDimUnit_i, \timeStamp}$ is also dependent on the LoD levels of other units $\left\{ \LoDLevel_{\threeDimUnit_j, \timeStamp - 1} \left| j \ne i \right. \right\} $,
which we omit in \Cref{eq:unit_sensitivity} for brevity.
The evaluation of $\threeDimAdaptive_{\threeDimUnit_i, \timeStamp}$ in \Cref{eq:unit_sensitivity} is computationally expensive,
so we propose a neural evaluation, as will be detailed in~\Cref{sec:method:neural}.

\newcommand{\vs}{\vspace{0.2em}}
\newcommand{\hs}{\hspace{0.03em}}

\begin{figure*}
	\centering
	\rotatebox[origin=c]{90}{\hspace{0.1em}Volume CT Data\hspace{10.4em} Displaced Mesh\hspace{11.2em} Triangle Mesh}
	\subfloat[current/target]{
	\begin{minipage}{0.17\linewidth}
		\includegraphics[width=\linewidth]{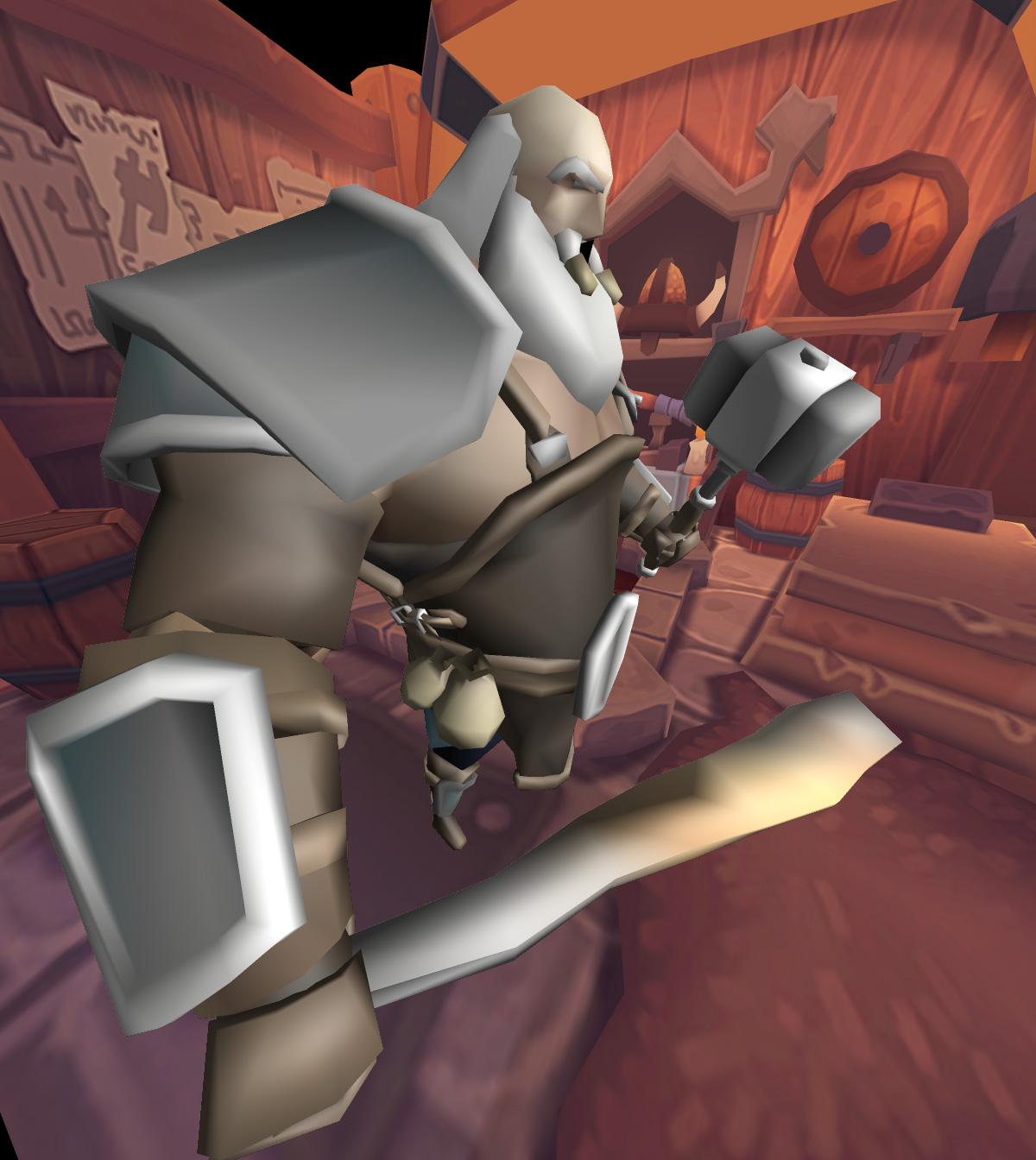}\vs

		\includegraphics[width=\linewidth]{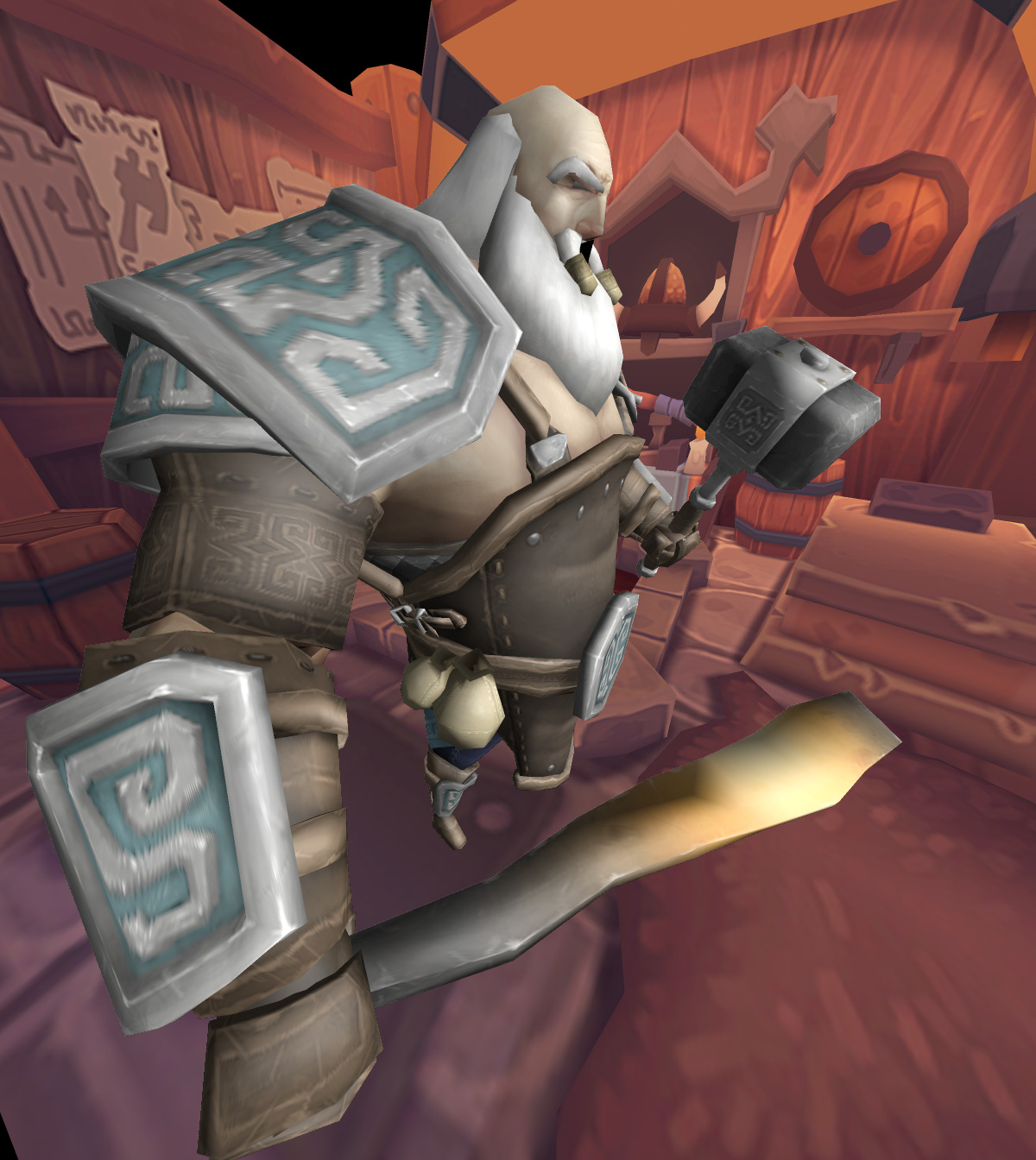}\vspace{0.4em}

		\includegraphics[width=\linewidth]{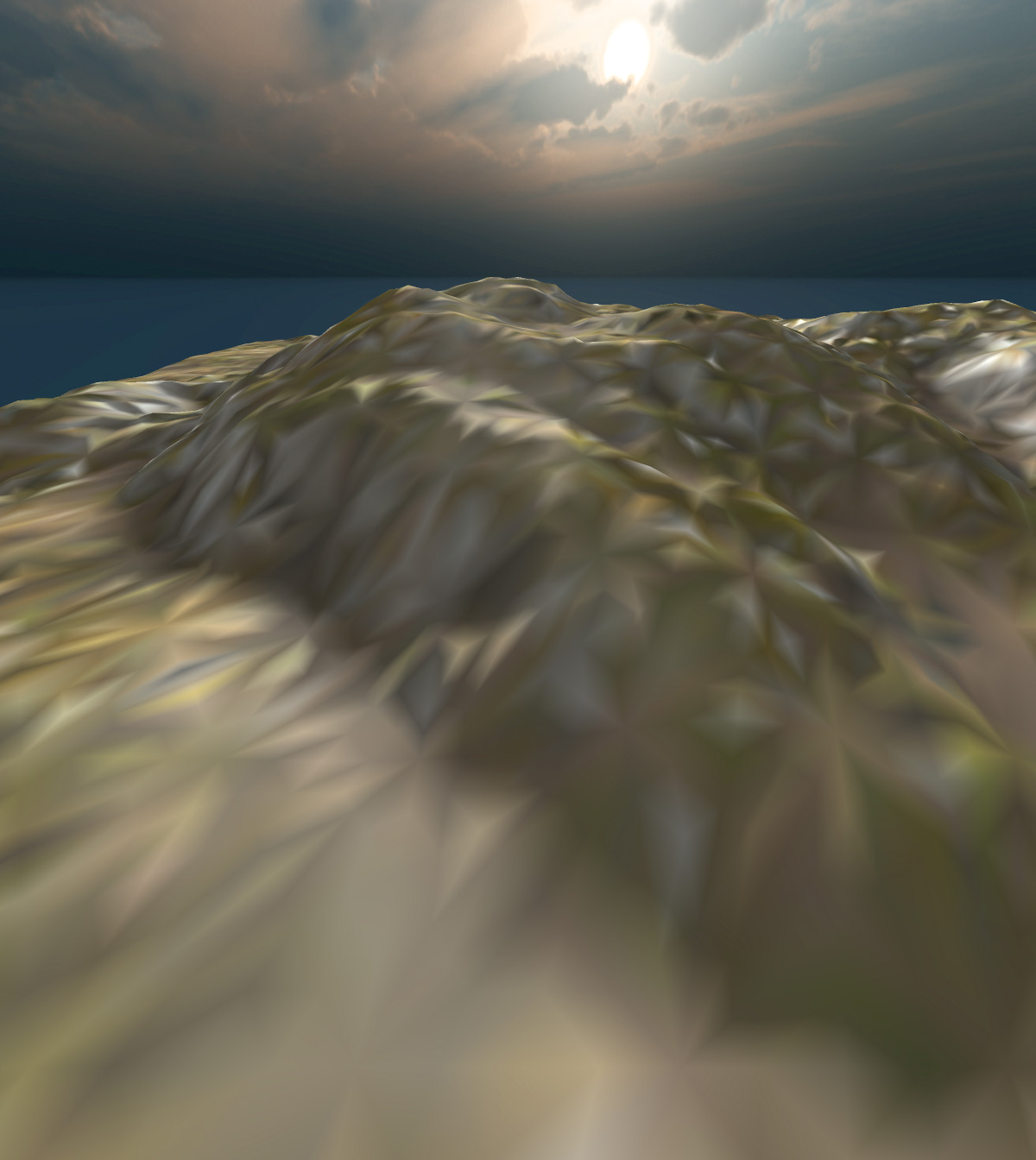}\vs

		\includegraphics[width=\linewidth]{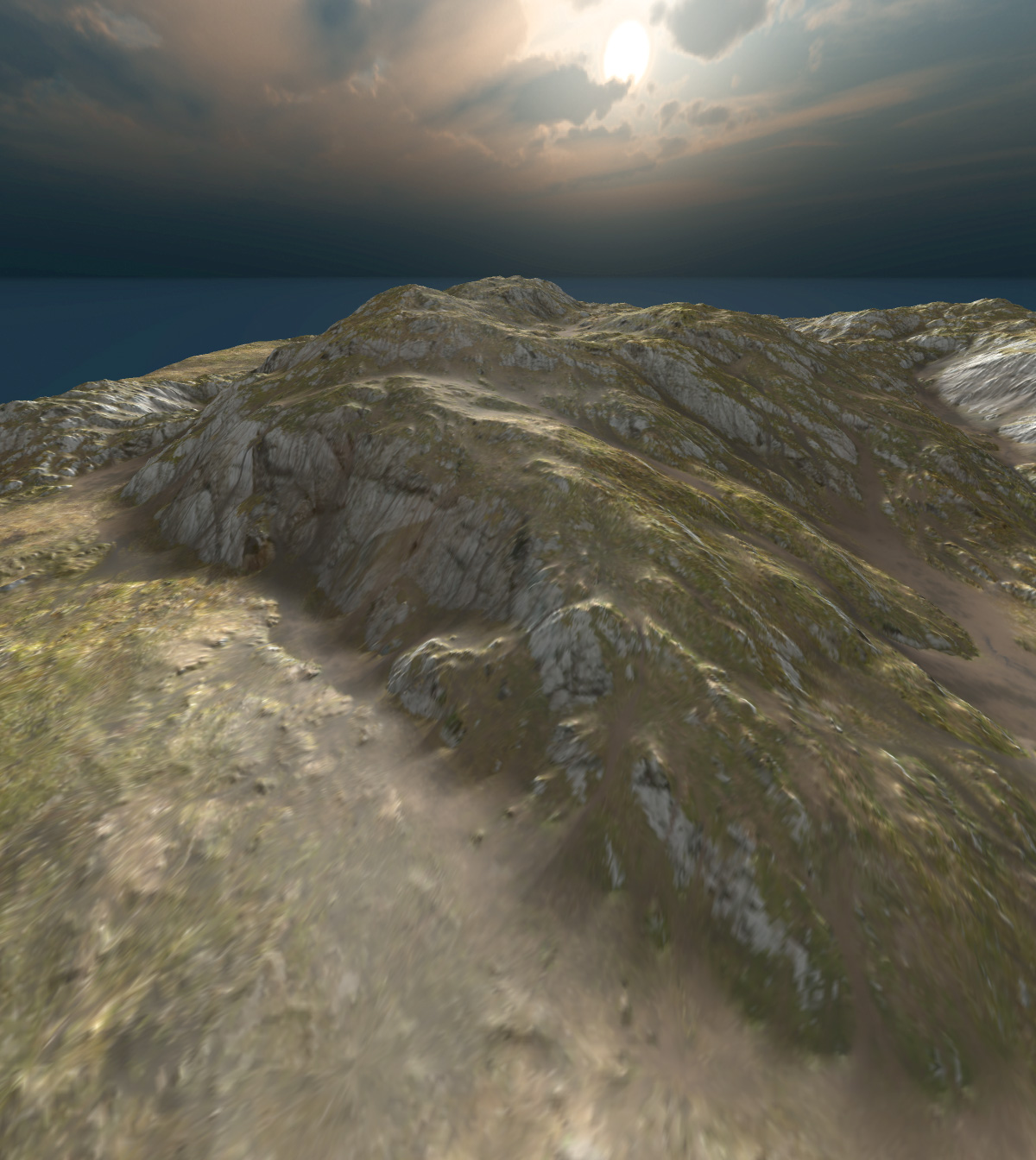}\vspace{0.4em}

		\includegraphics[width=\linewidth]{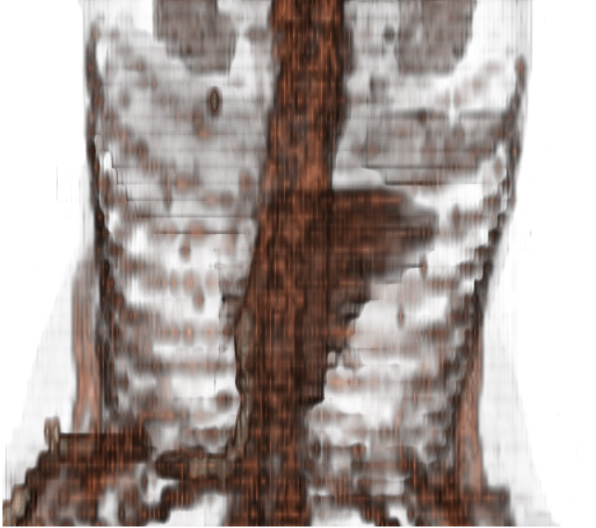}\vs

		\includegraphics[width=\linewidth]{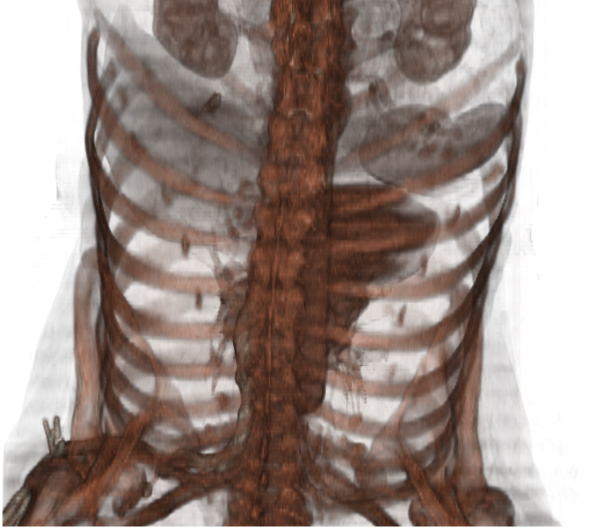}
	\end{minipage}
	\label{fig:3dAll:currentTarget}
	}
	\hspace{0.2em}
	\subfloat[$\eccentricity$]{
	\begin{minipage}{0.17\linewidth}
		\includegraphics[width=\linewidth]{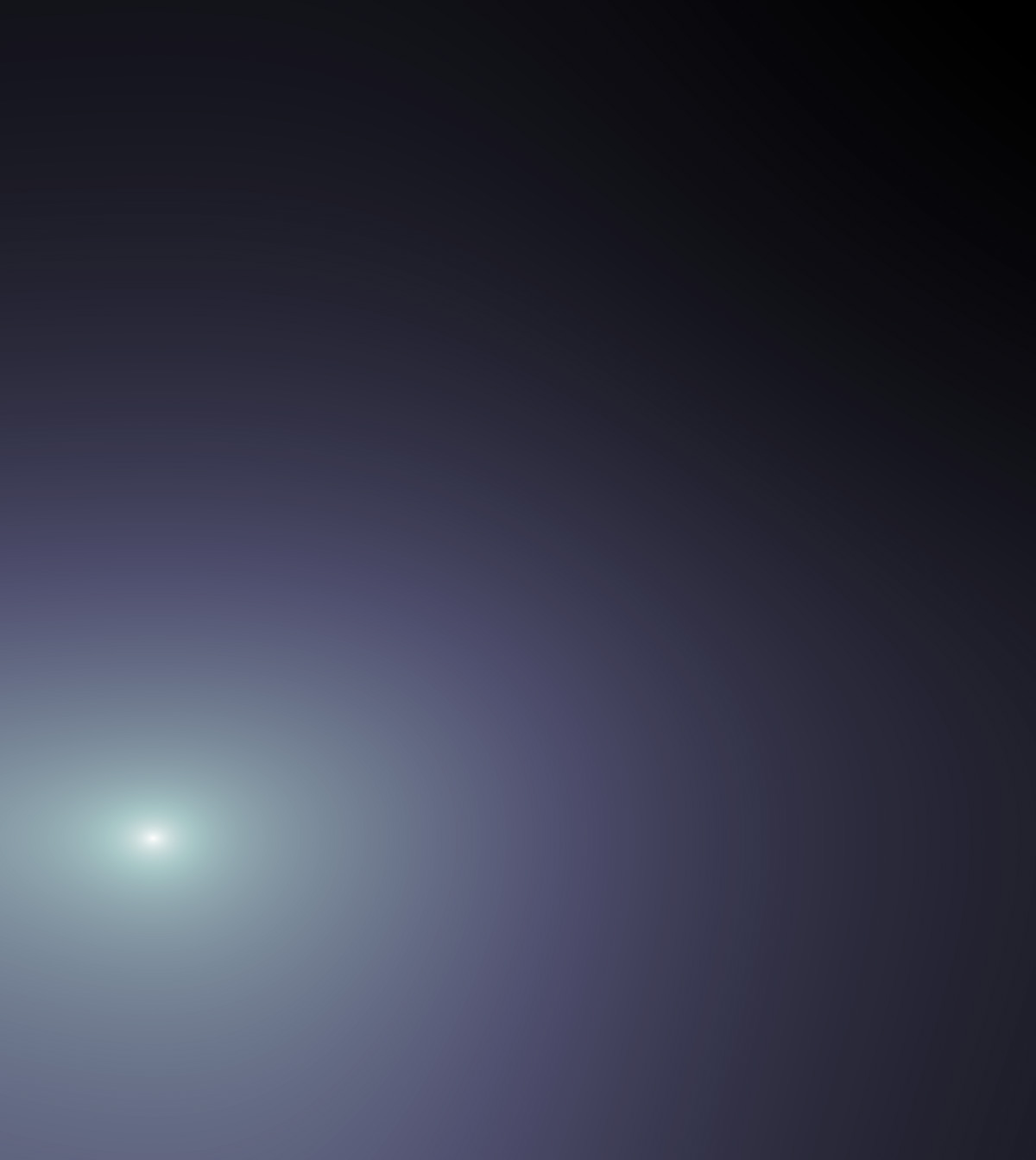}\vs

		\includegraphics[width=\linewidth]{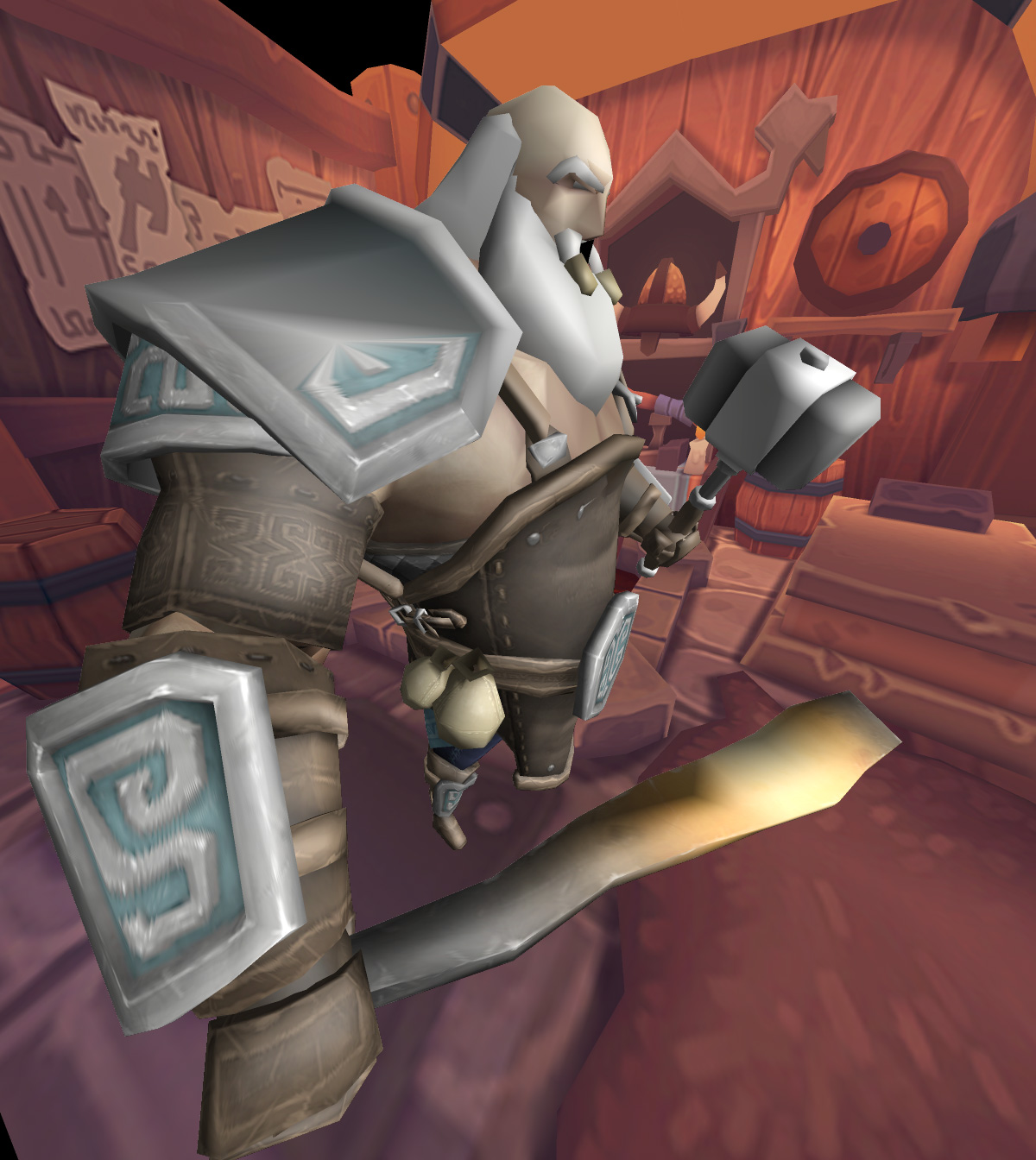}\vspace{0.4em}

		\includegraphics[width=\linewidth]{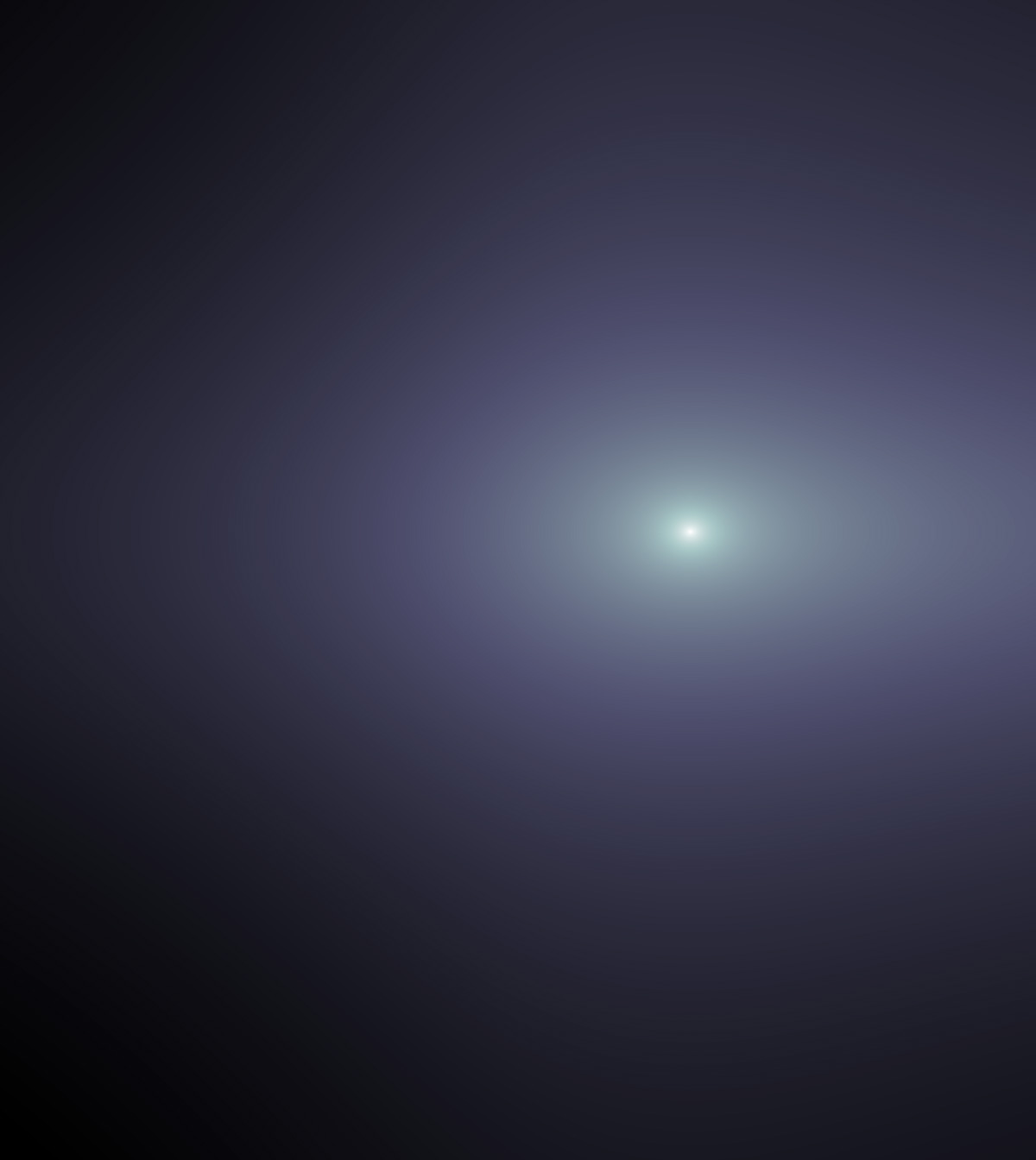}\vs

		\includegraphics[width=\linewidth]{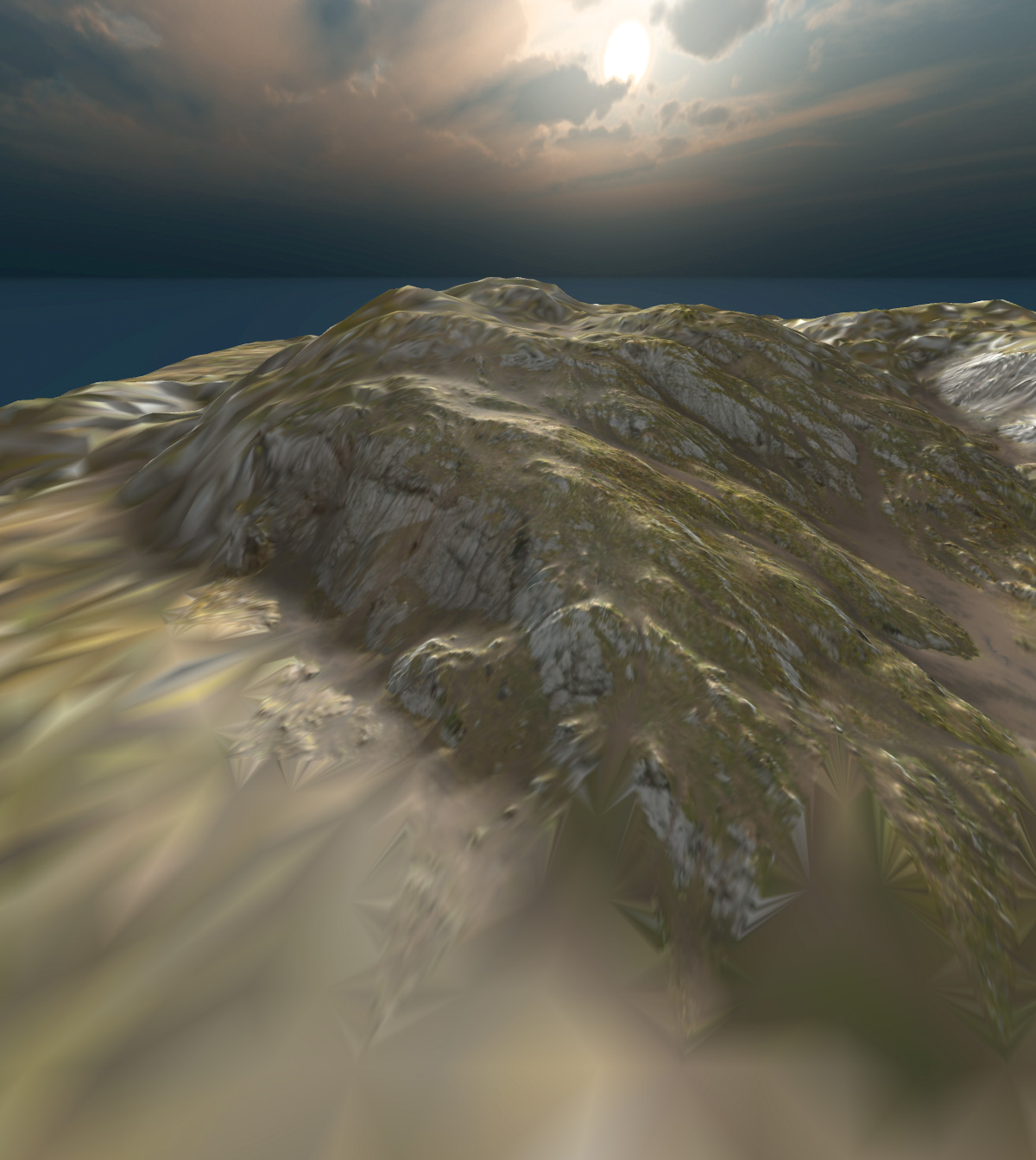}\vspace{0.4em}

		\includegraphics[width=\linewidth]{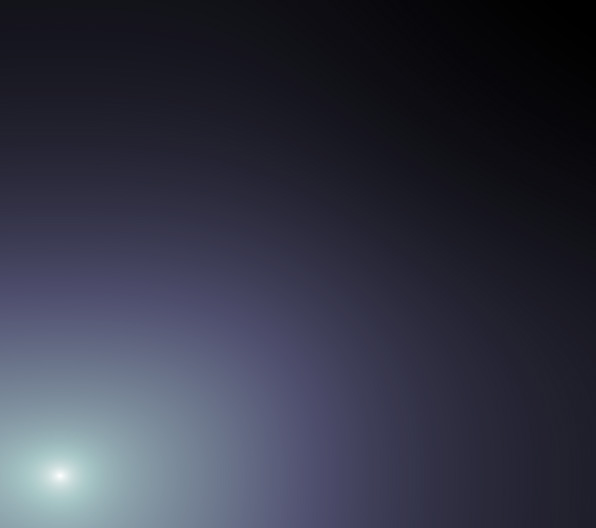}\vs

		\includegraphics[width=\linewidth]{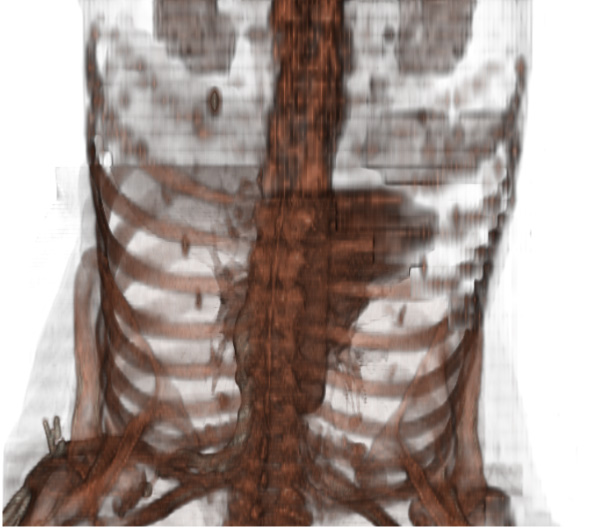}
	\end{minipage}
	\label{fig:3dAll:ecc}
	}
	\hs
	\subfloat[$\poppingIntensity$]{
	\begin{minipage}{0.17\linewidth}
		\includegraphics[width=\linewidth]{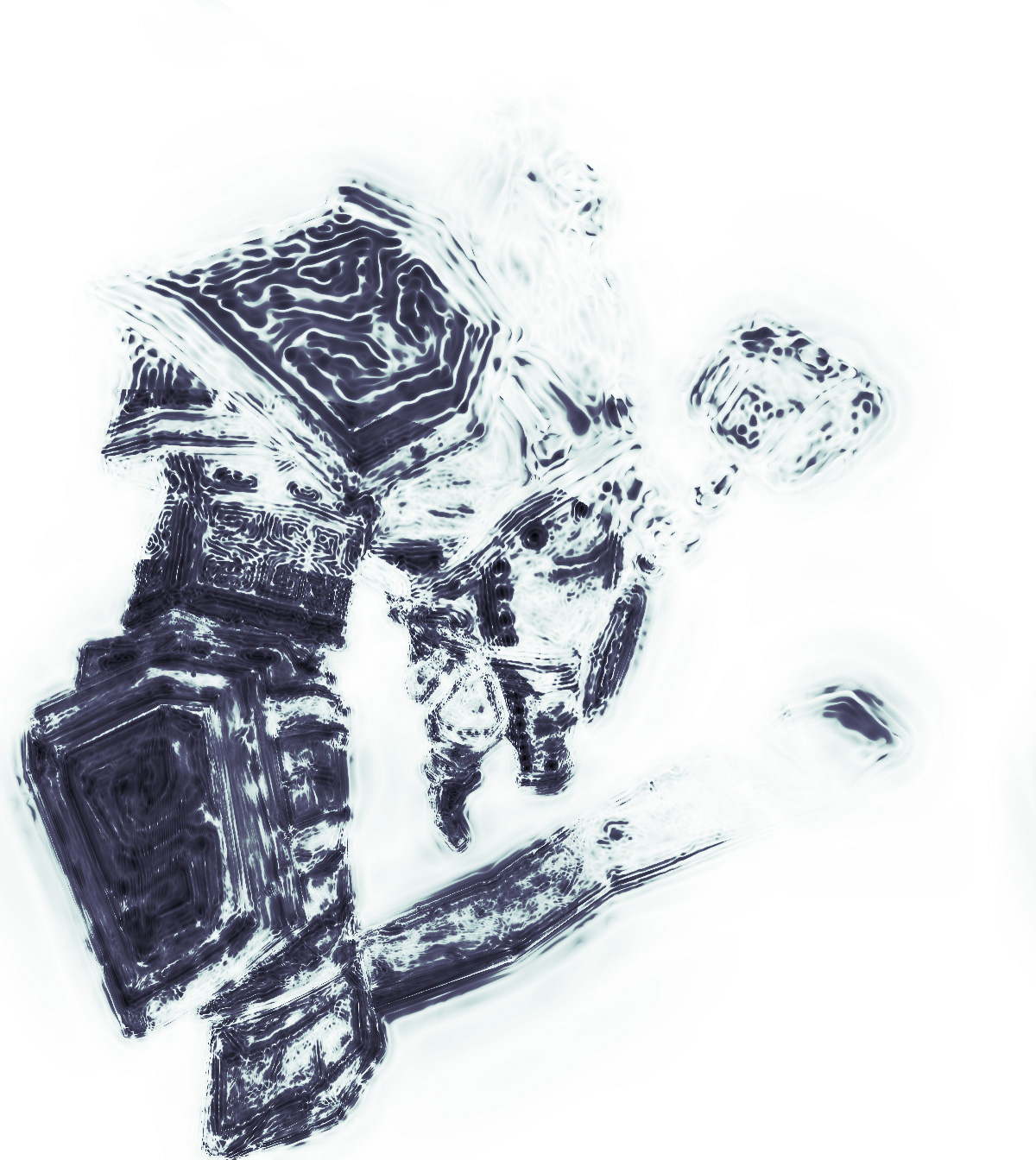}\vs

		\includegraphics[width=\linewidth]{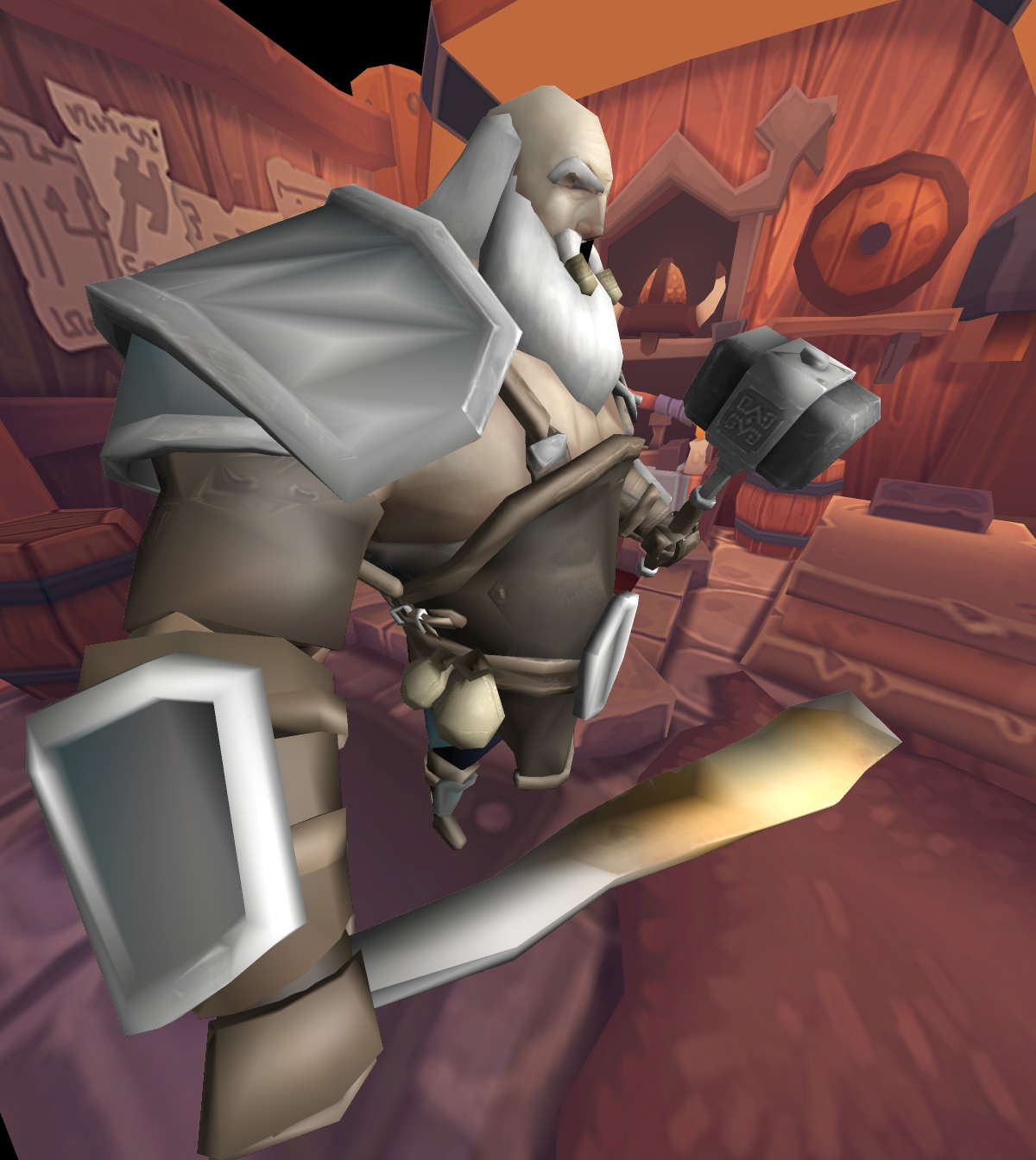}\vspace{0.4em}

		\includegraphics[width=\linewidth]{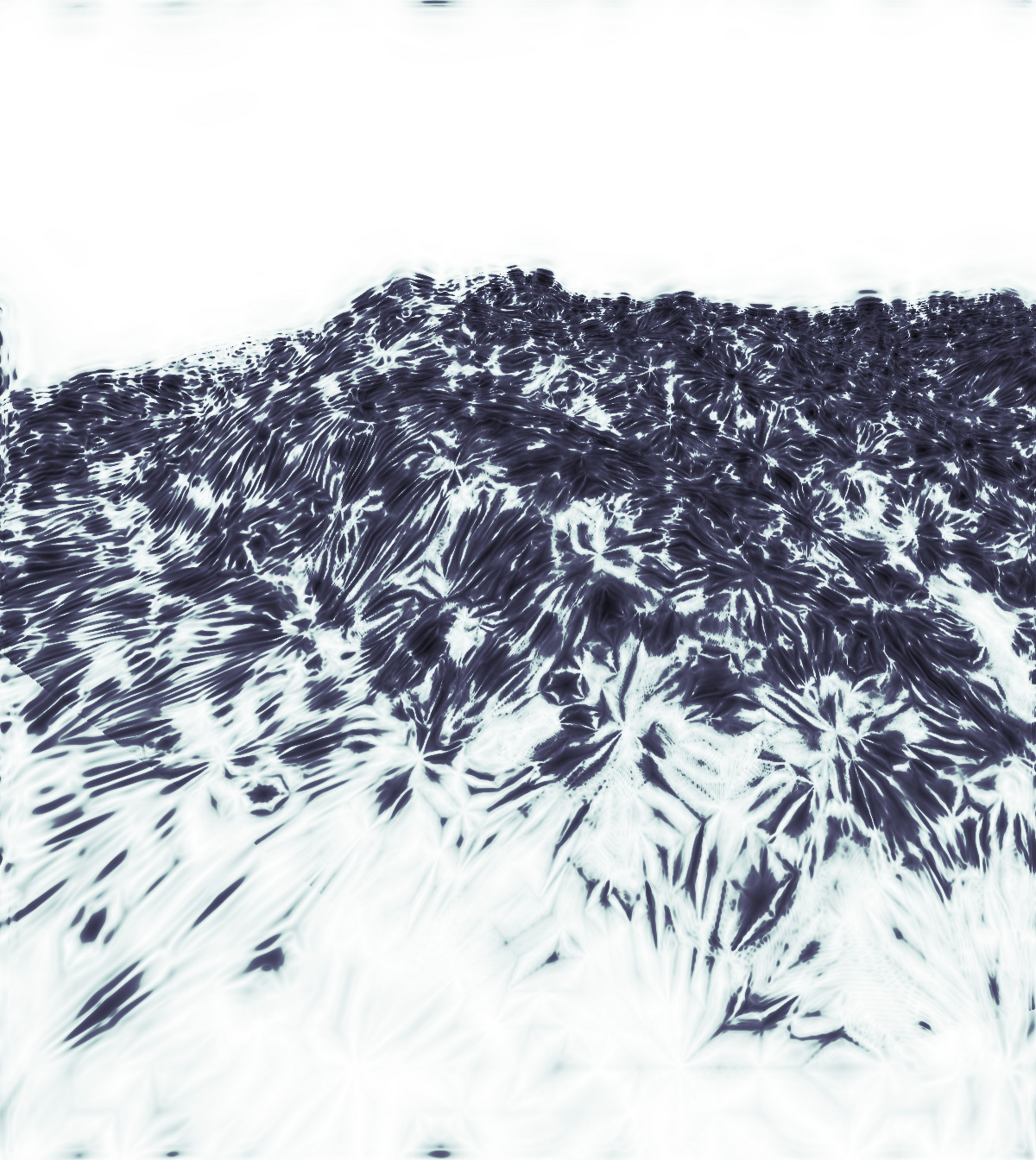}\vs

		\includegraphics[width=\linewidth]{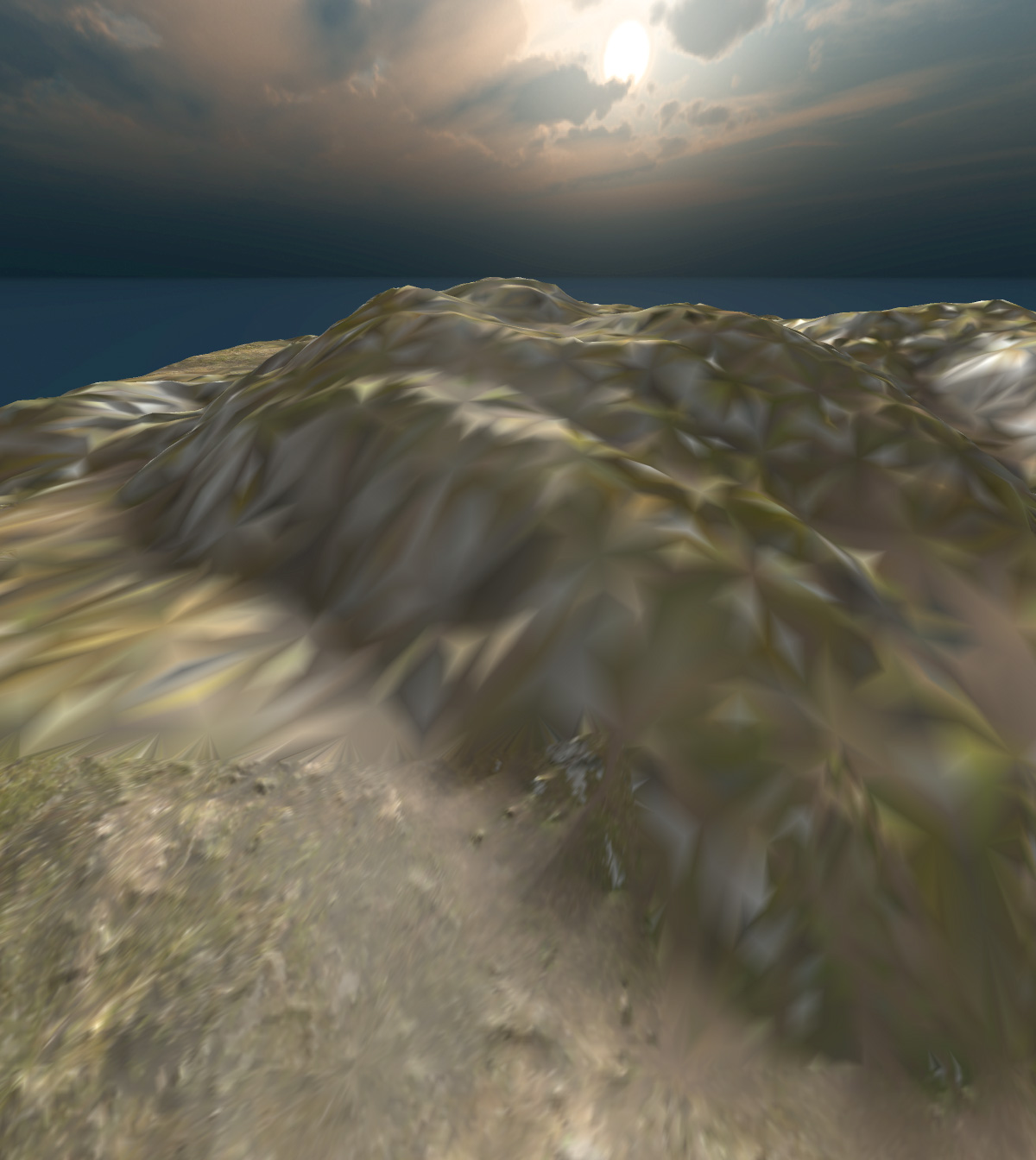}\vspace{0.4em}

		\includegraphics[width=\linewidth]{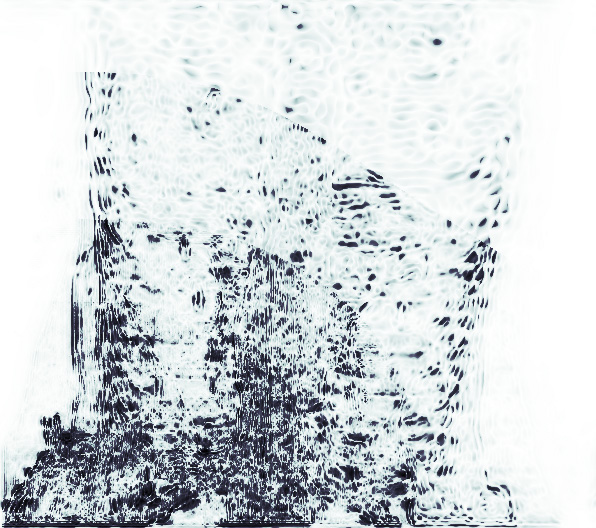}\vs

		\includegraphics[width=\linewidth]{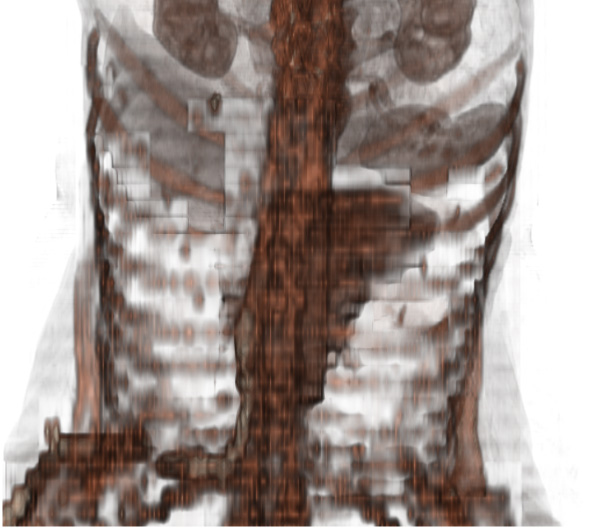}
	\end{minipage}
	\label{fig:3dAll:popping}
	}
	\hspace{0.2em}
	\subfloat[fixation]{
	\begin{minipage}{0.17\linewidth}
		\includegraphics[width=\linewidth]{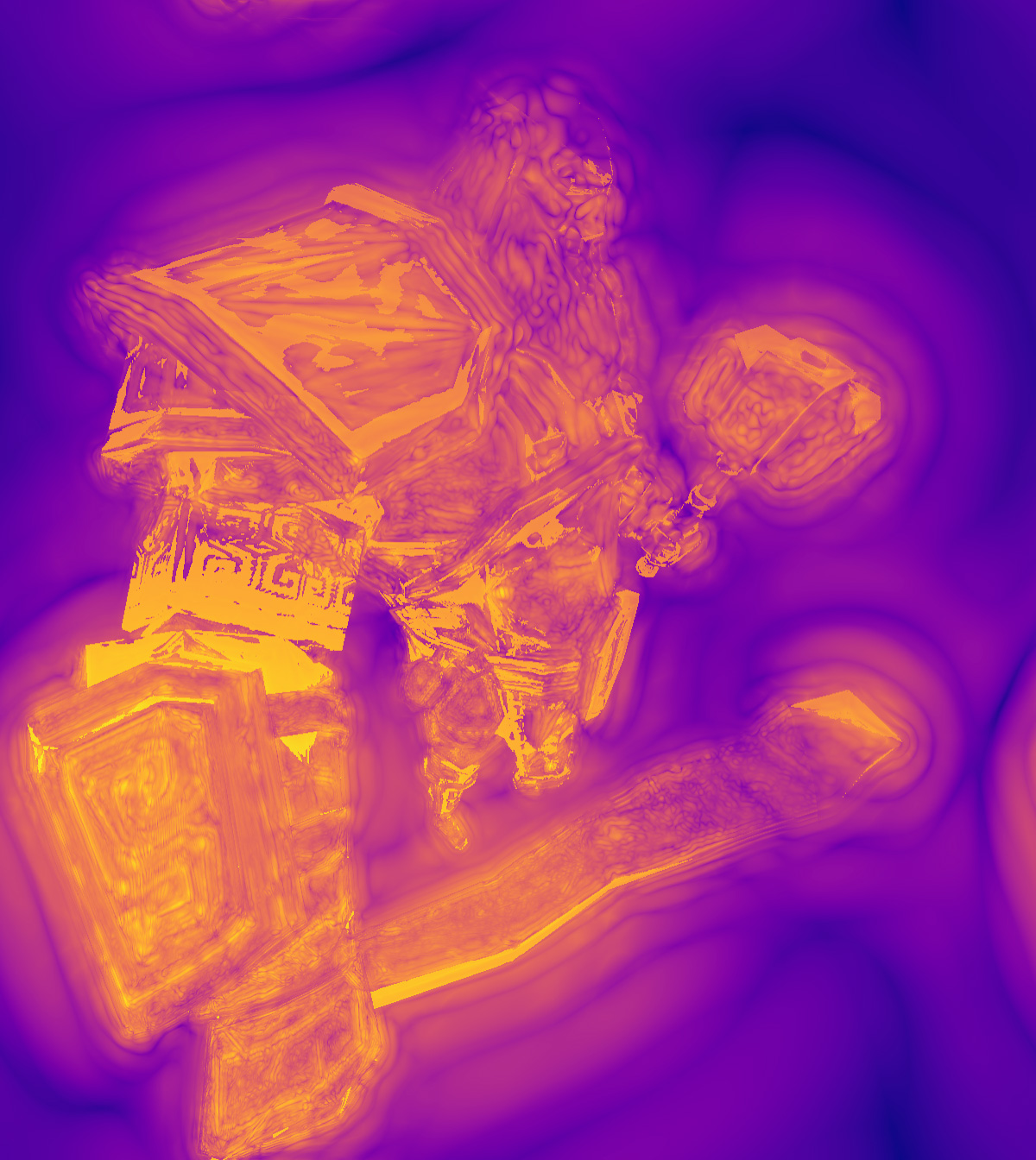}\vs

		\includegraphics[width=\linewidth]{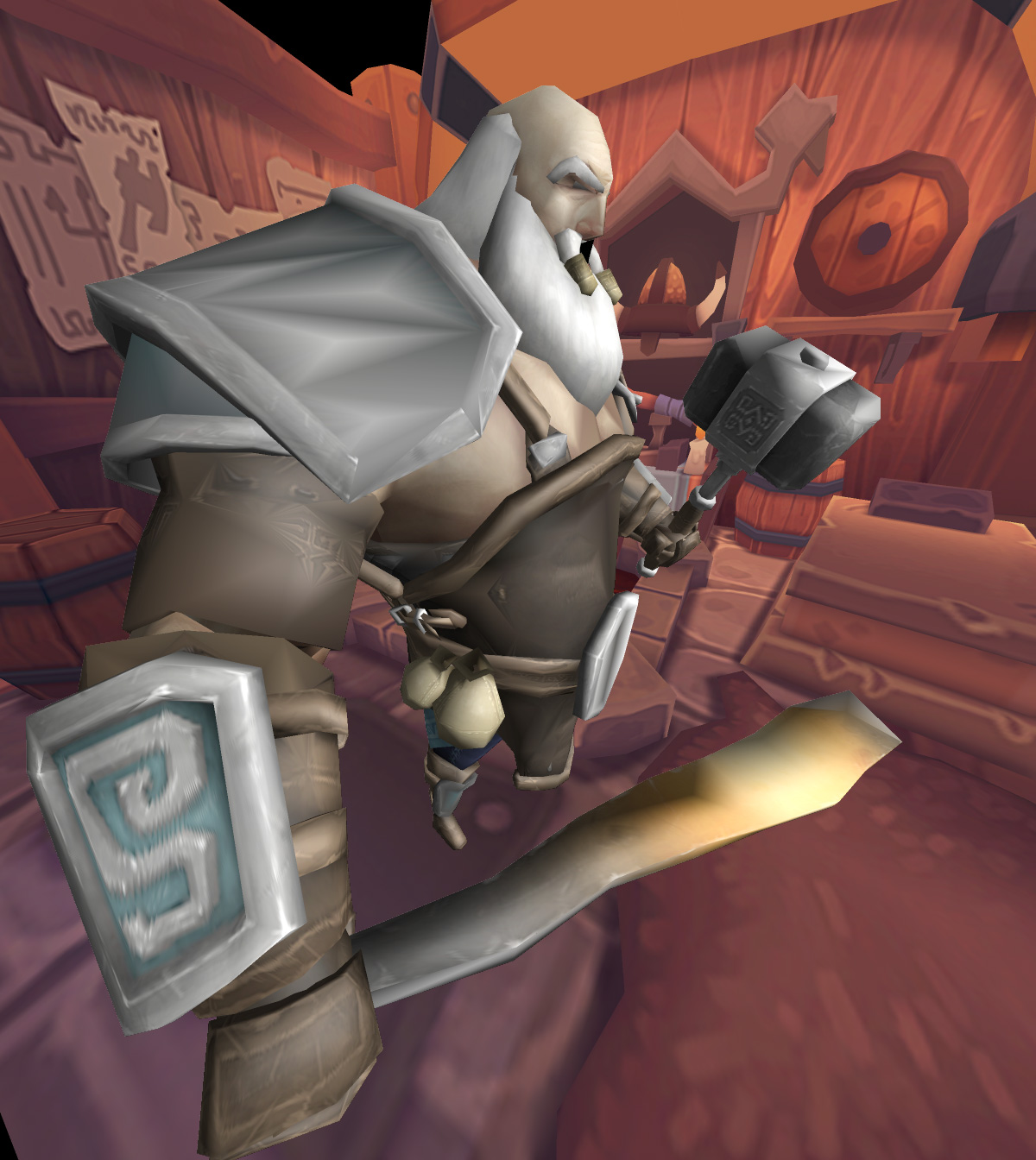}\vspace{0.4em}

		\includegraphics[width=\linewidth]{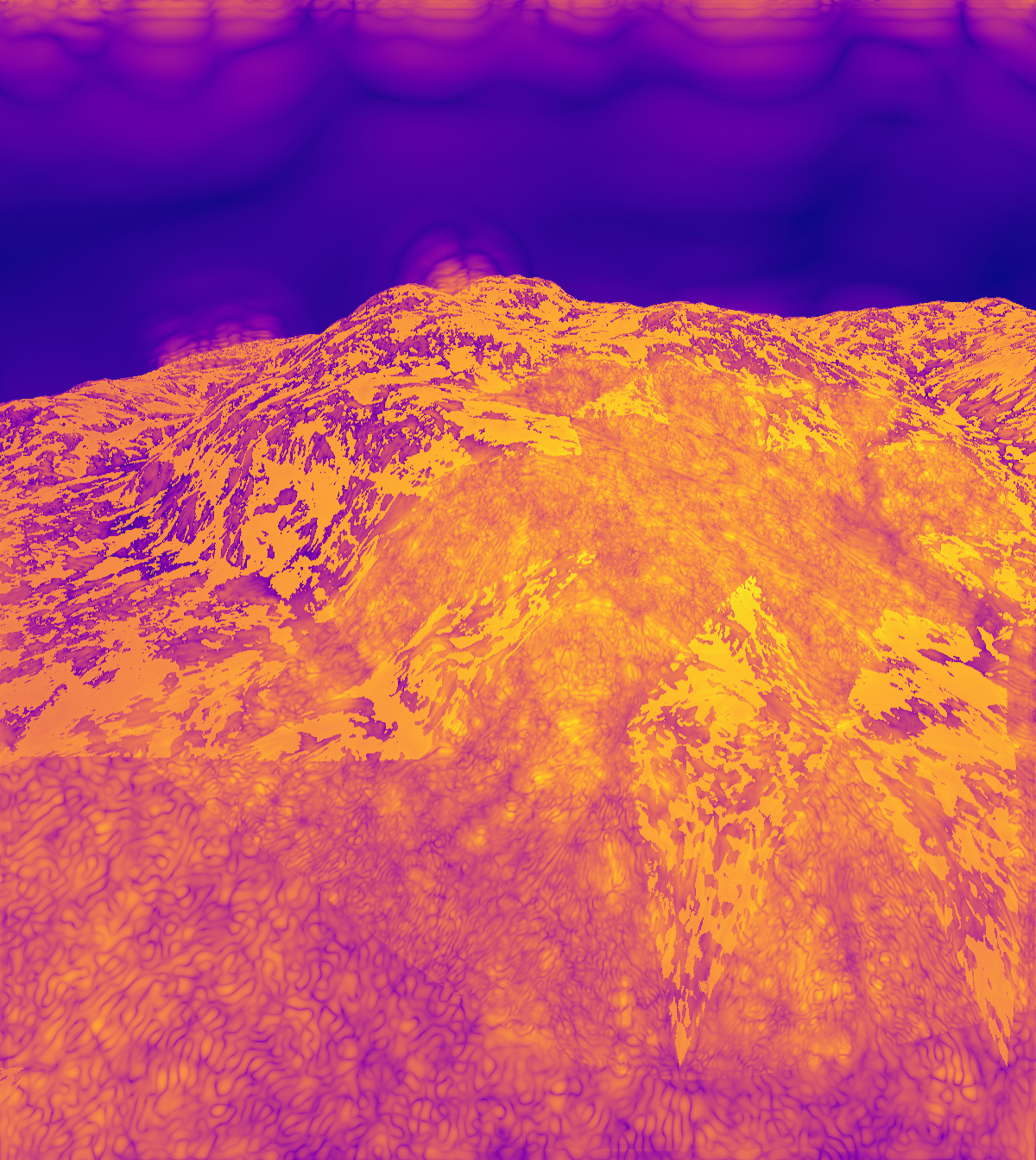}\vs

		\includegraphics[width=\linewidth]{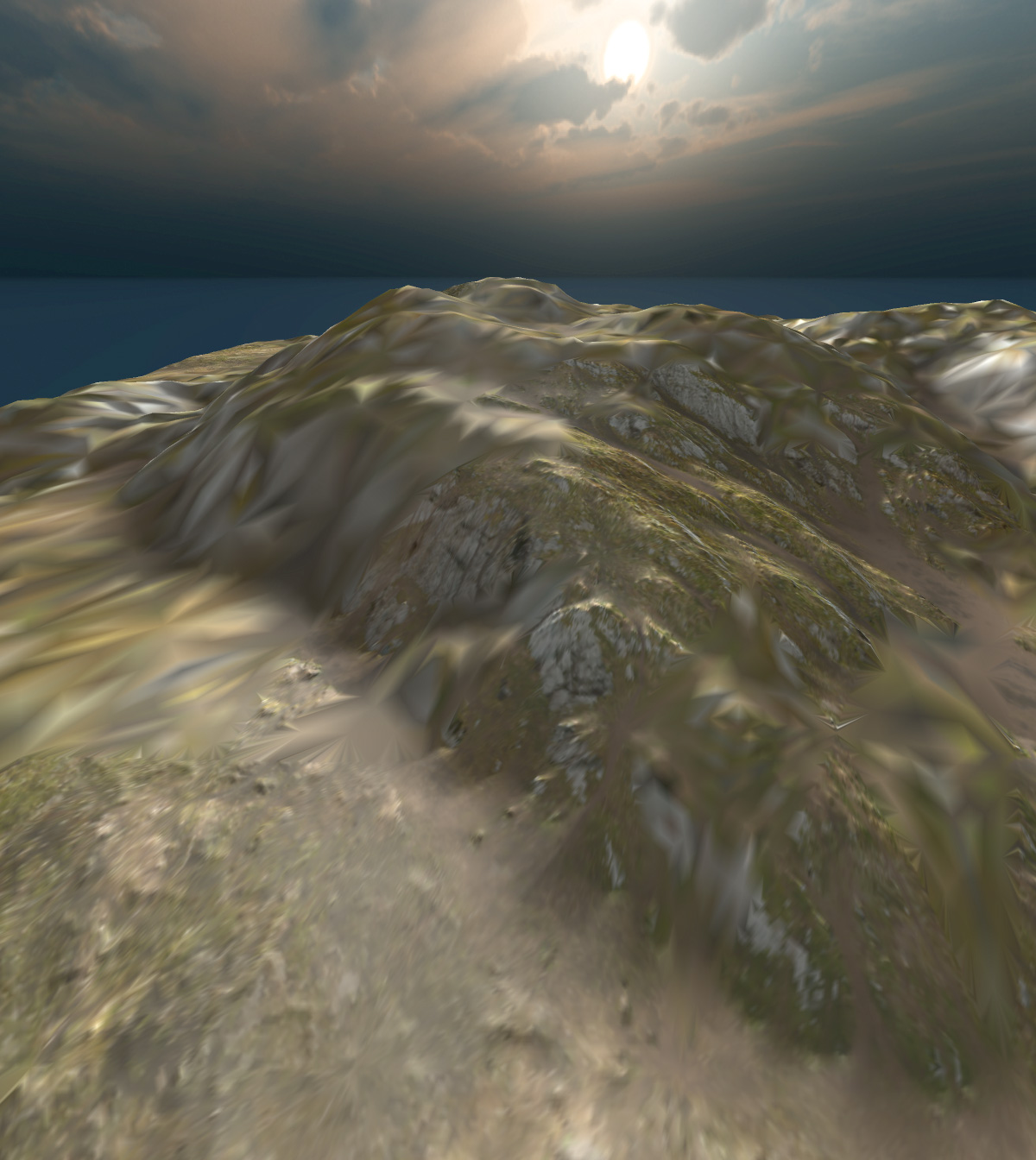}\vspace{0.4em}

		\includegraphics[width=\linewidth]{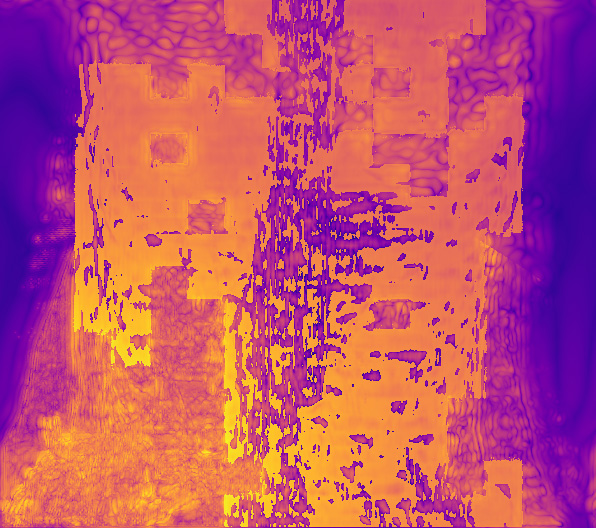}\vs

		\includegraphics[width=\linewidth]{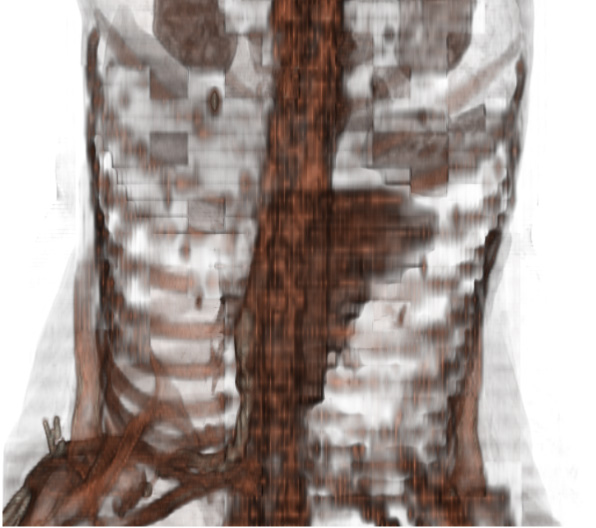}
	\end{minipage}
	\label{fig:3dAll:fix}
	}
	\hs
	\subfloat[saccade]{
	\begin{minipage}{0.17\linewidth}
		\includegraphics[width=\linewidth]{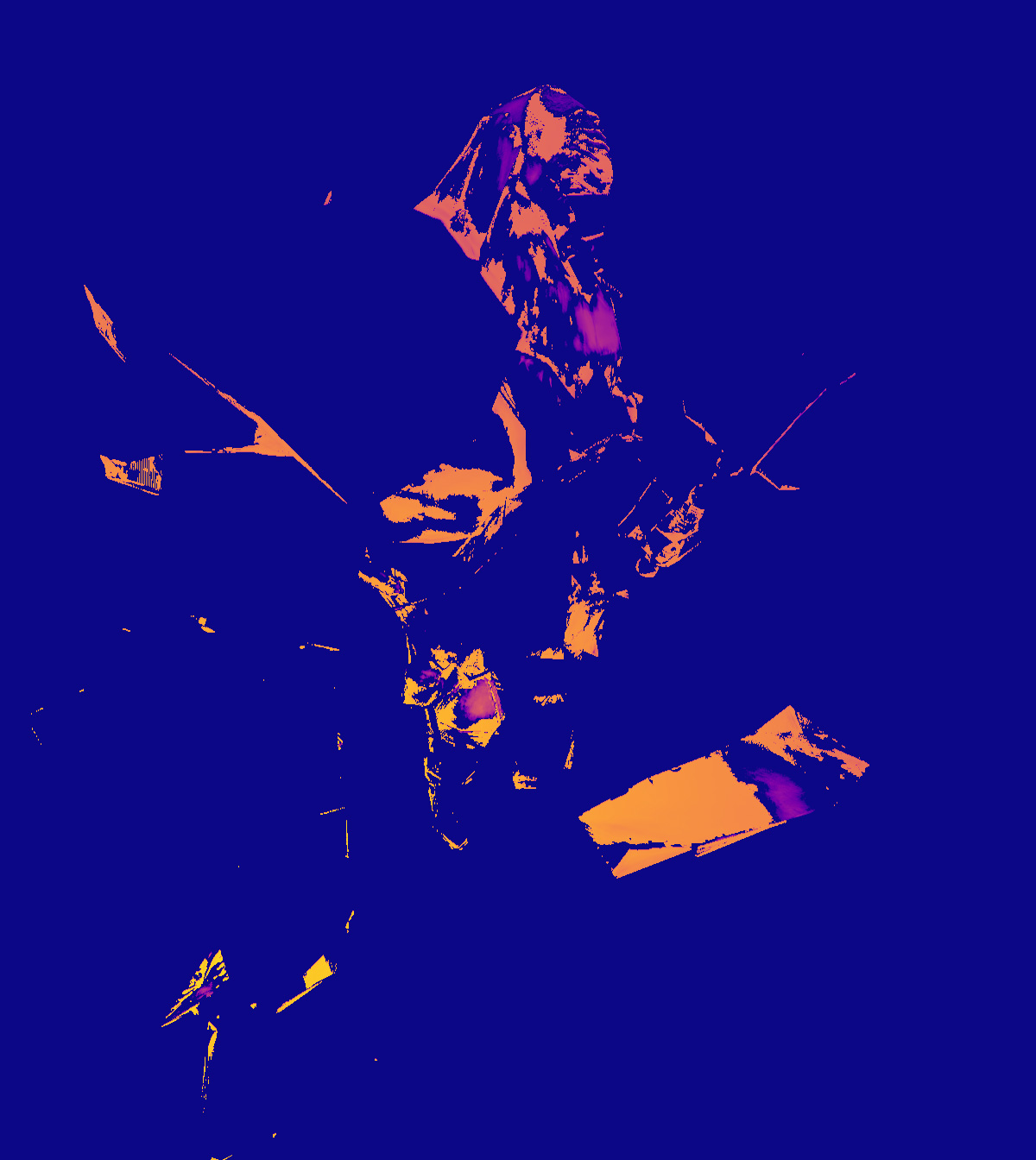}\vs

		\includegraphics[width=\linewidth]{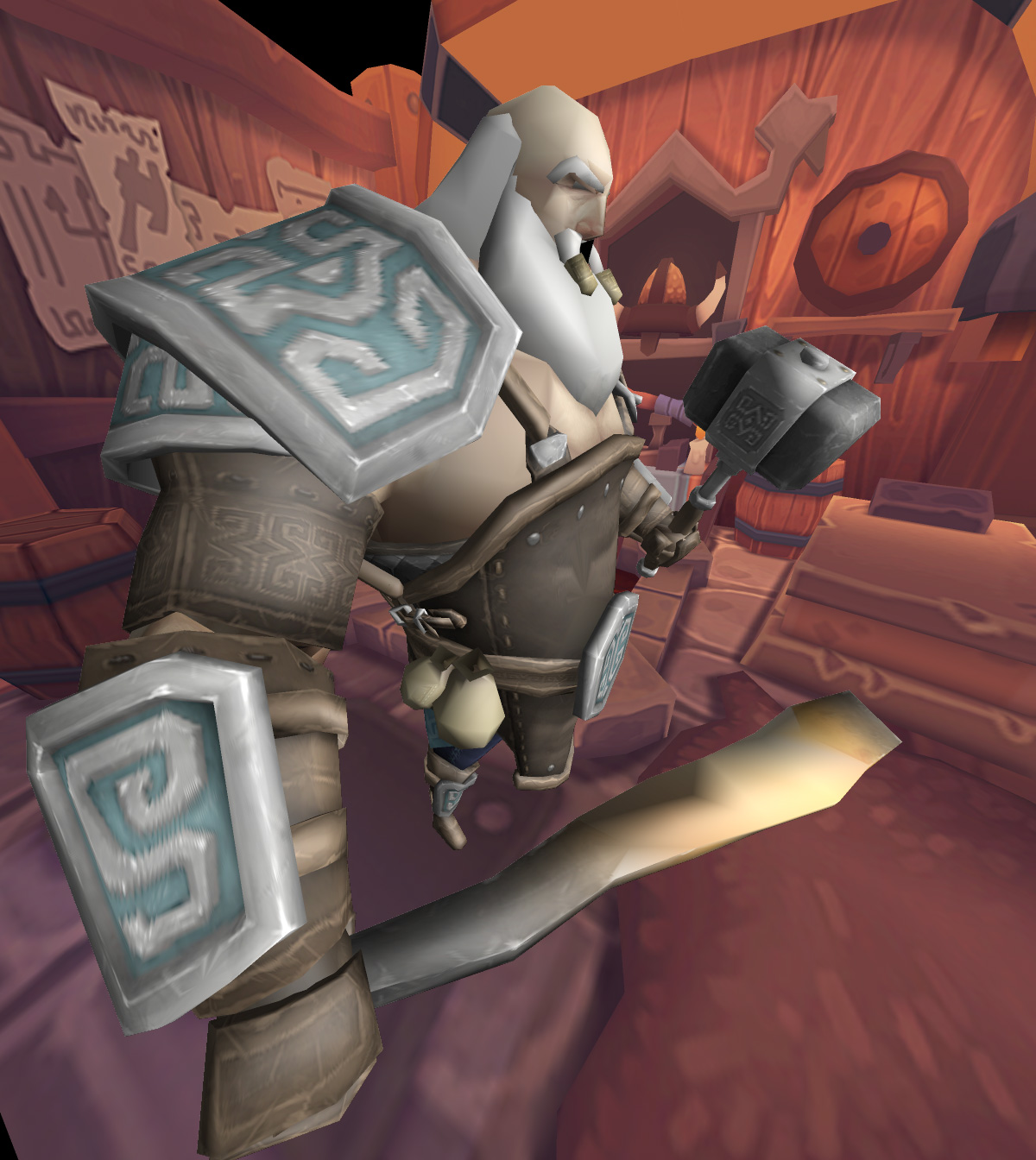}\vspace{0.4em}

		\includegraphics[width=\linewidth]{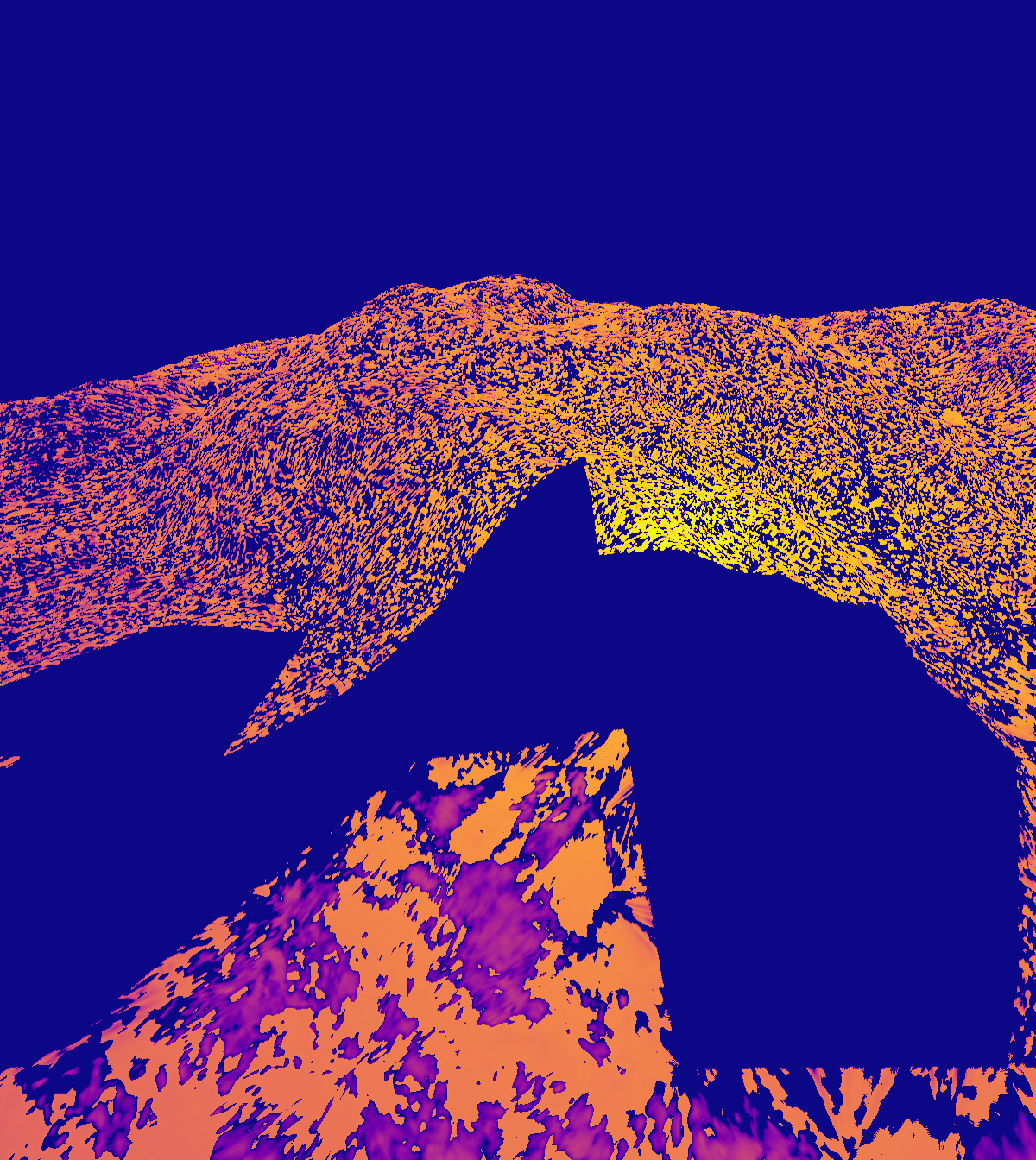}\vs

		\includegraphics[width=\linewidth]{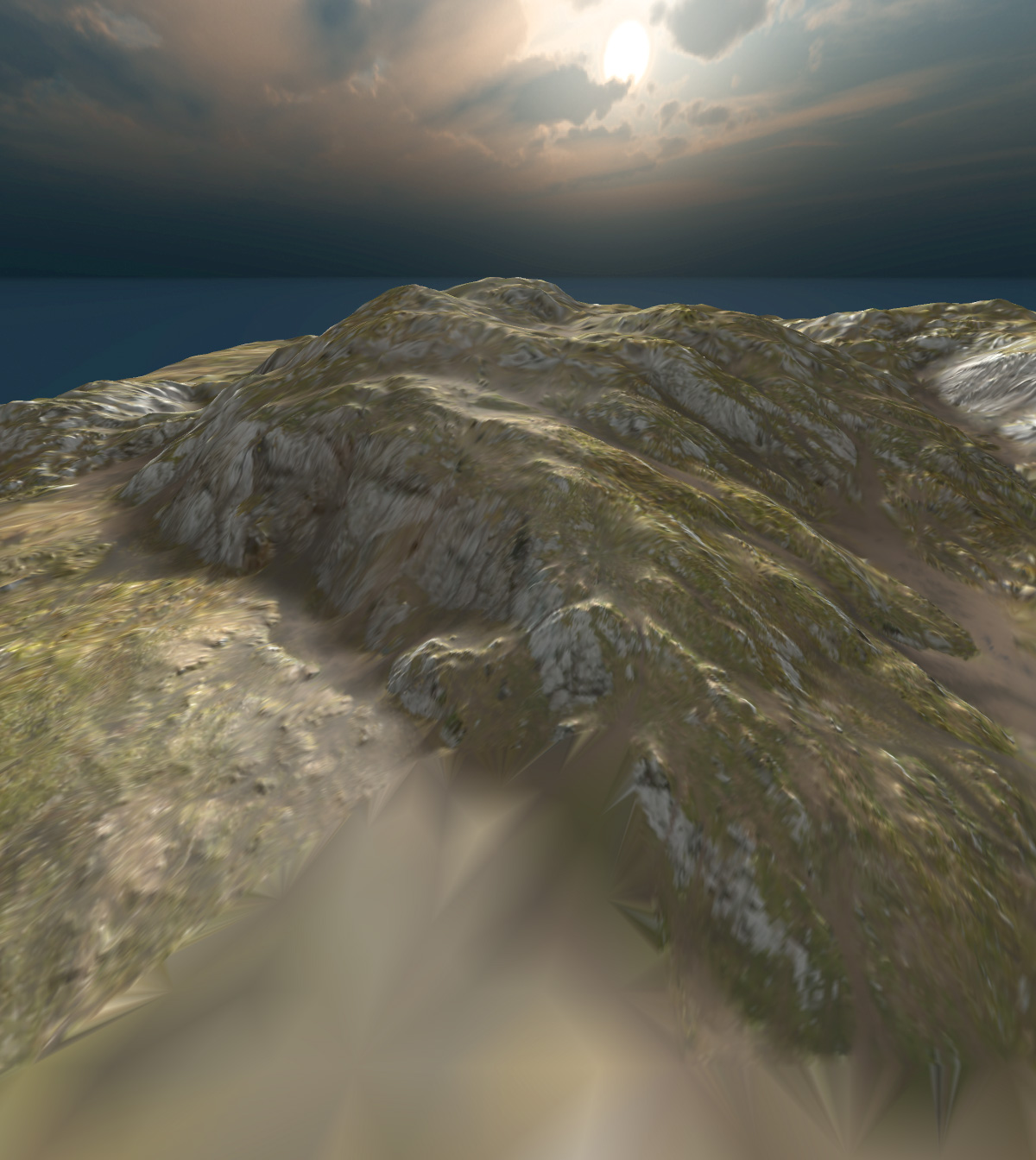}\vspace{0.4em}

		\includegraphics[width=\linewidth]{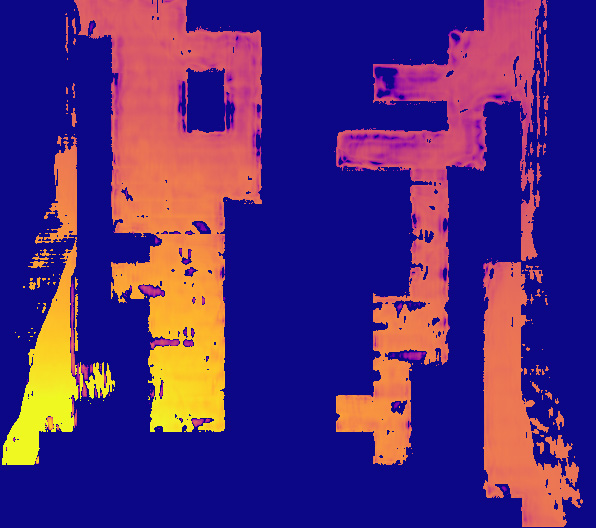}\vs

		\includegraphics[width=\linewidth]{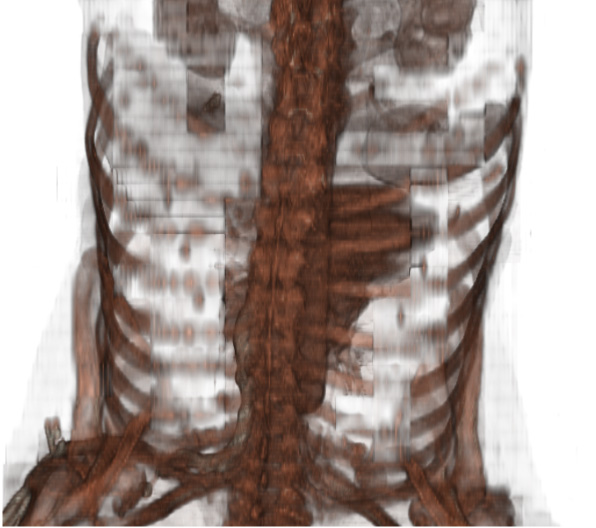}
	\end{minipage}
	\label{fig:3dAll:saccade}
	}

\Caption{Results for various 3D data types.}
{%
From top to bottom, the 3D assets are triangle mesh, displacement-based terrain, and CT scanned volume.
In each group of images, \protect\subref{fig:3dAll:currentTarget} shows \new{the semi-transmitted scenes at edge on top and full scenes at cloud on bottom}.
\protect\subref{fig:3dAll:ecc}/\protect\subref{fig:3dAll:popping} is the importance and corresponding updated rendering by considering only eccentricity \new{$\eccentricity$}/temporal consistency \new{$\poppingIntensity$}. Brighter means higher importance.
\protect\subref{fig:3dAll:fix}/\protect\subref{fig:3dAll:saccade} is the update if the gaze fixes/saccades considering both \protect\subref{fig:3dAll:ecc}\new{$\eccentricity$} and \protect\subref{fig:3dAll:popping}\new{$\poppingIntensity$}. Similar to \Cref{fig:teaser}, their first rows indicate estimated perceptual quality (brighter means stronger spatio-temporal artifacts or worse quality).

}
\label{fig:3dAll}
\end{figure*}

\paragraph{Cloud-based streaming}

Finally, we evaluate the perceptual quality {\it per bit} by updating the LoD level of a unit,
\begin{align}
\label{eqn:streaming_weight}
\weight_{\threeDimUnit_i, \timeStamp}(\LoDLevel_{\threeDimUnit_i, \timeStamp})
= 
\frac{
\threeDimAdaptive_{\threeDimUnit_i, \timeStamp}\left(\LoDLevel_{\threeDimUnit_i, \timeStamp}, \gazeVec_{\timeStamp-1}, \twoThreeDimMap_{\timeStamp - 1}\right)
}
{
\dataSize_{\threeDimUnit_i}(\LoDLevel_{\threeDimUnit_i, \timeStamp - 1}, \LoDLevel_{\threeDimUnit_i, \timeStamp})
}
,
\end{align}
where $\dataSize$ is the data volume difference by updating $\threeDimUnit_i$'s LoD level
from $\LoDLevel_{\threeDimUnit_i, \timeStamp - 1}$ at time $\timeStamp - 1$
to $\LoDLevel_{\threeDimUnit_i, \timeStamp}$ at time $\timeStamp$.

The \new{edge} side is a device with limited network bandwidth and storage.
Therefore, the 3D scene is fully stored on cloud side,
including all LoD levels of all units.
In runtime at time frame $\timeStamp - 1$,
the edge side sends the gaze position $\gazeVec_{\timeStamp - 1}$.
The cloud side knows the LoD levels of all units on the edge side as the edge side acknowledges the received network packages containing the 3D data.
According to that,
the cloud will compute the perceptual quality per bits for updating LoD levels of every unit accordingly,
with the updates holding highest $\weight$s to achieve the best improvement of perceptual quality,
given the constraint on the available update package size $\updatesize$.
\begin{align}
\begin{split}
\mathop{\mathrm{argmax}}_{\left\{\LoDLevel_{\threeDimUnit_i, \timeStamp}\right\}} &\sum_i\weight_{\threeDimUnit_i, \timeStamp}(\LoDLevel_{\threeDimUnit_i, \timeStamp}) \\
s.t. & \sum_i \dataSize_{\threeDimUnit_i}(\LoDLevel_{\threeDimUnit_i, \timeStamp- 1}, \LoDLevel_{\threeDimUnit_i, \timeStamp}) \leq \updatesize.
\label{eq:finalBandwidth}
\end{split}
\end{align}

\subsection{Neural Acceleration}
\label{sec:method:neural}
In a practical implementation, the cloud-side \new{needs} to compute
\Cref{eq:unit_sensitivity} for each update.
Although the cloud can afford more computation than the edge, the heavy frequency domain decomposition for individual LoD and frames in \Cref{eq:discrete_importance} may cause intolerable latencies on the end users.
Thus, for each scene, we train a multilayer perceptron (MLP) neural network for the fast prediction of $\threeDimAdaptive_{\threeDimUnit_i}$.

Specifically, with a camera view $\cameraVec_{\timeStamp-1}$ and gaze position $\gazeVec_{\timeStamp-1}$ as input, the network learns to compute
\begin{align}
    \network\left(\cameraVec_{\timeStamp-1}, \gazeVec_{\timeStamp-1}\right) = \left\{
        \threeDimAdaptive_{\threeDimUnit_i}\left(
			\LoDLevel_{\threeDimUnit_i, \timeStamp}, 
			\gazeVec_{\timeStamp-1}, 
			\twoThreeDimMap_{\timeStamp - 1}
        \right) \right\}
\label{eqn:neural}
\end{align}
for all $\threeDimUnit_i$ and possible $\LoDLevel_{\threeDimUnit_i, \timeStamp}$.
The camera view $\cameraVec_{\timeStamp-1}$ contains 3 vectors, which are coordinates of camera position, camera direction and up direction.
In the input, there is also a flag indicating whether a saccade is detected.
The current LOD state is not in the input since the network always predicts for all LODs, which will be used by the cloud to make the streaming decision.
Although we do not provide the mapping
$\twoThreeDimMap_{\timeStamp - 1}$ explicitly to the network, the network also infers it from $\cameraVec_{\timeStamp-1}$ once trained on a specific scene.
The diversity of projected areas of various units $\threeDimUnit_i$ cause
$\threeDimAdaptive_{\threeDimUnit_i}$ to have no upper bound, making it difficult to optimize the network. To combat this, we normalize
$\threeDimAdaptive_{\threeDimUnit_i}$ by the pixel counts of individual $\threeDimAdaptive_{\threeDimUnit_i}$ in the projected screen space.
After normalization, the value is bounded by the maximum value of \Cref{eq:poppingIntensity}.
The cloud stores the network and retrieves the fast-inferred $\threeDimAdaptive_{\threeDimUnit_i}$ for streaming decision (\Cref{eqn:streaming_weight,eq:finalBandwidth}),
so the cloud doesn't need to render the actual image for the decision.

\new{We report specific implementation details such as network architecture, loss
function, and dataset generation in \Cref{sec:implementation} and the performance/precision analysis in \Cref{sec:results:perf,sec:results:approximation}. 
Please also refer to our code repository for the actual implementation of the network structure.} We evaluate both performance gains and prediction precision of our neural network in \Cref{sec:results:perf,sec:results:approximation}.
\new{Change the super parameters like the $\omega$ will require retrain the neural network. To avoid retraining the network when changing the $\weightEccPop$, one solution is to train 2 networks to predict the $\eccentricity$ and $\poppingIntensity$ separately}.

\section{Implementation}
\label{sec:implementation}

\paragraph{3D Assets}
\label{sec:asset}

Our current method assumes we can set the LoD of each $\threeDimUnit$ independently from one another.
While this works for asset types with more independent units such as point clouds \cite{Rusinkiewicz:2000:QMP,Schutz:2020:PRR}, volumes, and crowd agents, it might introduce artifacts for others, such as cracks or T-junctions for triangle meshes.
To ensure quality without introducing complex implementation, we currently use vertex colors instead of textures, and build the mesh hierarchies by sub-sampling existing vertices without changing their positions.
The displaced mesh (terrain scene from \Cref{fig:3dAll}) is handled in a similar way, except that additionally a height map is also sampled for vertices.
For volume data, the LOD is created by sub-sampling the voxels.

Solving \Cref{eq:finalBandwidth} is equivalent to the knapsack problem, which is NP-Complete.
For this reason, we use a greedy approximation, which always selects the unit with the largest $\weight_{\threeDimUnit_i, \timeStamp}(\LoDLevel_{\threeDimUnit_i, \timeStamp})$ in \Cref{eqn:streaming_weight} if the bandwidth allows so.

\paragraph{System}

We simulate the streaming network via ZeroMQ library (\url{https://zeromq.org/}). The rendering system was implemented with OpenGL. 
The system runs on a PC with Intel i9-9880H 8-core CPU and with 32GB of RAM, and an NVIDIA RTX 2080 graphics card.
For the VR headset, we use an HTC Vive Pro Eye with an integrated eye tracker.
It has a $110$ degree FOV and $1440\times1600$ spatial resolution ($14$ pixels per degree, i.e., $7$ cycles per degree for $\displayBand$).
Our system runs at 90-FPS with neural acceleration.

\paragraph{Neural acceleration}
We prepare the dataset for the neural network in
\Cref{sec:method:neural} by first sampling short sequences of camera and/or gaze
movements inside a 3D scene at 90-FPS. 
The views are freely controlled by the HMD users.
Then the offline computed (via \Cref{eq:unit_sensitivity}) $\{\left(\cameraVec_{\timeStamp-1}, \gazeVec_{\timeStamp-1}\right),\threeDimAdaptive_{\threeDimUnit_i}\left(
			\LoDLevel_{\threeDimUnit_i, \timeStamp}, 
			\gazeVec_{\timeStamp-1}, 
			\twoThreeDimMap_{\timeStamp - 1}
        \right)\}$ pairs are used to train the network. 
Since the adjacent frames are likely to have similar data, we sample 4 frames per second uniformly and compute the ground truth
$\threeDimAdaptive_{\threeDimUnit_i}$ for all $\threeDimUnit_i$ and LOD
levels offline.
There are 16 sequences, and each sequence lasts for about 30 seconds and contains $\sim 120$ samples.
We use 15 sequences as the training set and the remaining one as the test set.

The neural network is fully-connected and consists of 3 inner-layers with
$(100, 1000, 1000)$ neurons, each with \emph{ReLU} activations, and an output
layer with a \emph{Sigmoid} activation. The network was trained with \emph{L1}
loss for $100$ epochs.
In the training, we set the learning rate to $0.001$ and batch size to $128$. we used an SGD optimizer with momentum = 0.9.
For the neural network in the user study (\Cref{sec:evaluation:study:live}), since we have $\sim 2000$ samples and each sample contains $722$ (triangles) $\times 4$ (LOD) $\approx 3000$ scores.
The sample size, higher than the parameter numbers, \new{avoids} biased overfitting. \new{We refer the readers to our open source implementation for reproduction details.}

\paragraph{Parameters}
We determine the optimal spatio-temporal balancing $\weightEccPop$ via a pilot user study. Users reported $3.0$ as the most plausible experience.
In our experiment, the Weber's law adjustment for low intensities $\weberWeight$ was set to $10$.
With low network bandwidths, the peripheral content may not receive the priority of transmission.
Thus, we included a pedestal constant $2.0$ to the $\eccentricity$ for each pixel.
\section{Evaluation: User Study}
\label{sec:evaluation:study}
We compare different progressive streaming methods with different usages of perceptual mechanisms under the same network bandwidth -
uniformly increasing the LoD of each $\threeDimUnit$ based on visibility (\uniformCondition),
``foveated'' streaming without considering dynamic consistency (\eccCondition, a 3D version of \cite{Romero:2018:FSV}),
and our method (\oursCondition).
We experimented with two display environments,
active HMD-based 6 DoF (degrees of freedom) navigation (\Cref{sec:evaluation:study:live}) for temporal consistency evaluation only
and passive screen-based observation (\Cref{sec:evaluation:study:video}) for evaluation of both visual temporal consistency and static visual quality.

\subsection{Eye-Tracked Study}
\label{sec:evaluation:study:live}
We first conducted a user study with an eye-tracked VR HMD to assess our method's effectiveness on perceived temporal consistency.

\begin{figure}[t]
\centering	
  \subfloat[VR setting]{
    \includegraphics[width=0.4\linewidth]{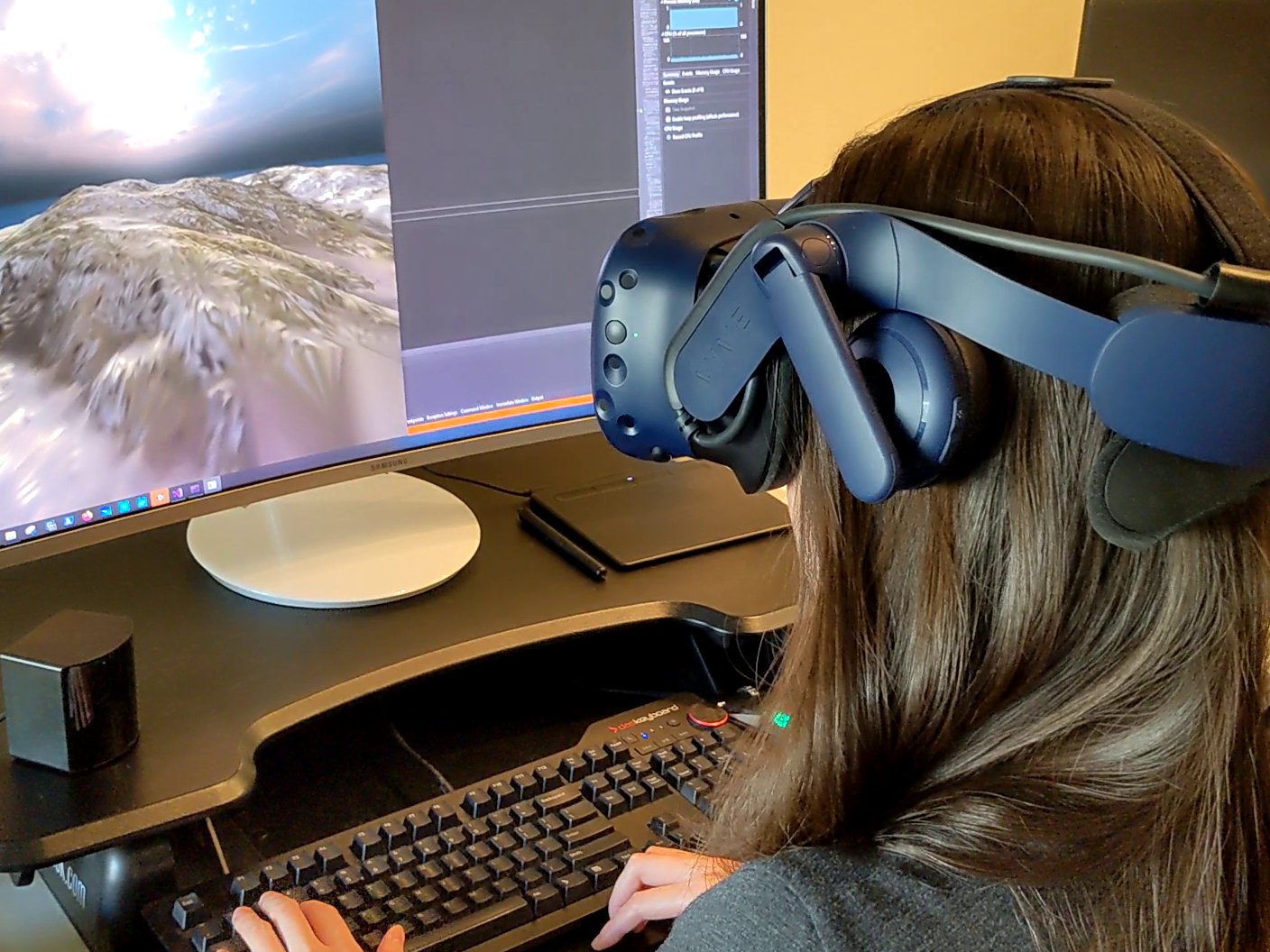}
    \label{fig:study:setting:vr}
  }
  \subfloat[Video setting]{
    \includegraphics[width=0.4\linewidth]{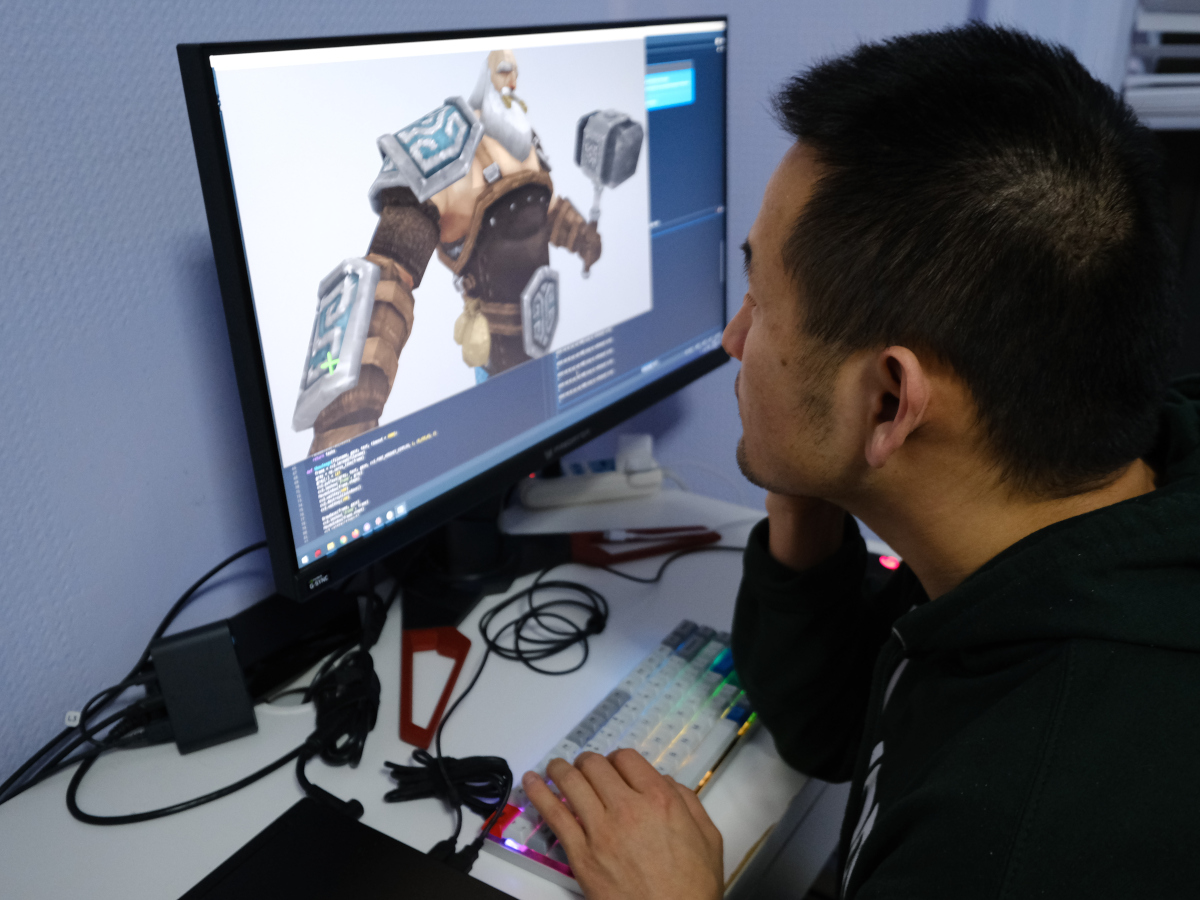}
    \label{fig:study:setting:screen}
  }
	\Caption{The setting of our user study.}
        {%
Both \subref{fig:study:setting:vr} and \subref{fig:study:setting:screen} were captured from parts of our user studies.
        }
	\label{fig:user_study_scene}
	\vspace{-2em}
\end{figure}

\paragraph{Stimulus and setup}
We used a terrain displacement scene from \Cref{fig:3dAll} in this experiment. Based on users' real-time head and gaze motions, the stimuli is a sequence of simulated progressive streaming updates (from a coarse to adaptively finer level).
During the study, users wore an HTC Vive Pro Eye Headset and remained seated. They were instructed to freely observe the virtual scene and keep their fingers on the keyboard to make selections after each trial, as shown in \Cref{fig:study:setting:vr}.
Eight users (age $23-31$, $1$ female) participated in the study. 
To accommodate the scene data size and study duration, we experimented with a simulated 5G network bandwidth locally.

\paragraph{Task}
Due to the limited access to participants during COVID-19, we aim to maximize trial randomness and numbers to ensure objective measurements. 
We designed a two-alternative-forced-choice (2AFC) experiment. Each trial consists of a pair of conditions among the three (\uniformCondition/\eccCondition/\oursCondition). In each condition, the stimulus was initiated with the coarsest LoD. The participants were then instructed to freely observe the environment that was streamed and updated with the corresponding method. Each condition lasted for $10$ seconds. At the end of each trial, the participants used the keyboard to select which one of the two conditions appeared more smoothly and comfortably updated with fewer artifacts over the entire duration. 

Each experiment began with a warm-up session to familiarize the participants with the interface and task, followed by $24$ counter-balanced and randomly ordered trials.
Thus, each pair of conditions was evaluated $8$ times per participant.
To minimize fatigue, a 2-second break was enforced between trials.

\begin{figure*}
\centering
\subfloat[VR eye-tracked temporal consistency]{
\includegraphics[height=3.8cm]{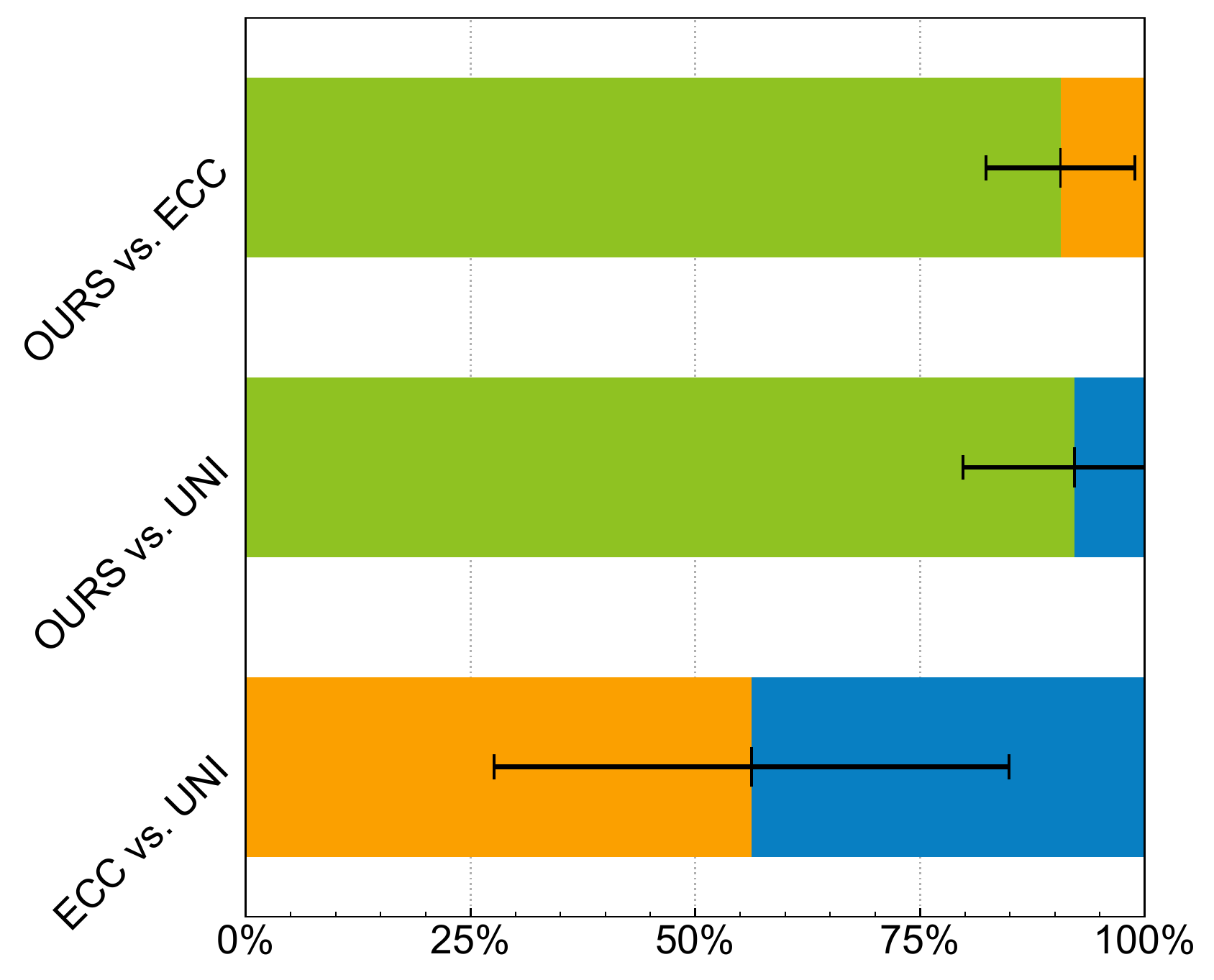}
\label{fig:study:vr}
}
\subfloat[video-based temporal consistency]{
\includegraphics[height=3.8cm]{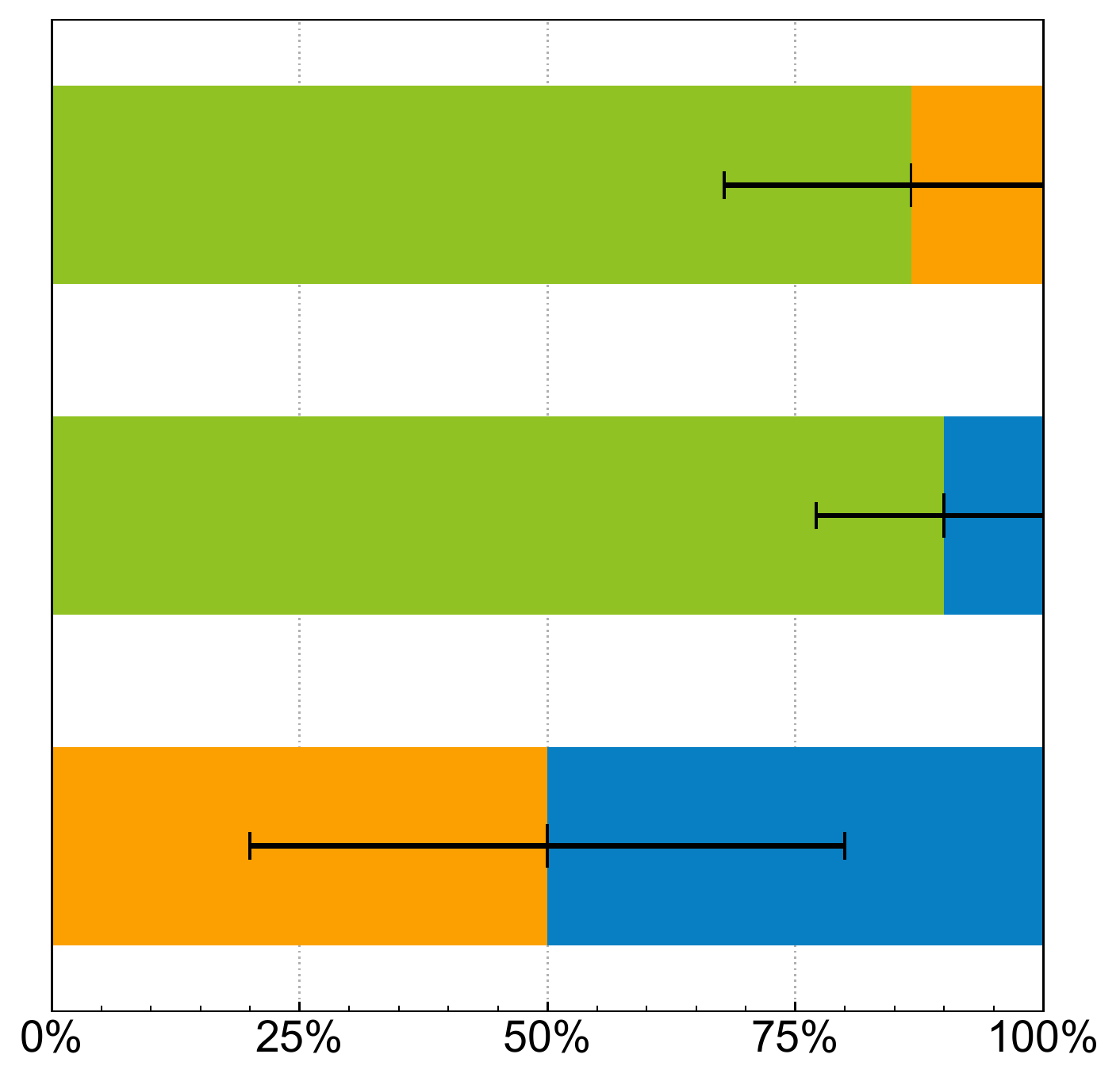}
\label{fig:study:video:popping}
}
\subfloat[video-based quality]{
\includegraphics[height=3.8cm]{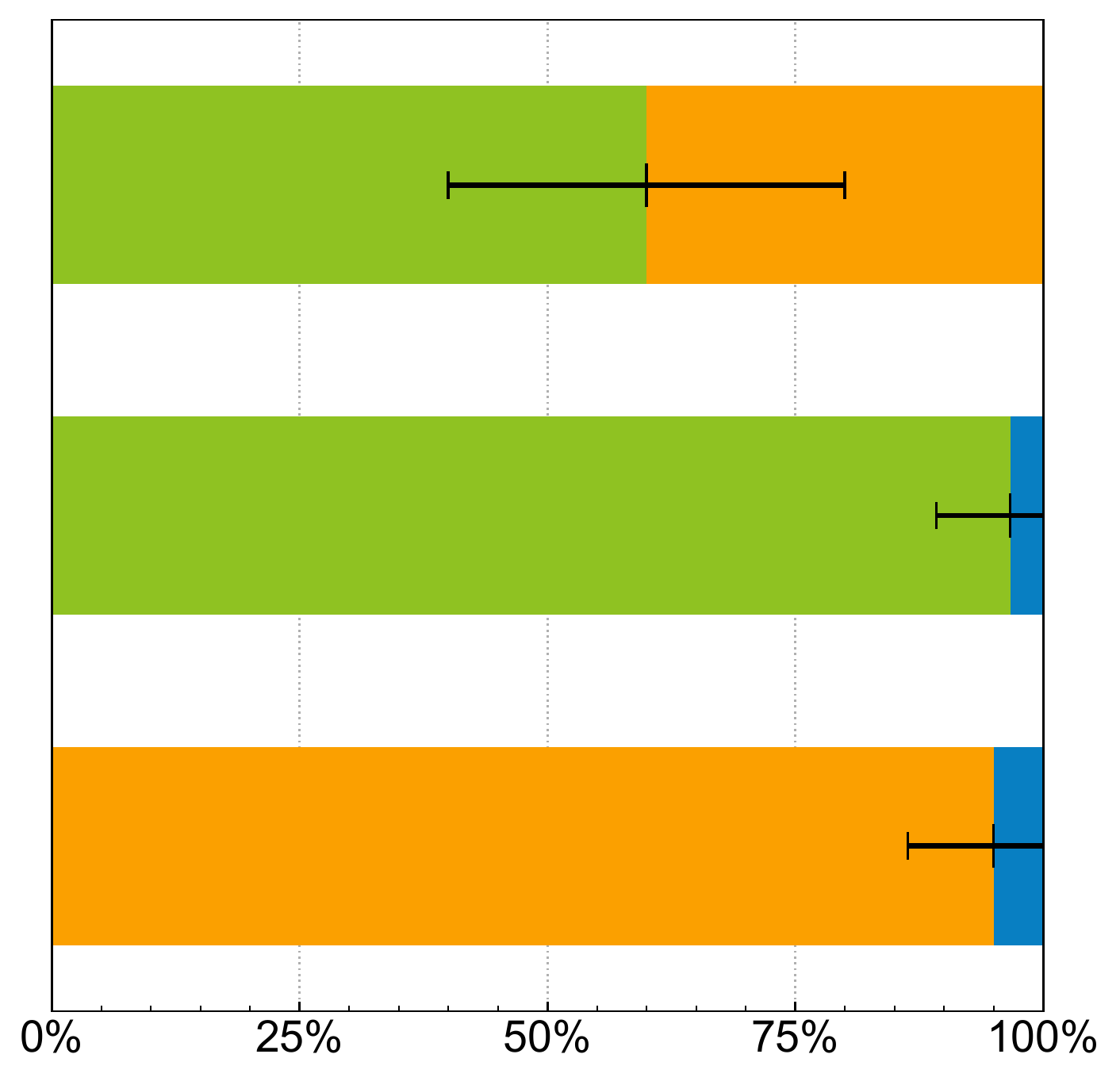}
\label{fig:study:video:quality}
}
\Caption{User study results.}
{%
We compare pair-wise 2AFC user vote percentage in both HMD and traditional display studies from \protect\Cref{sec:evaluation:study}. The green/orange/blue bar indicates the percentage voting for \oursCondition/\eccCondition/\uniformCondition and the errorbars indicates the standard deviations.
}
\label{fig:study}
\end{figure*}

\paragraph{Results}
\Cref{fig:study:vr} plots the results of the pair-wise percentages that participants indicated as the preferred method.
Specifically, \oursCondition{} was voted as significantly more preferred than both \eccCondition{} ($90.6\%\pm8.3\%$) and \uniformCondition{} ($92.2\%\pm12.4\%$).
With a significance level established at $p=0.05$ and power $>.999$, binomial tests indicate the difference is significant for both comparisons ($p<.001$).
\eccCondition{} showed marginally higher voting rate than \uniformCondition{} ($56.3\%\pm28.6\%$). We did not observe significance with a binomial test ($p=.38$).

\paragraph{Discussion}
The results above showed our method's significant temporal quality preference over the alternative solutions: under the same network conditions, our method could stream 3D assets more smoothly by reducing popping artifacts.
However, this shall be achieved without compromising spatial-visual quality. Our initial study design considered several evaluation metrics, including visual fidelity to the reference, temporal smoothness, and perceptual comfort.
We planned to evaluate them both individually and collectively: ask participants to focus on individual metric (such as visual fidelity or temporal smoothness only), and additional post-study comments (such as why they choose one condition over another).

However, during our pilot runs, we noticed several problems with this initial design.
Some participants commented that it was difficult for them to focus on more than one aspect of the evaluation criteria, especially for those with less VR or gaming experiences.
Meanwhile, visual quality depends on the temporal integration of multiple instead of individual frames. The integration also depends on the exact view path, which can be freely moved by the participants and thus may differ across pairs of conditions for a valid comparison.
To address these issues, we decided to focus on only temporal smoothness, which tends to be less view-path dependent than spatial quality, for this free navigation HMD experiment, and leave \new{per-frame} visual quality evaluation by fixing the gazes, as described in the following experiment.

\subsection{Screen-Based Study}
\label{sec:evaluation:study:video}

Due to the aforementioned design challenges and extra difficulties of physically distributing the eye-tracked VR display, we further simulate the study procedure in \Cref{sec:evaluation:study:live} to a remote setting with recorded stimuli. This ensures a thorough evaluation of both temporal smoothness and spatial quality.

Specifically, we recorded videos by randomly pre-defined gaze paths.
During the study, users were remotely monitored and instructed to keep one eye open, fix their head positions, and gaze on the green cross and follow the same procedure as \Cref{sec:evaluation:study:live}. 

\paragraph{Stimulus and setup}

To evaluate a different data type from the eye-tracked study, we chose the triangle mesh scene in \Cref{fig:3dAll} for this study.
The visual stimuli were rendered with $1920 \times 1080$ resolution and $60$ degree of vertical field of view,
so that the $\displayBand$ is the same for every participant.
In the monitored remote study, participants first input their computer resolution and screen size. Our study protocol would automatically \new{compute and inform} participants of the correct eye-display distance. The user then fixed their head at the reported distance cross accordingly. Twelve users (age $24-41$, $5$ females) participated in this study. All users had normal or corrected-to-normal vision.

\paragraph{Tasks}
The experiment was conducted in two phases: video-based stimuli for temporal smoothness and static frame-based stimuli for visual quality.
The task of the \textit{first} phase was similar to \Cref{sec:evaluation:study:live}.
The users observed two sequentially played stimuli from the three conditions.
After each trial, they selected the one with preferred temporal consistency (i.e., fewer artifacts).
In the \textit{second} phase, each trial consisted of two static images randomly sampled from the sequences in the first phase at the same timestamp.
In each trial, a full-quality rendering (second row of \Cref{fig:3dAll:currentTarget}) was first shown for 3 seconds as the quality reference.
Then, 2 images of the 3 conditions were sequentially displayed for 3 seconds.
After each trial, the participants selected the one that appeared to be closer to the reference in visual quality.
Both tasks contain $15$ counter-balanced and randomly ordered trials.
Thus, each pair of conditions was evaluated $5$ times per participant.
\new{
\paragraph{Task Design Rationale}
Fixing gaze in 2AFC experiment has been practiced in assessing retinal acuity and perceived visual quality in VR \cite{Patney:2016:TFR, Sun:2017:PGF, Tursun:2019:LCA}. 
In fact, the human visual perception is limited during natural, active viewing conditions \cite{Cohen:2020:LCA}. 
Further, because of the suppressed vision during saccades, the visual quality is primarily determined by the frames at gaze fixations. For this reason, the current experiment where the gaze is fixed may generally represent free viewing conditions where fixations are connected by saccades.}

\paragraph{Results}
\Cref{fig:study:video:popping,fig:study:video:quality} plot the results.
For temporal consistency (phase-1), a consistent trend to \Cref{sec:evaluation:study:live} was observed: \oursCondition{} had a significant higher ration of voting over the alternatives ($86.7\%\pm18.8\%$ against \eccCondition{} and $90.0\%\pm12.9\%$ against \uniformCondition{}).
With a significance level established at $p=0.05$ and power $>.999$, binomial tests indicate the difference is significant for both comparisons ($p<.001$).
\eccCondition{}, however, did not show major preference gain over \uniformCondition ($50.0\%\pm30\%,p=1.0$).

On the extended visual quality (phase-2), marginally higher voting rate on \oursCondition{} was observed over \eccCondition{} ($60.0\%\pm20\%,p=.16$). On the other hand, with a significance level established at $p=0.05$ and power $>.999$, both \oursCondition{} and \eccCondition{} were voted to have significantly improved quality than \uniformCondition{} ($96.7\%\pm7.4\%$ and $95.0\%\pm8.7\%$ respectively, $p<.001$ for both).

\paragraph{Discussion}
This study simulates and extends \Cref{sec:evaluation:study:live} to a remote screen-based setting.
Beyond the agreeing observation of temporal consistency, the phase-2 study also indicated that \oursCondition{} preserves the high visual fidelity as \eccCondition.
The experiments with both eye-tracked VR and traditional monitors show that our method significantly improves the perceived temporal consistency (i.e., minimizing popping artifacts)
without compromising the visual quality.

\section{Evaluation: Objective Analysis}
\label{sec:evaluation:analysis}
\begin{figure}
\centering
	\subfloat[static scene]{
		\includegraphics[height=3.2cm]{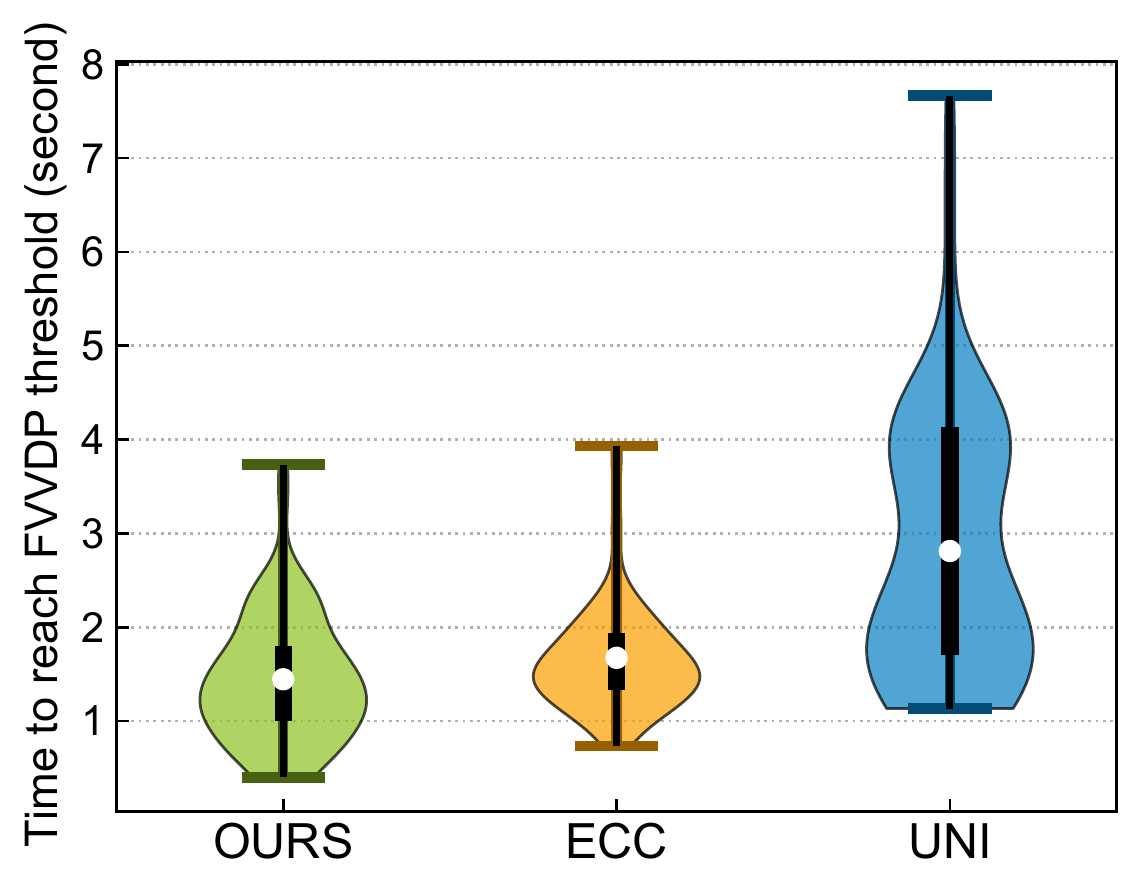}
		\label{fig:fvvdp:static}
	}
	\subfloat[\new{dynamic scene}]{
		\includegraphics[height=3.2cm]{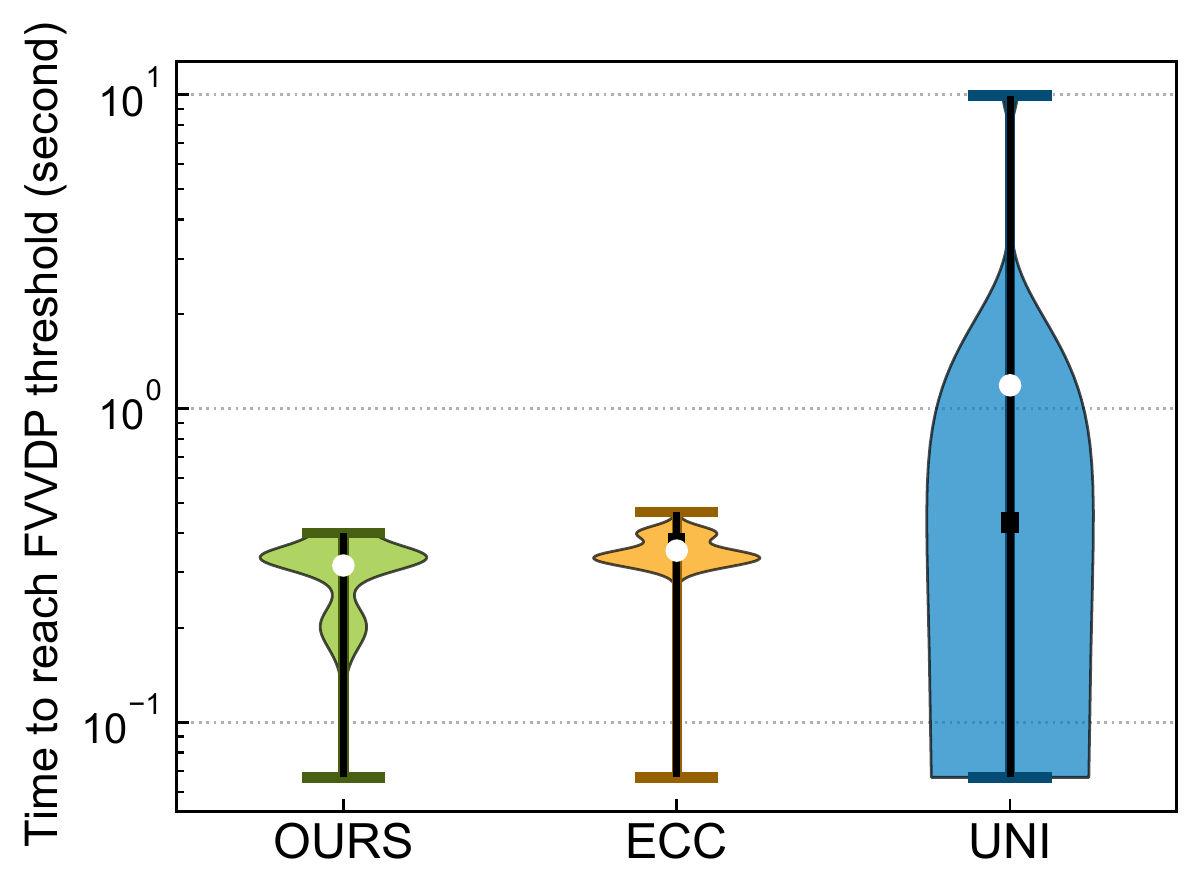}
		\label{fig:fvvdp:dynamic}
	}
	\Caption{The statistics of the time FOVVideoVDP first reaches threshold per sequence among all users \new{(\Cref{sec:results:fovvideovdp})}. }
	{\new{The sub-figure \subref{fig:fvvdp:static}/\subref{fig:fvvdp:dynamic} shows the results with static/dynamic scenes.} The violin plot shows mean values, 1st/3rd quartiles, and whiskers min/max. Lower values indicate better streaming efficiency in additively improving the quality.}
	\label{fig:fvvdp}
\end{figure}

\subsection{Visual Quality}
\label{sec:results:fovvideovdp}

\Cref{sec:evaluation:study} shows our significant benefits of subjective quality judgement in terms of temporal consistency and artifact reduction.
For further objective evaluation considering dynamic gaze adaptation stimuli (\Cref{eqn:adaptive}), we conducted a series of analytical experiments.

\paragraph{Static scene}

Recently, a thoroughly considered metric, FovVideoVDP \cite{Mantiuk:2021:FVD}, was proposed to measure video quality considering both spatio-temporal and foveated effects. We compute the FovVideoVDP with rendering on the fully transmitted asset as a reference.
The analysis comprises full $10$-second sequences from each user (one sample from each tri-condition group, $8\times8$ in total) of the eye-tracked study.
Since edge-cloud streaming is a temporally procedural quality enhancement, we measure the timing when FovVideoVDP reaches a shared threshold. Intuitively, this measures the efficiency of the streaming achieving high quality. In the experiment, we chose the threshold as $6.5$ to ensure all conditions, even the slowest, can reach it with the limited trial durations.
As shown in \Cref{fig:fvvdp:static}, the average time of \oursCondition{} ($1.446\pm0.669$) and \eccCondition{} ($1.674\pm0.589$) are significantly shorter than \uniformCondition{} ($2.813\pm1.429$). 
Pairwise t-tests show the difference between all pairs are significant ($p<.001$ for \oursCondition vs. \uniformCondition and \eccCondition vs. \uniformCondition, $p = .032$ for \oursCondition vs. \eccCondition).

\begin{figure}[b]
	\centering	
	\includegraphics[width=0.96\linewidth]{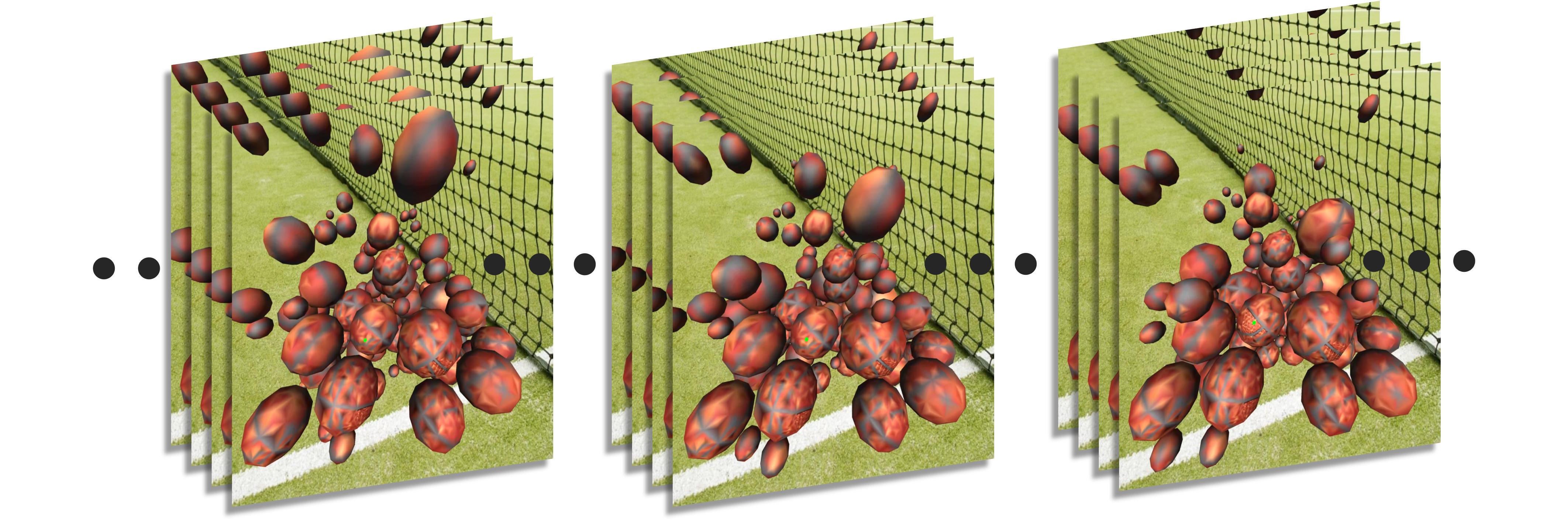} 
	\Caption{\new{Example frames of the dynamic scene.}}
	{%
	\new{The animation is composed of free-falling soft balls. Each image represents a frame along time. Please refer to the supplementary video for the full visualization.}
	}
	\label{fig:dynamic_scene}
\end{figure}

\paragraph{Dynamic scene}
\new{
We further preformed an experiment on a dynamic scene which consists of an animation of a pile of falling soft balls as shown in \Cref{fig:dynamic_scene}. We refer the complementary video for full animation.
The gaze is from the eye-tracked study with the same setting as the static scene study, except that the dynamic assets and threshold as $6.7$. 
As shown in \Cref{fig:fvvdp:dynamic}, the average time of \oursCondition{} ($0.316\pm0.069$) and \eccCondition{} ($0.352\pm0.049$) are significantly shorter than \uniformCondition{} ($1.184\pm2.548$). 
Pairwise t-tests show the difference between all pairs are significant ($p=.008$ for \oursCondition vs. \uniformCondition, $p=.011$ for \eccCondition vs. \uniformCondition, and $p<.001$ for \oursCondition vs. \eccCondition).
}

The above analysis on the eye-tracked gaze data revealed a consistent observation to the subjective studies in \Cref{sec:evaluation:study}. 
That is, \oursCondition and \eccCondition deliver high visual quality more efficiently than \uniformCondition. Meanwhile, comparing with \eccCondition, \oursCondition does not compromise the visual quality with the benefits of improved temporal consistency.

\subsection{System Performance}
\label{sec:results:perf}

\paragraph{Performance gain from neural acceleration}
We conducted a performance analysis of the neural acceleration effectiveness. With all the scenes and gaze trajectories, it takes an average of $389.28 \pm 14.44$ ms without the neural acceleration. In comparison, the neural network accelerates the computation to $20.30 \pm 0.89$ ms.

\begin{figure}[t]
  \subfloat[illustration]{
    \includegraphics[height=0.24\linewidth]{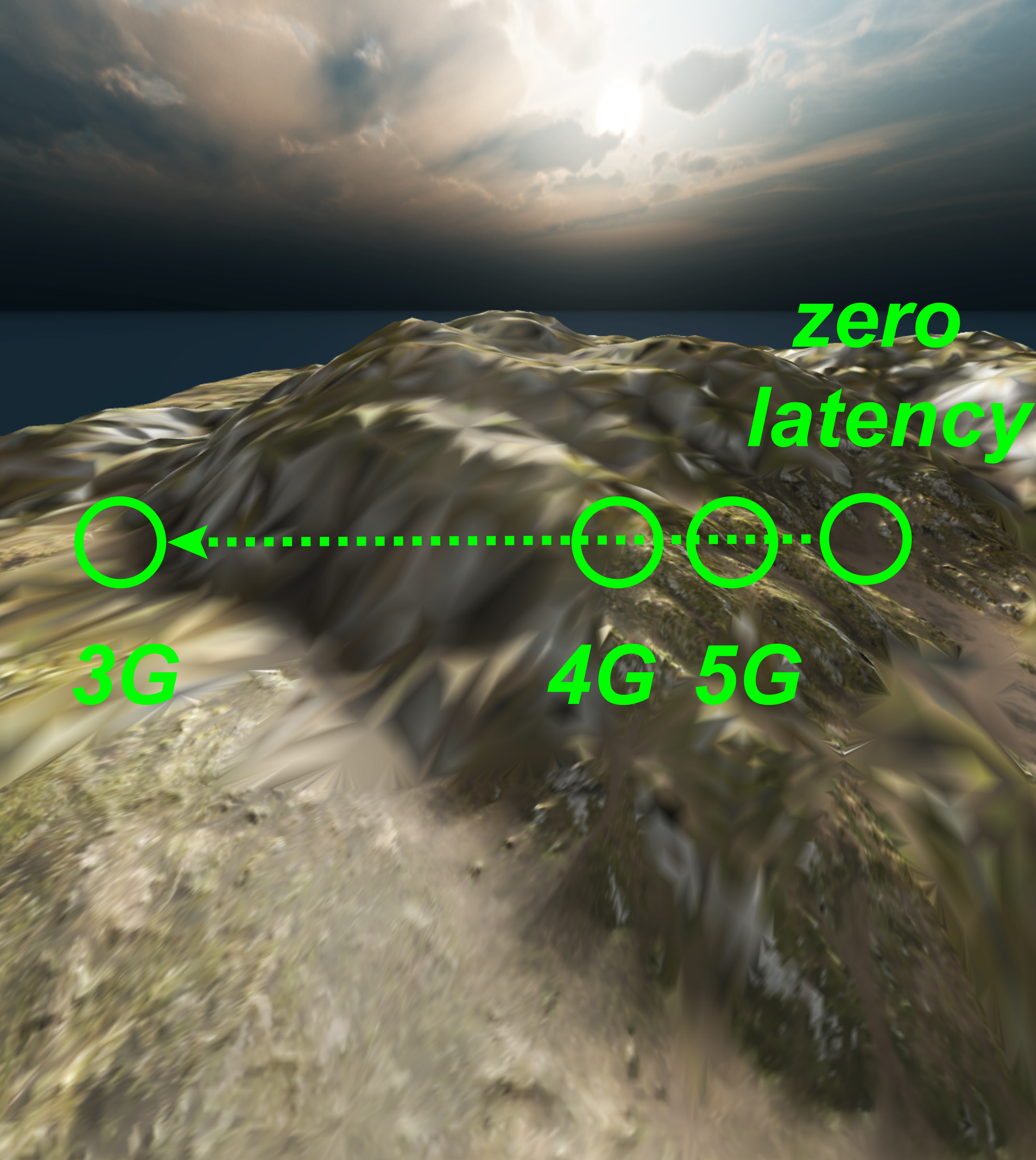}
    \label{fig:bandwidth:illustration}
  }
  \subfloat[quality gain w.r.t. bandwidth]{
    \includegraphics[height=0.24\linewidth]{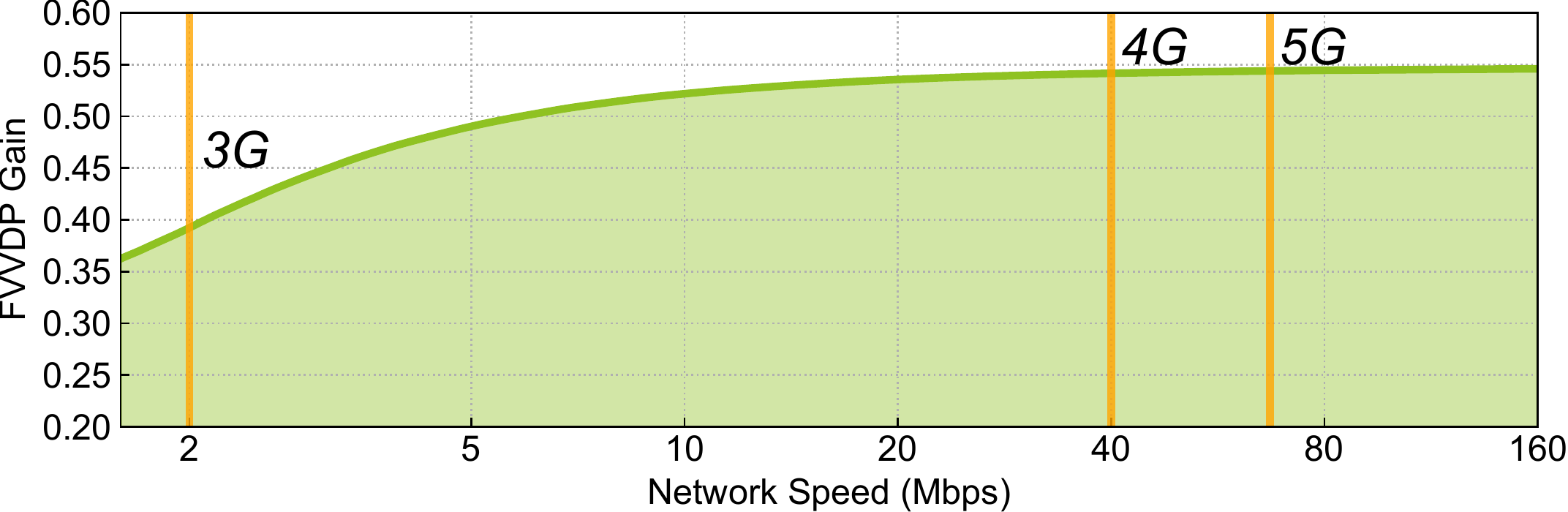}
    \label{fig:bandwidth:quality}
  }
	\Caption{Extendability test with varied network bandwidths.}
  {%
\subref{fig:bandwidth:illustration} shows a gaze trajectory from right to left, starting at the position circled by {\em zero latency}.
The 3G/4G/5G circles indicate the gaze positions for which the required data packets arrive.
\subref{fig:bandwidth:quality} shows the FoVVideoVDP gain of \oursCondition from \uniformCondition along with varied simulated network conditions.
The significant quality elevation in \oursCondition is only subtly affected by gaze transmission latency when the speed is lower than 10Mbps.
}
	\label{fig:bandwidth}
\end{figure}

\begin{figure}[h]
\centering
\includegraphics[clip,width=0.8\linewidth]{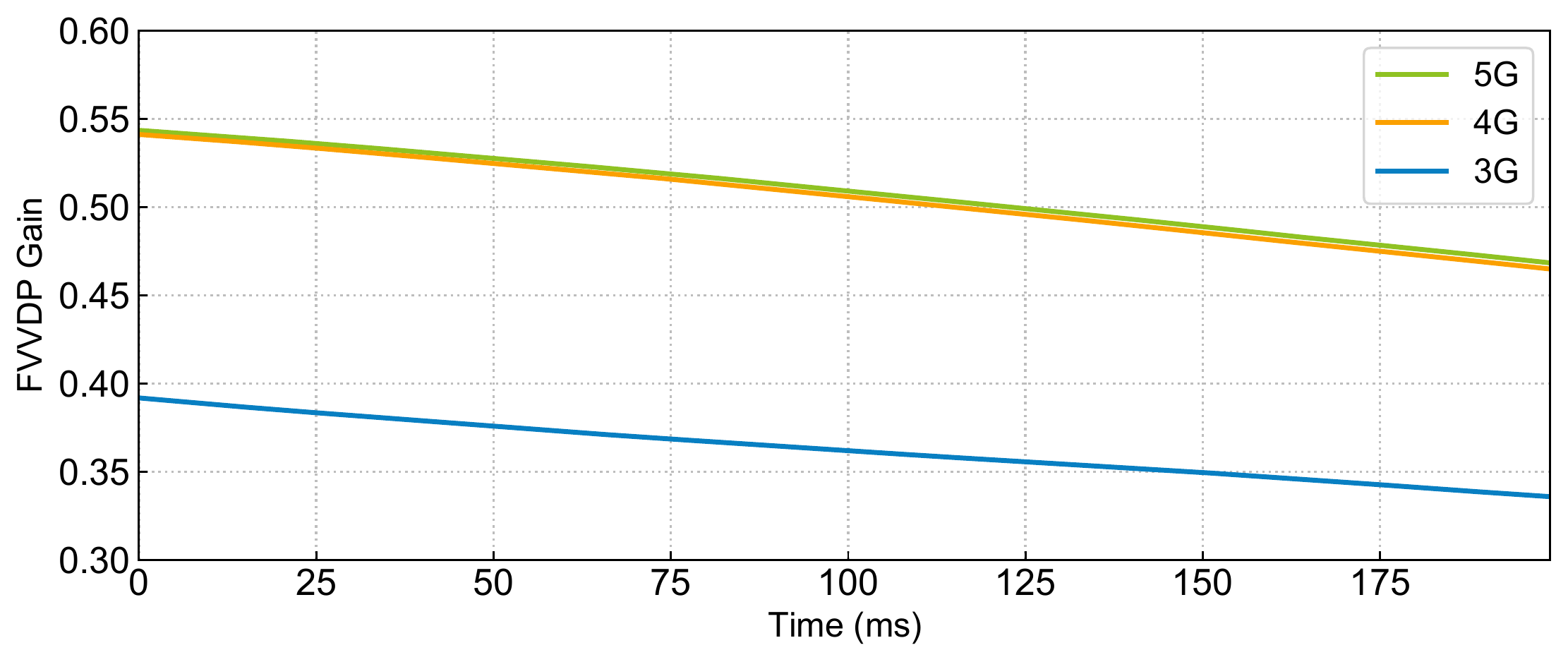}
\Caption{Pressure test with artificially introduced network latencies.}
{
  X-/Y- axis indicates the introduced latency/corresponding average FVVDP values.
  Across all network conditions, the low FVVDP changes demonstrate our method's robust benefits under real-world scenarios.
}
\label{fig:latency}
\end{figure}

\paragraph{Cloud/edge latency}
\label{sec:evaluation:analysis:latency}
Large-scale (e.g., 3D) data transmission inevitably introduces latency between the cloud and the edge due to the underlying network.
That is, the gaze analyzed in the cloud may be outdated when the edge receives the additive asset.
To validate the impact under real-world network conditions, we conduct a simulated experiment with varied transmission speeds ranging from 3G to 5G+. With a recorded gaze motion trajectory, we compute the FoVVideoVDP gain between \oursCondition and \uniformCondition.
By assuming 100KB packets, the approximated network speeds for 3G ($\sim$2Mbps) / 4G ($\sim$40Mbps) / 5G ($\sim$67Mbps) are referred from \cite{Riiser:2013:CPB,Raca:2020:BTG}.
As seen in \Cref{fig:bandwidth}, \oursCondition shows significantly elevated spatio-temporal quality (i.e., higher FVVDP) compared with \uniformCondition under all network conditions.
The elevation slightly decreases when network speed is slower than 10Mbps.
The analysis demonstrates our approach's consistent outperform than \uniformCondition under both modern (4G and 5G) and legacy (3G - 4G) network conditions.
Using the same data, we further perform \new{a} pressure experiment with artificially created latencies (from e.g., pings, database retrieval, etc.) with 3G/4G/5G network speeds.
\Cref{fig:latency} shows the result.
Our method is only subtly affected by latencies across all conditions.
Under 4G and 5G network, \oursCondition can reach 0.5 in JODs compared with \uniformCondition when the latency is lower than 100ms.
The 4G and 5G results are almost identical because their speeds are close.

\begin{figure}[tbh]
\includegraphics[width=0.96\linewidth]{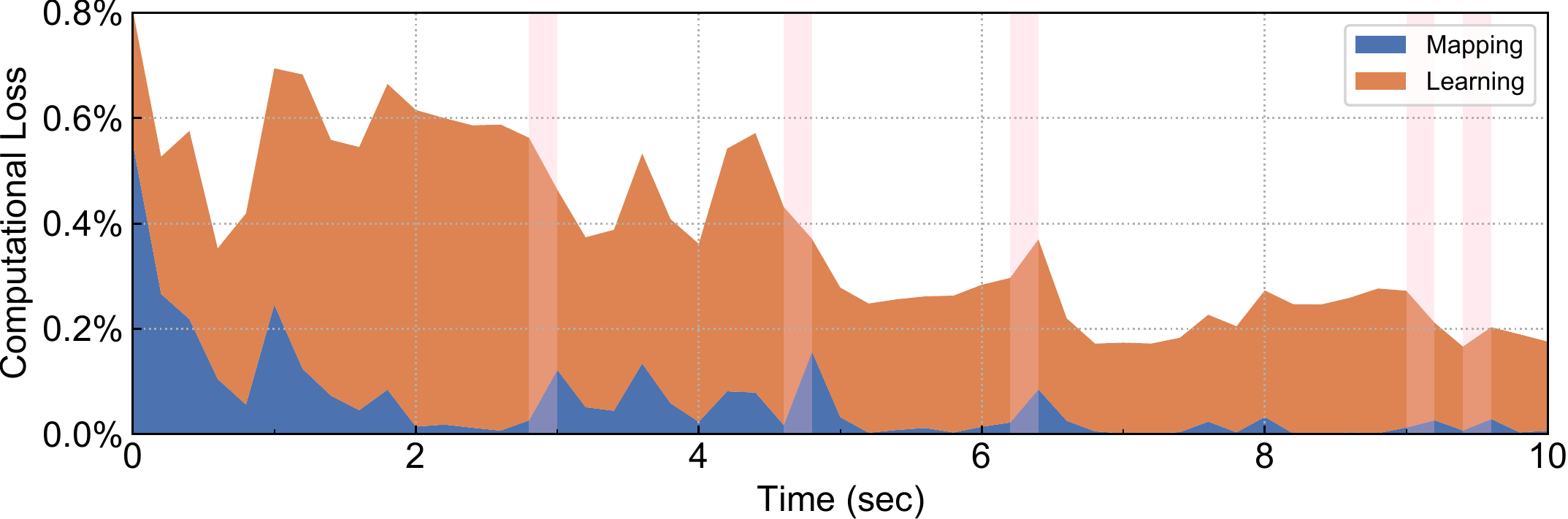}\label{fig:approximation_error}
\Caption{Approximation error visualization.}
{%
Areas highlighted by pink indicates that there is a saccade. Due to the 2D to 3D mapping (\Cref{sec:method:3d}, blue area) and neural acceleration (\Cref{sec:method:neural}, orange area), the final importance prediction may contain approximation errors. However, their overall error is below $1\%$.
}
\label{fig:error_evaluation}
\end{figure}

\paragraph{Approximation errors}
\label{sec:results:approximation}
Unlike 2D frame streaming, our system faces challenges of the complex 3D structures including connectivity and sparsity.
Although our method analytically transforms screen-space perceptual models to 3D (\Cref{sec:method:3d}) with neural acceleration (\Cref{sec:method:neural}), each of them may introduce numerical loss.
In this section, we also analyzed the potential error.
Specifically, we compute the MSE loss among the image-based per-pixel update, per-triangle update with/without neural acceleration.
\Cref{fig:error_evaluation} visualizes an example sequence.
The experiments showed that the loss is relatively minor and worth the performance gain enabled by the neural network.

\section{Conclusion}
\label{sec:conclusion}
We present a method for high-quality 3D immersive streaming with gaze-contingent perceptual optimization.
Compared with 2D frame-based streaming systems, our 3D streaming method enables low-latency interaction, low cloud overload, and consistent viewing.
Our evaluation demonstrates that our system delivers a statistically significant reduction of temporal artifacts without compromising the visual quality.
 
\paragraph{Limitations and Future Work}

The two main parts of our method, modeling human perception for visual importance in 2D, and mapping the 2D importance to 3D for rendering and streaming,
are largely orthogonal, as various \new{perceptual} factors can be considered for the 2D importance, before mapping it to 3D (\Cref{sec:method:3d}).
This paper focuses on combining foveation and saccade as the main perceptual mechanisms to optimize spatial quality and temporal smoothness (\Cref{eqn:adaptive}).
Other possibilities include applying other perceptual phenomena to compute 2D importance, such as color, saliency/attention \cite{Hillaire:2011:DAR,Watson:1986:WVP}, and masking \cite{Ferwerda:1997:MVM,Griffin:2015:ETC,Lavoue:2018:PET}, as well as applying our 2D importance estimation for streaming 2D and 360-degree videos \cite{Corbillon:2017:VAN}.
The modeling, implementation, and evaluation of all these possibilities are promising future works, beyond the scope of a single paper.

In our system and analysis, we presumed an ideal and stable network environment without bandwidth sharing. Moreover, the data transmission was considered as individual packages of 100KB without compression and decomposition between the cloud and the edge \cite{Levoy:1995:PJM,Cohen-Or:1999:DCS}.
As shown in \Cref{sec:evaluation:analysis:latency}, all these factors may introduce additional latencies of the timely gaze tracking data sharing between the cloud and the edge, causing approximation errors. \new{There are several super parameters in our method like the spatio-temporal balancing $\weightEccPop$ and the Weber’s law adjustment for low intensities $\omega$. 
One interesting future work is to adaptively optimize these super parameters for each scene.}

The neural network's prediction precision is shown to support local 6DoF motions, as shown in \Cref{sec:results:approximation}.
However, its capability of supporting very large motions such as in a flight simulation may require larger training data volume. Similarly, they are currently trained with static scenes. Predicting high accuracy results would also increase data samples due to introducing time as an additional input dimension.

To address the quality drop due to the gaze information latency, we foresee gaze motion prediction with machine learning may shed light on further reducing the perceptual latency in low bandwidth and unstable networks.
In our current method, the rendering updates immediately upon receiving the relevant assets.
\new{Our framework \textit{mitigates }the perceived flickering. For instance, during saccadic movement, we reduce the future (after the gaze lands) perceived flickering by taking advantage of the change blindness.
However, targeting at progressive streaming systems, the temporal artifacts are not to be completely eliminated. }
Instead, a locally adaptive rendering may increase temporal consistency at the cost of delayed full quality.
Recent research on balancing quality and latency \cite{Li:2020:TSP} could make the system adaptable to various users. 

One important application scenario that may require 3D assets (instead of 2D frames) is AR.
High resolution and high FoV AR displays are yet to be available.
Thus, we conducted the experiments with VR.
In the future, considering not only virtual assets but also its interaction with physical surroundings may inspire more efficient transmission.

\section*{Acknowledgments}
We would like to thank our colleagues from New York University for their help in this research.
This work was supported in part by: NSF awards CNS-1229185, CCF-1533564, CNS-1544753, CNS-1730396, CNS-1828576, CNS-1626098;
and a generous gift from Adobe, Inc.
C. T. Silva and Q. Sun are partially supported by the DARPA PTG program.
Any opinions, findings, and conclusions or recommendations expressed in this material are those of the authors and do not necessarily reflect the views of DARPA.

\bibliographystyle{abbrv-doi}
{
\bibliography{paper,misc}

\begin{thebibliography}{10}

\bibitem{Albert:2017:LRF}
R.~Albert, A.~Patney, D.~Luebke, and J.~Kim.
\newblock Latency requirements for foveated rendering in virtual reality.
\newblock {\em ACM Trans. Appl. Percept.}, 14(4), Sept. 2017. doi: {{%
10\hspace{.1pt}\discretionary{.}{%
}{.}\hspace{.4pt}1145\discretionary{/}{%
}{/}3127589}}


\bibitem{Arabadzhiyska:2017:SLP}
E.~Arabadzhiyska, O.~T. Tursun, K.~Myszkowski, H.-P. Seidel, and P.~Didyk.
\newblock Saccade landing position prediction for gaze-contingent rendering.
\newblock {\em ACM Trans. Graph.}, 36(4), July 2017. doi: {{%
10\hspace{.1pt}\discretionary{.}{%
}{.}\hspace{.4pt}1145\discretionary{/}{%
}{/}3072959\hspace{.1pt}\discretionary{.}{%
}{.}\hspace{.4pt}3073642}}


\bibitem{Barten:1990:IQM}
P.~G.~J. Barten.
\newblock Evaluation of subjective image quality with the square-root integral
  method.
\newblock {\em J. Opt. Soc. Am. A}, 7(10):2024--2031, Oct 1990. doi: {{%
10\hspace{.1pt}\discretionary{.}{%
}{.}\hspace{.4pt}1364\discretionary{/}{%
}{/}JOSAA\hspace{.1pt}\discretionary{.}{%
}{.}\hspace{.4pt}7\hspace{.1pt}\discretionary{.}{%
}{.}\hspace{.4pt}002024}}


\bibitem{Brown:2021:EDM}
R.~Brown, V.~DuTell, B.~Walter, R.~Rosenholtz, P.~Shirley, M.~McGuire, and
  D.~Luebke.
\newblock Efficient dataflow modeling of peripheral encoding in the human
  visual system, 2021.

\bibitem{Bruckert:2021:WLM}
A.~Bruckert, M.~Christie, and O.~L. Meur.
\newblock Where to look at the movies : Analyzing visual attention to
  understand movie editing, 2021.

\bibitem{Cohen:2020:LCA}
M.~A. Cohen, T.~L. Botch, and C.~E. Robertson.
\newblock The limits of color awareness during active, real-world vision.
\newblock {\em Proceedings of the National Academy of Sciences},
  117(24):13821--13827, 2020. doi: {{%
10\hspace{.1pt}\discretionary{.}{%
}{.}\hspace{.4pt}1073\discretionary{/}{%
}{/}pnas\hspace{.1pt}\discretionary{.}{%
}{.}\hspace{.4pt}1922294117}}


\bibitem{Cohen-Or:1999:DCS}
D.~Cohen-Or, Y.~Mann, and S.~Fleishman.
\newblock Deep compression for streaming texture intensive animations.
\newblock In {\em SIGGRAPH '99}, pp. 261--267, 1999. doi: {{%
10\hspace{.1pt}\discretionary{.}{%
}{.}\hspace{.4pt}1145\discretionary{/}{%
}{/}311535\hspace{.1pt}\discretionary{.}{%
}{.}\hspace{.4pt}311564}}


\bibitem{Corbillon:2017:VAN}
X.~{Corbillon}, G.~{Simon}, A.~{Devlic}, and J.~{Chakareski}.
\newblock Viewport-adaptive navigable 360-degree video delivery.
\newblock In {\em 2017 IEEE International Conference on Communications (ICC)},
  pp. 1--7, 2017. doi: {{%
10\hspace{.1pt}\discretionary{.}{%
}{.}\hspace{.4pt}1109\discretionary{/}{%
}{/}ICC\hspace{.1pt}\discretionary{.}{%
}{.}\hspace{.4pt}2017\hspace{.1pt}\discretionary{.}{%
}{.}\hspace{.4pt}7996611}}


\bibitem{Elliott:1995:VAC}
D.~B. Elliott, K.~C. Yang, and D.~Whitaker.
\newblock Visual acuity changes throughout adulthood in normal, healthy eyes:
  seeing beyond 6/6.
\newblock {\em Optometry and vision science : official publication of the
  American Academy of Optometry}, 72 3:186--91, 1995.

\bibitem{Ferwerda:1997:MVM}
J.~A. Ferwerda, P.~Shirley, S.~N. Pattanaik, and D.~P. Greenberg.
\newblock A model of visual masking for computer graphics.
\newblock In {\em SIGGRAPH '97}, pp. 143--152, 1997.

\bibitem{Griffin:2015:ETC}
W.~Griffin and M.~Olano.
\newblock Evaluating texture compression masking effects using objective image
  quality assessment metrics.
\newblock {\em IEEE Transactions on Visualization and Computer Graphics},
  21(8):970--979, 2015.

\bibitem{Guenter:2012:F3G}
B.~Guenter, M.~Finch, S.~Drucker, D.~Tan, and J.~Snyder.
\newblock Foveated 3d graphics.
\newblock {\em ACM Trans. Graph.}, 31(6), Nov. 2012. doi: {{%
10\hspace{.1pt}\discretionary{.}{%
}{.}\hspace{.4pt}1145\discretionary{/}{%
}{/}2366145\hspace{.1pt}\discretionary{.}{%
}{.}\hspace{.4pt}2366183}}


\bibitem{Gutterman:2020:MMSys}
C.~Gutterman, B.~Fridman, T.~Gilliland, Y.~Hu, and G.~Zussman.
\newblock Stallion: Video adaptation algorithm for low-latency video streaming.
\newblock In {\em Proceedings of the 11th ACM Multimedia Systems Conference},
  MMSys '20, p. 327–332. Association for Computing Machinery, New York, NY,
  USA, 2020. doi: {{%
10\hspace{.1pt}\discretionary{.}{%
}{.}\hspace{.4pt}1145\discretionary{/}{%
}{/}3339825\hspace{.1pt}\discretionary{.}{%
}{.}\hspace{.4pt}3397044}}


\bibitem{Hillaire:2011:DAR}
S.~{Hillaire}, A.~{Lecuyer}, T.~{Regia-Corte}, R.~{Cozot}, J.~{Royan}, and
  G.~{Breton}.
\newblock Design and application of real-time visual attention model for the
  exploration of 3d virtual environments.
\newblock {\em IEEE Transactions on Visualization and Computer Graphics},
  18(3):356--368, 2012. doi: {{%
10\hspace{.1pt}\discretionary{.}{%
}{.}\hspace{.4pt}1109\discretionary{/}{%
}{/}TVCG\hspace{.1pt}\discretionary{.}{%
}{.}\hspace{.4pt}2011\hspace{.1pt}\discretionary{.}{%
}{.}\hspace{.4pt}154}}


\bibitem{Hladky:2019:COS}
J.~Hladky, H.-P. Seidel, and M.~Steinberger.
\newblock The camera offset space: Real-time potentially visible set
  computations for streaming rendering.
\newblock {\em ACM Trans. Graph.}, 38(6), Nov. 2019. doi: {{%
10\hspace{.1pt}\discretionary{.}{%
}{.}\hspace{.4pt}1145\discretionary{/}{%
}{/}3355089\hspace{.1pt}\discretionary{.}{%
}{.}\hspace{.4pt}3356530}}


\bibitem{Simon:2020:CPS}
H.~Hristova, G.~Simon, V.~Swaminathan, and S.~Petrangeli.
\newblock 3cps: A novel supercompression for the delivery of 3d object
  textures, 2020. doi: {{%
10\hspace{.1pt}\discretionary{.}{%
}{.}\hspace{.4pt}1145\discretionary{/}{%
}{/}3339825\hspace{.1pt}\discretionary{.}{%
}{.}\hspace{.4pt}3391860}}


\bibitem{Hu:2019:RSS}
P.~Hu, Q.~Sun, P.~Didyk, L.-Y. Wei, and A.~E. Kaufman.
\newblock Reducing simulator sickness with perceptual camera control.
\newblock {\em ACM Trans. Graph.}, 38(6), Nov. 2019. doi: {{%
10\hspace{.1pt}\discretionary{.}{%
}{.}\hspace{.4pt}1145\discretionary{/}{%
}{/}3355089\hspace{.1pt}\discretionary{.}{%
}{.}\hspace{.4pt}3356490}}


\bibitem{Ibbotson:2011:VPS}
M.~Ibbotson and B.~Krekelberg.
\newblock Visual perception and saccadic eye movements.
\newblock {\em Current Opinion in Neurobiology}, 21(4):553--558, 2011.
\newblock Sensory and motor systems. doi: {{%
10\hspace{.1pt}\discretionary{.}{%
}{.}\hspace{.4pt}1016\discretionary{/}{%
}{/}j\hspace{.1pt}\discretionary{.}{%
}{.}\hspace{.4pt}conb\hspace{.1pt}\discretionary{.}{%
}{.}\hspace{.4pt}2011\hspace{.1pt}\discretionary{.}{%
}{.}\hspace{.4pt}05\hspace{.1pt}\discretionary{.}{%
}{.}\hspace{.4pt}012}}


\bibitem{Kaplanyan:2019:DNR}
A.~S. Kaplanyan, A.~Sochenov, T.~Leimk\"{u}hler, M.~Okunev, T.~Goodall, and
  G.~Rufo.
\newblock Deepfovea: Neural reconstruction for foveated rendering and video
  compression using learned statistics of natural videos.
\newblock {\em ACM Trans. Graph.}, 38(6), Nov. 2019. doi: {{%
10\hspace{.1pt}\discretionary{.}{%
}{.}\hspace{.4pt}1145\discretionary{/}{%
}{/}3355089\hspace{.1pt}\discretionary{.}{%
}{.}\hspace{.4pt}3356557}}


\bibitem{Knoell:2011:SPP}
J.~Knoell, P.~Binda, M.~C. Morrone, and F.~Bremmer.
\newblock Spatiotemporal profile of peri-saccadic contrast sensitivity.
\newblock {\em Journal of vision}, 11(14):15--15, 2011.

\bibitem{Konrad:2019:GCO}
R.~Konrad, A.~Angelopoulos, and G.~Wetzstein.
\newblock Gaze-contingent ocular parallax rendering for virtual reality.
\newblock {\em ACM Trans. Graph.}, 39, 2020.

\bibitem{Koulieris:2019:NED}
G.~A. Koulieris, K.~Akşit, M.~Stengel, R.~K. Mantiuk, K.~Mania, and
  C.~Richardt.
\newblock Near-eye display and tracking technologies for virtual and augmented
  reality.
\newblock {\em Computer Graphics Forum}, 38(2):493--519, 2019. doi: {{%
10\hspace{.1pt}\discretionary{.}{%
}{.}\hspace{.4pt}1111\discretionary{/}{%
}{/}cgf\hspace{.1pt}\discretionary{.}{%
}{.}\hspace{.4pt}13654}}


\bibitem{Krajancich:2020:ODP}
B.~Krajancich, P.~Kellnhofer, and G.~Wetzstein.
\newblock Optimizing depth perception in virtual and augmented reality through
  gaze-contingent stereo rendering.
\newblock {\em ACM Trans. Graph.}, 39, 2020.

\bibitem{Lavoue:2018:PET}
G.~Lavou{\'e}, M.~Langer, A.~Peytavie, and P.~Poulin.
\newblock A psychophysical evaluation of texture compression masking effects.
\newblock {\em IEEE transactions on visualization and computer graphics},
  25(2):1336--1346, 2018.

\bibitem{Levoy:1995:PJM}
M.~Levoy.
\newblock Polygon-assisted jpeg and mpeg compression of synthetic images.
\newblock In {\em SIGGRAPH ’95}, pp. 21--28, 1995. doi: {{%
10\hspace{.1pt}\discretionary{.}{%
}{.}\hspace{.4pt}1145\discretionary{/}{%
}{/}218380\hspace{.1pt}\discretionary{.}{%
}{.}\hspace{.4pt}218392}}


\bibitem{Li:2020:TSP}
M.~Li, Y.~Wang, and D.~Ramanan.
\newblock Towards streaming perception.
\newblock In {\em ECCV}, 2020.

\bibitem{Luebke:2002:LD3}
D.~Luebke, M.~Reddy, J.~D. Cohen, A.~Varshney, B.~Watson, and R.~Huebner.
\newblock {\em Level of Detail for 3D Graphics}.
\newblock Morgan Kaufmann Publishers Inc., San Francisco, CA, USA, 2002.

\bibitem{Luidolt:2020:GDS}
L.~R. Luidolt, M.~Wimmer, and K.~Kr{\"o}sl.
\newblock Gaze-dependent simulation of light perception in virtual reality.
\newblock {\em IEEE Transactions on Visualization and Computer Graphics},
  26(12):3557--3567, 2020. doi: {{%
10\hspace{.1pt}\discretionary{.}{%
}{.}\hspace{.4pt}1109\discretionary{/}{%
}{/}TVCG\hspace{.1pt}\discretionary{.}{%
}{.}\hspace{.4pt}2020\hspace{.1pt}\discretionary{.}{%
}{.}\hspace{.4pt}3023604}}


\bibitem{Mantiuk:2021:FVD}
R.~K. Mantiuk, G.~Denes, A.~Chapiro, A.~Kaplanyan, G.~Rufo, R.~Bachy, T.~Lian,
  and A.~Patney.
\newblock Fovvideovdp: A visible difference predictor for wide field-of-view
  video.
\newblock {\em ACM Trans. Graph.}, 40(4), July 2021. doi: {{%
10\hspace{.1pt}\discretionary{.}{%
}{.}\hspace{.4pt}1145\discretionary{/}{%
}{/}3450626\hspace{.1pt}\discretionary{.}{%
}{.}\hspace{.4pt}3459831}}


\bibitem{Murphy:2001:GCL}
H.~Murphy and A.~Duchowski.
\newblock Gaze-contingent level of detail rendering.
\newblock {\em EuroGraphics}, 2001, 01 2001.

\bibitem{Patney:2016:TFR}
A.~Patney, M.~Salvi, J.~Kim, A.~Kaplanyan, C.~Wyman, N.~Benty, D.~Luebke, and
  A.~Lefohn.
\newblock Towards foveated rendering for gaze-tracked virtual reality.
\newblock {\em ACM Trans. Graph.}, 35(6), Nov. 2016. doi: {{%
10\hspace{.1pt}\discretionary{.}{%
}{.}\hspace{.4pt}1145\discretionary{/}{%
}{/}2980179\hspace{.1pt}\discretionary{.}{%
}{.}\hspace{.4pt}2980246}}


\bibitem{Peli:1990:CCI}
E.~Peli.
\newblock Contrast in complex images.
\newblock {\em J. Opt. Soc. Am. A}, 7(10):2032--2040, Oct 1990. doi: {{%
10\hspace{.1pt}\discretionary{.}{%
}{.}\hspace{.4pt}1364\discretionary{/}{%
}{/}JOSAA\hspace{.1pt}\discretionary{.}{%
}{.}\hspace{.4pt}7\hspace{.1pt}\discretionary{.}{%
}{.}\hspace{.4pt}002032}}


\bibitem{Petrangeli:2019:DAS}
S.~Petrangeli, G.~Simon, H.~Wang, and V.~Swaminathan.
\newblock Dynamic adaptive streaming for augmented reality applications.
\newblock In {\em 2019 IEEE International Symposium on Multimedia (ISM)}, pp.
  56--567, 2019.

\bibitem{Parraga:2005:EAS}
C.~Párraga, T.~Troscianko, and D.~Tolhurst.
\newblock The effects of amplitude-spectrum statistics on foveal and peripheral
  discrimination of changes in natural images, and a multi-resolution model.
\newblock {\em Vision Research}, 45(25):3145 -- 3168, 2005. doi: {{%
10\hspace{.1pt}\discretionary{.}{%
}{.}\hspace{.4pt}1016\discretionary{/}{%
}{/}j\hspace{.1pt}\discretionary{.}{%
}{.}\hspace{.4pt}visres\hspace{.1pt}\discretionary{.}{%
}{.}\hspace{.4pt}2005\hspace{.1pt}\discretionary{.}{%
}{.}\hspace{.4pt}08\hspace{.1pt}\discretionary{.}{%
}{.}\hspace{.4pt}006}}


\bibitem{Raca:2020:BTG}
D.~Raca, D.~Leahy, C.~J. Sreenan, and J.~J. Quinlan.
\newblock Beyond throughput, the next generation: A 5g dataset with channel and
  context metrics.
\newblock In {\em MMSys '20}, pp. 303--308, 2020. doi: {{%
10\hspace{.1pt}\discretionary{.}{%
}{.}\hspace{.4pt}1145\discretionary{/}{%
}{/}3339825\hspace{.1pt}\discretionary{.}{%
}{.}\hspace{.4pt}3394938}}


\bibitem{Rashidi:2020:OVS}
S.~Rashidi, K.~Ehinger, A.~Turpin, and L.~Kulik.
\newblock Optimal visual search based on a model of target detectability in
  natural images.
\newblock In H.~Larochelle, M.~Ranzato, R.~Hadsell, M.~F. Balcan, and H.~Lin,
  eds., {\em Advances in Neural Information Processing Systems}, vol.~33, pp.
  9288--9299. Curran Associates, Inc., 2020.

\bibitem{Riiser:2013:CPB}
H.~Riiser, P.~Vigmostad, C.~Griwodz, and P.~Halvorsen.
\newblock Commute path bandwidth traces from 3g networks: Analysis and
  applications.
\newblock In {\em Proceedings of the 4th ACM Multimedia Systems Conference},
  MMSys '13, pp. 114--118. Association for Computing Machinery, New York, NY,
  USA, 2013. doi: {{%
10\hspace{.1pt}\discretionary{.}{%
}{.}\hspace{.4pt}1145\discretionary{/}{%
}{/}2483977\hspace{.1pt}\discretionary{.}{%
}{.}\hspace{.4pt}2483991}}


\bibitem{Romero:2018:FSV}
M.~F. Romero-Rond\'{o}n, L.~Sassatelli, F.~Precioso, and R.~Aparicio-Pardo.
\newblock Foveated streaming of virtual reality videos.
\newblock In {\em Proceedings of the 9th ACM Multimedia Systems Conference},
  MMSys '18, pp. 494--497, 2018. doi: {{%
10\hspace{.1pt}\discretionary{.}{%
}{.}\hspace{.4pt}1145\discretionary{/}{%
}{/}3204949\hspace{.1pt}\discretionary{.}{%
}{.}\hspace{.4pt}3208114}}


\bibitem{Rusinkiewicz:2000:QMP}
S.~Rusinkiewicz and M.~Levoy.
\newblock Qsplat: A multiresolution point rendering system for large meshes.
\newblock In {\em SIGGRAPH '00}, pp. 343--352, 2000. doi: {{%
10\hspace{.1pt}\discretionary{.}{%
}{.}\hspace{.4pt}1145\discretionary{/}{%
}{/}344779\hspace{.1pt}\discretionary{.}{%
}{.}\hspace{.4pt}344940}}


\bibitem{Schutz:2020:PRR}
M.~Sch\"{u}tz, G.~Mandlburger, J.~Otepka, and M.~Wimmer.
\newblock Progressive real-time rendering of one billion points without
  hierarchical acceleration structures.
\newblock {\em Computer Graphics Forum}, 39(2):51--64, 2020. doi: {{%
10\hspace{.1pt}\discretionary{.}{%
}{.}\hspace{.4pt}1111\discretionary{/}{%
}{/}cgf\hspace{.1pt}\discretionary{.}{%
}{.}\hspace{.4pt}13911}}


\bibitem{Schwarz:2009:PVP}
M.~Schwarz and M.~Stamminger.
\newblock On predicting visual popping in dynamic scenes.
\newblock In {\em APGV '09}, pp. 93--100, 2009. doi: {{%
10\hspace{.1pt}\discretionary{.}{%
}{.}\hspace{.4pt}1145\discretionary{/}{%
}{/}1620993\hspace{.1pt}\discretionary{.}{%
}{.}\hspace{.4pt}1621012}}


\bibitem{Serrano:2019:MPR}
A.~Serrano, I.~Kim, Z.~Chen, S.~DiVerdi, D.~Gutierrez, A.~Hertzmann, and
  B.~Masia.
\newblock Motion parallax for 360$^{\circ}$ rgbd video.
\newblock {\em IEEE Transactions on Visualization and Computer Graphics},
  25(5):1817--1827, May 2019. doi: {{%
10\hspace{.1pt}\discretionary{.}{%
}{.}\hspace{.4pt}1109\discretionary{/}{%
}{/}TVCG\hspace{.1pt}\discretionary{.}{%
}{.}\hspace{.4pt}2019\hspace{.1pt}\discretionary{.}{%
}{.}\hspace{.4pt}2898757}}


\bibitem{Serrano:2017:MEC}
A.~Serrano, V.~Sitzmann, J.~Ruiz-Borau, G.~Wetzstein, D.~Gutierrez, and
  B.~Masia.
\newblock Movie editing and cognitive event segmentation in virtual reality
  video.
\newblock {\em ACM Trans. Graph.}, 36(4), July 2017. doi: {{%
10\hspace{.1pt}\discretionary{.}{%
}{.}\hspace{.4pt}1145\discretionary{/}{%
}{/}3072959\hspace{.1pt}\discretionary{.}{%
}{.}\hspace{.4pt}3073668}}


\bibitem{Serrano:2009:LBM}
C.~Serrano, B.~Garriga, J.~Velasco, J.~Urbano, S.~Tenorio, and M.~Sierra.
\newblock Latency in broad-band mobile networks.
\newblock In {\em VTC Spring 2009 - IEEE 69th Vehicular Technology Conference},
  pp. 1--7, 2009. doi: {{%
10\hspace{.1pt}\discretionary{.}{%
}{.}\hspace{.4pt}1109\discretionary{/}{%
}{/}VETECS\hspace{.1pt}\discretionary{.}{%
}{.}\hspace{.4pt}2009\hspace{.1pt}\discretionary{.}{%
}{.}\hspace{.4pt}5073642}}


\bibitem{Shu:2019:FFR}
S.~Shi, V.~Gupta, and R.~Jana.
\newblock Freedom: Fast recovery enhanced vr delivery over mobile networks.
\newblock In {\em Proceedings of the 17th Annual International Conference on
  Mobile Systems, Applications, and Services}, MobiSys '19, p. 130–141.
  Association for Computing Machinery, New York, NY, USA, 2019. doi: {{%
10\hspace{.1pt}\discretionary{.}{%
}{.}\hspace{.4pt}1145\discretionary{/}{%
}{/}3307334\hspace{.1pt}\discretionary{.}{%
}{.}\hspace{.4pt}3326087}}


\bibitem{Shuai:2015:Icc}
Y.~Shuai, G.~Petrovic, and T.~Herfet.
\newblock Olac: An open-loop controller for low-latency adaptive video
  streaming.
\newblock In {\em 2015 IEEE International Conference on Communications (ICC)},
  pp. 6874--6879, 2015. doi: {{%
10\hspace{.1pt}\discretionary{.}{%
}{.}\hspace{.4pt}1109\discretionary{/}{%
}{/}ICC\hspace{.1pt}\discretionary{.}{%
}{.}\hspace{.4pt}2015\hspace{.1pt}\discretionary{.}{%
}{.}\hspace{.4pt}7249421}}


\bibitem{Simon:2019:SST}
G.~Simon, S.~Petrangeli, N.~Carr, and V.~Swaminathan.
\newblock Streaming a sequence of textures for adaptive 3d scene delivery.
\newblock In {\em 2019 IEEE Conference on Virtual Reality and 3D User
  Interfaces (VR)}, pp. 1159--1160, 2019.

\bibitem{Sitzmann:2019:SVH}
V.~Sitzmann, A.~Serrano, A.~Pavel, M.~Agrawala, D.~Gutierrez, B.~Masia, and
  G.~Wetzstein.
\newblock Saliency in vr: How do people explore virtual environments?
\newblock {\em IEEE transactions on visualization and computer graphics},
  24(4):1633--1642, April 2018. doi: {{%
10\hspace{.1pt}\discretionary{.}{%
}{.}\hspace{.4pt}1109\discretionary{/}{%
}{/}TVCG\hspace{.1pt}\discretionary{.}{%
}{.}\hspace{.4pt}2018\hspace{.1pt}\discretionary{.}{%
}{.}\hspace{.4pt}2793599}}


\bibitem{Stengel:2021:DDS}
M.~Stengel, Z.~Majercik, B.~Boudaoud, and M.~McGuire.
\newblock A distributed, decoupled system for losslessly streaming dynamic
  light probes to thin clients.
\newblock In {\em Proceedings of the 12th ACM Multimedia Systems Conference},
  MMSys '21, p. 159–172. Association for Computing Machinery, New York, NY,
  USA, 2021. doi: {{%
10\hspace{.1pt}\discretionary{.}{%
}{.}\hspace{.4pt}1145\discretionary{/}{%
}{/}3458305\hspace{.1pt}\discretionary{.}{%
}{.}\hspace{.4pt}3463379}}


\bibitem{Sun:2017:PGF}
Q.~Sun, F.-C. Huang, J.~Kim, L.-Y. Wei, D.~Luebke, and A.~Kaufman.
\newblock Perceptually-guided foveation for light field displays.
\newblock {\em ACM Trans. Graph.}, 36(6), Nov. 2017. doi: {{%
10\hspace{.1pt}\discretionary{.}{%
}{.}\hspace{.4pt}1145\discretionary{/}{%
}{/}3130800\hspace{.1pt}\discretionary{.}{%
}{.}\hspace{.4pt}3130807}}


\bibitem{Sun:2018:TVR}
Q.~Sun, A.~Patney, L.-Y. Wei, O.~Shapira, J.~Lu, P.~Asente, S.~Zhu, M.~Mcguire,
  D.~Luebke, and A.~Kaufman.
\newblock Towards virtual reality infinite walking: Dynamic saccadic
  redirection.
\newblock {\em ACM Trans. Graph.}, 37(4), July 2018. doi: {{%
10\hspace{.1pt}\discretionary{.}{%
}{.}\hspace{.4pt}1145\discretionary{/}{%
}{/}3197517\hspace{.1pt}\discretionary{.}{%
}{.}\hspace{.4pt}3201294}}


\bibitem{Tursun:2019:LCA}
O.~T. Tursun, E.~Arabadzhiyska-Koleva, M.~Wernikowski, R.~Mantiuk, H.-P.
  Seidel, K.~Myszkowski, and P.~Didyk.
\newblock Luminance-contrast-aware foveated rendering.
\newblock {\em ACM Trans. Graph.}, 38(4), July 2019. doi: {{%
10\hspace{.1pt}\discretionary{.}{%
}{.}\hspace{.4pt}1145\discretionary{/}{%
}{/}3306346\hspace{.1pt}\discretionary{.}{%
}{.}\hspace{.4pt}3322985}}


\bibitem{Walton:2021:BBR}
D.~R. Walton, R.~K.~D. Anjos, S.~Friston, D.~Swapp, K.~Ak\c{s}it, A.~Steed, and
  T.~Ritschel.
\newblock Beyond blur: Real-time ventral metamers for foveated rendering.
\newblock {\em ACM Trans. Graph.}, 40(4), jul 2021. doi: {{%
10\hspace{.1pt}\discretionary{.}{%
}{.}\hspace{.4pt}1145\discretionary{/}{%
}{/}3450626\hspace{.1pt}\discretionary{.}{%
}{.}\hspace{.4pt}3459943}}


\bibitem{Wang:2020:FIR}
L.~Wang, R.~Li, X.~Shi, L.-Q. Yan, and Z.~Li.
\newblock Foveated instant radiosity.
\newblock In {\em 2020 IEEE International Symposium on Mixed and Augmented
  Reality (ISMAR)}, pp. 1--11, 2020. doi: {{%
10\hspace{.1pt}\discretionary{.}{%
}{.}\hspace{.4pt}1109\discretionary{/}{%
}{/}ISMAR50242\hspace{.1pt}\discretionary{.}{%
}{.}\hspace{.4pt}2020\hspace{.1pt}\discretionary{.}{%
}{.}\hspace{.4pt}00017}}


\bibitem{Wang:1997:AFE}
Y.-Z. Wang, A.~Bradley, and L.~N. Thibos.
\newblock Aliased frequencies enable the discrimination of compound gratings in
  peripheral vision.
\newblock {\em Vision Research}, 37(3):283--290, 1997. doi: {{%
10\hspace{.1pt}\discretionary{.}{%
}{.}\hspace{.4pt}1016\discretionary{/}{%
}{/}S0042\discretionary{%
}{-}{-}6989\discretionary{%
}{(}{(}96\discretionary{)}{%
}{)}00160\discretionary{%
}{-}{-}5}}


\bibitem{Watson:2014:AFH}
A.~B. Watson.
\newblock {A formula for human retinal ganglion cell receptive field density as
  a function of visual field location}.
\newblock {\em Journal of Vision}, 14(7):15--15, 06 2014. doi: {{%
10\hspace{.1pt}\discretionary{.}{%
}{.}\hspace{.4pt}1167\discretionary{/}{%
}{/}14\hspace{.1pt}\discretionary{.}{%
}{.}\hspace{.4pt}7\hspace{.1pt}\discretionary{.}{%
}{.}\hspace{.4pt}15}}


\bibitem{Watson:1986:WVP}
A.~B. Watson, A.~J. Ahumada, and J.~E. Farrell.
\newblock Window of visibility: a psychophysical theory of fidelity in
  time-sampled visual motion displays.
\newblock {\em J. Opt. Soc. Am. A}, 3(3):300--307, Mar 1986. doi: {{%
10\hspace{.1pt}\discretionary{.}{%
}{.}\hspace{.4pt}1364\discretionary{/}{%
}{/}JOSAA\hspace{.1pt}\discretionary{.}{%
}{.}\hspace{.4pt}3\hspace{.1pt}\discretionary{.}{%
}{.}\hspace{.4pt}000300}}


\bibitem{Weier:2017:PAR}
M.~Weier, M.~Stengel, T.~Roth, P.~Didyk, E.~Eisemann, M.~Eisemann,
  S.~Grogorick, A.~Hinkenjann, E.~Kruijff, M.~Magnor, et~al.
\newblock Perception-driven accelerated rendering.
\newblock In {\em Computer Graphics Forum}, vol.~36, pp. 611--643, 2017.

\bibitem{Zhang:2021:DVP}
W.~Zhang, F.~Qian, B.~Han, and P.~Hui.
\newblock Deepvista: 16k panoramic cinema on your mobile device.
\newblock In {\em Proceedings of the Web Conference 2021}, WWW '21, p.
  2232–2244. Association for Computing Machinery, New York, NY, USA, 2021.
  doi: {{%
10\hspace{.1pt}\discretionary{.}{%
}{.}\hspace{.4pt}1145\discretionary{/}{%
}{/}3442381\hspace{.1pt}\discretionary{.}{%
}{.}\hspace{.4pt}3449829}}


\end{thebibliography}
}

\appendix
\normalsize

\section{Spatial Visual Acuity}
\label{sec:sup:foveation}
The human vision is foveated. 
Watson \cite{Watson:2014:AFH} proposed a formula that approximates midget ganglion cells density as a function of retinal eccentricity $r = \sqrt{x^2+y^2}$, where $x$ and $y$ are the eccentricity degree in horizontal and vertical direction respectively, for the meridian type $m$:
\begin{align}
	\begin{split}
		\rho (r,m) = &2
		\rho_{cone} \left(1+\frac{r}{41.03} \right)^{-1} \times \\
		&\left[ a_m \left( 1+\frac{r}{r_{2,m}} \right)^{-2} + (1-a_m)\exp\left(-\frac{r}{r_{e,m}}\right) \right],
		\label{eqn:eccentricity}
	\end{split}
\end{align}
where $\rho_{cone}=14804.6deg^{-2}$ is the density of cone cell at fovea and $a_m,r_{2,m},r_{e,m}$ are all fitting constants along the four meridians of the visual field.
And the cell-wise spacing is calculated by \cite{Sun:2017:PGF}:
\begin{align}
	\sigma(x,y) = \frac{1}{r}
	\sqrt{
		\frac{2}{\sqrt3} 
		\left(
		\frac{x^2}{\rho(r,1)} + \frac{y^2}{\rho(r,2)}
		\right)
	}
	\label{eq:retinal_spacing}
\end{align}
Then, given the eccentricity, the importance function $\eccentricityMask$ of a given eccentricity $\imageSpaceVec=(x,y)$ is modeled as \cite{Sun:2017:PGF}:
\begin{equation}
	\label{eqn:eccentricity_mask}
	\eccentricityMask(\imageSpaceVec) = 0.5\sigma(\imageSpaceVec)^{-1}.
\end{equation}

\section{Perceiving Static Stimuli}
\label{sec:sup:static_vision_model}
The visual sensitivity $\csf$ of a certain frequency $\contentFreq$ and illumination $\generalLuminance$ is determined by \cite{Barten:1990:IQM}:
\begin{align}
	\csf(\contentFreq,\generalLuminance) &= a\contentFreq e^{-b\contentFreq} \sqrt{1+ce^{b\contentFreq}},
	\label{eq:static_sensitivity}
\end{align}
where
$a = \frac{540(1+0.7/\generalLuminance)^{-0.2}}{1+\frac{1}{1+\contentFreq/3}}$,
$b = 0.3(1+100/\generalLuminance)^{0.15}$
and $c = 0.06$.

Because illumination of commodity displays is largely insensitive to the displaying content,
we assume a constant illumination $\generalLuminance$ for all frequencies.
Given an image $\image$,
its visual sensitivity is the integral of all frequencies weighted by their amplitude,
\begin{align}
	\int \csf\left(\lvert \contentFreqVec \rvert,\luminance\right)
	\lvert \contentAmp\left(\contentFreqVec \right) \rvert \mathrm{d}\contentFreqVec
	\label{eq:integral}
\end{align}
where $\contentFreqVec=(\contentFreq_\imageSpaceX,\contentFreq_\imageSpaceY)$ is the two-dimensional frequency of $\image$,
and $\contentAmp$ is the amplitude of $\contentFreqVec$.

\begin{figure}[htb]
\centering
	\includegraphics[width=0.6\linewidth]{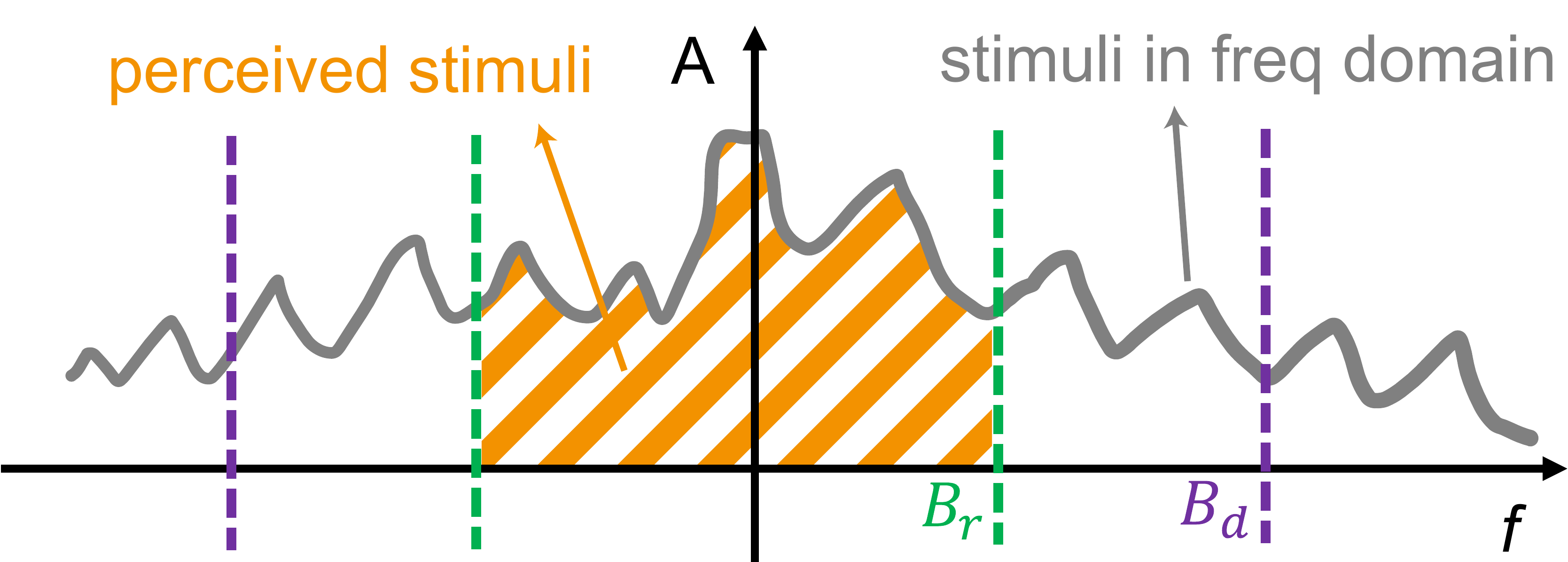}
	\Caption{Perceived static visual stimuli.}
        {%
        The perceived stimulus (crossed area) is an integration of frequency amplitudes.
        Its integral domain is bounded by the densities of display pixels ($\displayBand$) and retinal receptors ($\retinalBand$).
        }
    \label{fig:integration_illustration}
\end{figure}

As shown in \Cref{fig:integration_illustration}, because of the finite display angular resolution and the finite retinal receptor densities, the non-zero frequency in \Cref{eq:integral} lies in a finite domain as well,
\begin{align}
	\int_{\lvert \contentFreqVec \rvert < \min\left(\displayBand, \retinalBandSup\right)} \csf\left(\lvert \contentFreqVec \rvert,\luminance\right)
	\lvert \contentAmp\left( \contentFreqVec \right) \rvert \mathrm{d}\contentFreqVec,
	\label{eq:integralFreqClamped}
\end{align}
where $\displayBand$ is the display band from the pixel density and eye-panel distance ($0.5$ cycle per pixel (CPP) in our display);
$\retinalBandSup$ is the supremum of the foveated retinal band.

The amplitude $\contentAmp$ in \Cref{eq:integralFreqClamped} will become spatially-variant if combined with foveated vision (\Cref{eqn:eccentricity}).
That requires the evaluation of spatially-variant spectrum,
such as windowed Fourier or wavelet,
which is prohibitively expensive in a real-time system.
Therefore, we evaluate local contrast $\pointContrast$, which can be more efficiently computed than $\contentAmp$, of every pixel:
\begin{align}
	& \specturmIntegralClampedOverall(\gazeVec,\image) \triangleq \int_{\imageSpaceVec \in \image}\overbracket{\int_{\lvert \contentFreqVec \rvert < \clampedBand(\gazeVec,\imageSpaceVec)}\csf\left(\contentFreqVec,\luminance\right)\pointContrast(\imageSpaceVec,\contentFreqVec,\image)
		\mathrm{d}\mathbf\contentFreqVec}^{\specturmIntegralClamped(\gazeVec, \imageSpaceVec,\image)} \mathrm{d}\imageSpaceVec,
	\label{eq:integralSpatialVariant} \\
	& \clampedBand(\gazeVec,\imageSpaceVec) = \min\left(\displayBand,\retinalBand(\gazeVec,\imageSpaceVec)\right) \nonumber
\end{align}
where $\pointContrast$ is the local contrast of $\imageSpaceVec=(\imageSpaceX,\imageSpaceY)$ of $\image$ under the frequency $\contentFreqVec$.
$\gazeVec=(\gazeX,\gazeY)$ is the tracked gaze position on the screen space.
$\specturmIntegralClamped$ is the corresponding sensitivity value for a spatial position $\imageSpaceVec$,
thus the frequency is integrated within a narrower foveated retinal band $\retinalBand(\gazeVec,\imageSpaceVec) = \eccentricityMask(\gazeVec-\imageSpaceVec)$
according to the gaze position and spatial position as shown in~\Cref{eqn:eccentricity_mask}.
By applying clamping with $\retinalBand(\gazeVec,\imageSpaceVec)$, the eccentricity variances on the contrast sensitivity function are considered.
Note that our model is for importance-based decision than pixel rendering so we assume the visibility of certain spatial frequencies is defined by receptor density.
The peripheral vision can ``reason'' (other than directly perceive) higher-than-receptor-frequency information from the aliasing on the receptors \cite{Wang:1997:AFE}. Therefore, fully approximate the value by clamping the integration with the receptor frequency may cause loss of importance in the periphery. Thus, we introduced a constant $\weberWeight$ in \Cref{eq:poppingIntensity} that compensates the peripheral region with low importance.

\section{Bandpass filtering}
\label{sec:sup:bandpass}

To analytically compute the temporal consistency considering not only the content but the retinal receptors and display capability, we discretize the $\specturmIntegralClamped$ in \Cref{eq:integralSpatialVariant}. Specifically, we perform a series of bandpass filtering of $\image$ to first obtain the gaze- and content-aware pixel-wise sensitivity:
\begin{align}
    \specturmIntegralClamped(\gazeVec, \imageSpaceVec, \image) 
    \approx\sum_{i=0}^{\bandStepSize - 1}\csf\left(\contentFreqVec_i,\luminance\right)\pointContrast(\imageSpaceVec,\contentFreqVec_i,\image),
\label{eq:discrete_importance}
\end{align}
where we divide the frequency domain of integral in \Cref{eq:integralSpatialVariant} into $\bandStepSize$ bands, where $\contentFreqVec_i = 2^{i\frac{\clampedBand(\gazeVec,\imageSpaceVec)}{\bandStepSize}}$ are the representative frequencies located at the mid-point of every band (following \cite{Parraga:2005:EAS}).
$\image_i$ is the $\contentFreqVec_i$-filtered version of $\image$.
\Cref{fig:bandpass_filtered_images} visualizes $\specturmIntegralClamped$ across different bands $\contentFreqVec_i$.
The contrast $\pointContrast$ at point $\imageSpaceVec$ is defined as
\begin{align}
\pointContrast(\imageSpaceVec,\contentFreqVec,\image) = \frac{\contrastSensitivityFilter_{\contentFreqVec}(\imageSpaceVec,\image)}{\contrastSensitivityFilter_0(\imageSpaceVec,\image)}.
\label{eq:point_contrast}
\end{align}
Here $\contrastSensitivityFilter_{\contentFreqVec}$ is the approximated local frequency of the $\contentFreqVec$-filtered $\image$, as detailed in \cite{Peli:1990:CCI}, and $\contrastSensitivityFilter_{0}$ is $\contrastSensitivityFilter_{\contentFreqVec}$ with $\contentFreqVec_i = 0$.

\end{document}